\documentclass[aps,prb,prl,twocolumn,superscriptaddress,nopacs,amsmath,amssymb,nofootinbib]{revtex4-2}
\usepackage[utf8]{inputenc}
\usepackage{mathptmx}
\usepackage[T1]{fontenc}
\usepackage{float}

\usepackage{amssymb}
\usepackage{graphicx} 
\usepackage{esint}

\DeclareFontShape{OMX}{cmex}{m}{b}{<-> cmexb10}{}
\SetSymbolFont{largesymbols}{bold}{OMX}{cmex}{m}{b}

\usepackage{bm}

\usepackage{hyperref} 
\usepackage[nameinlink]{cleveref}
\crefname{subsection}{subsection}{subsections}

\usepackage{color}
\usepackage{booktabs}
\usepackage{makecell}
\usepackage{soul}
\usepackage{sidecap} 
\usepackage{gensymb} 

\usepackage{enumitem}
\setlist[itemize]{leftmargin=*}

\renewcommand\[{\begin{equation}}
\renewcommand\]{\end{equation}} 

\makeatother

\usepackage{babel}
\usepackage{physics}

\makeatletter
\providecommand*\bigcdot{\mathpalette\bigcdot@{.5}}
\providecommand*\bigcdot@[2]{\mathbin{\vcenter{\hbox{\scalebox{#2}{$\m@th#1\bullet$}}}}}
\makeatother

\usepackage{mathtools}
\usepackage{scalerel}

\makeatletter
\DeclareFontFamily{OMX}{MnSymbolE}{}
\DeclareSymbolFont{MnLargeSymbols}{OMX}{MnSymbolE}{m}{n}
\SetSymbolFont{MnLargeSymbols}{bold}{OMX}{MnSymbolE}{b}{n}
\DeclareFontShape{OMX}{MnSymbolE}{m}{n}{
    <-6>  MnSymbolE5
   <6-7>  MnSymbolE6
   <7-8>  MnSymbolE7
   <8-9>  MnSymbolE8
   <9-10> MnSymbolE9
  <10-12> MnSymbolE10
  <12->   MnSymbolE12
}{}
\DeclareFontShape{OMX}{MnSymbolE}{b}{n}{
    <-6>  MnSymbolE-Bold5
   <6-7>  MnSymbolE-Bold6
   <7-8>  MnSymbolE-Bold7
   <8-9>  MnSymbolE-Bold8
   <9-10> MnSymbolE-Bold9
  <10-12> MnSymbolE-Bold10
  <12->   MnSymbolE-Bold12
}{}

\let\llangle\@undefined
\let\rrangle\@undefined
\DeclareMathDelimiter{\llangle}{\mathopen}%
                     {MnLargeSymbols}{'164}{MnLargeSymbols}{'164}
\DeclareMathDelimiter{\rrangle}{\mathclose}%
                     {MnLargeSymbols}{'171}{MnLargeSymbols}{'171}
\makeatother

\renewcommand{\footnoterule}{\kern -1ex\rule{\linewidth}{0.5pt}\\\vspace{1ex}}

\usepackage[makeroom]{cancel} 

\usepackage{titlesec}
\titlespacing{\subsection}{0pt}{\baselineskip}{0.5\baselineskip}

\usepackage{accents}

\DeclareMathAlphabet{\mathcal}{OMS}{cmsy}{m}{n}

\usepackage{dsfont}

\usepackage[normalem]{ulem}
\usepackage{comment}
\usepackage{wrapfig}
\usepackage[title]{appendix}
\usepackage{orcidlink}

\usepackage{subfigure}

\usepackage{xr}
\makeatletter

\newcommand*{\addFileDependency}[1]{
\typeout{(#1)}
\@addtofilelist{#1}
\IfFileExists{#1}{}{\typeout{No file #1.}}
}\makeatother

\newcommand*{\myexternaldocument}[1]{%
\externaldocument{#1}%
\addFileDependency{#1.tex}%
\addFileDependency{#1.aux}%
}

\myexternaldocument{supplement}

\begin{document}
{
\footnote{Notice: This manuscript has been coauthored by UT-Battelle, LLC, under Contract No. DE-AC0500OR22725 with
the U.S. Department of Energy. The United States Government retains and the publisher, by accepting the article for
publication, acknowledges that the United States Government retains a non-exclusive, paid-up, irrevocable, world-wide
license to publish or reproduce the published form of this manuscript, or allow others to do so, for the United States
Government purposes. The Department of Energy will provide public access to these results of federally sponsored
research in accordance with the DOE Public Access Plan (\href{http://energy.gov/downloads/doe-public-access-plan}{http://energy.gov/downloads/doe-public-access-plan}).}
}

\title{Mechanism of charge transfer and electrostatic field fluctuations in complex metallic alloys}
\author{Wai-Ga D. Ho}
\affiliation{Department of Physics and National High Magnetic Field Laboratory, Florida State University, Tallahassee, FL, USA}
\author{Wasim Raja Mondal\orcidlink{0000-0002-5652-9785}}
\affiliation{Department of Physics and Astronomy, Middle Tennessee State University, Murfreesboro, Tennessee 37132, USA}
\author{Swarnava Ghosh\orcidlink{0000-0003-3800-5264}}
\affiliation{National Center for Computational Sciences, Oak Ridge National Laboratory, Oak Ridge, Tennessee 37830, USA}
\author{Hanna Terletska\orcidlink{0000-0001-6215-9664}}
\affiliation{Department of Physics and Astronomy, Middle Tennessee State University, Murfreesboro, Tennessee 37132, USA}
\author{Ka-Ming Tam\orcidlink{0000-0002-2273-000X}}
\affiliation{Department of Physics and Astronomy, Louisiana State University, Baton Rouge, LA 70803, USA}
\author{Mariia Karabin\orcidlink{0000-0003-0081-8497}}
\affiliation{Department of Physics and Astronomy, Middle Tennessee State University, Murfreesboro, Tennessee 37132, USA}
\author{Markus Eisenbach\orcidlink{0000-0001-8805-8327}}
\affiliation{National Center for Computational Sciences, Oak Ridge National Laboratory, Oak Ridge, Tennessee 37830, USA}
\author{Yang Wang\orcidlink{0000-0002-9837-5796}}
\affiliation{Pittsburgh Supercomputing Center, Carnegie Mellon University, Pittsburgh, Pennsylvania 15213, USA}
\author{Vladimir Dobrosavljevi\'{c}\orcidlink{0000-0002-7525-2662}}
\affiliation{Department of Physics and National High Magnetic Field Laboratory, Florida State University, Tallahassee, FL, USA}

\begin{abstract}


Complex metallic alloys exhibit rich disorder-driven electronic, magnetic, and vibrational behavior arising from strong chemical disorder, giving rise to unconventional structure-property relationships not typically observed in conventional crystalline solids. These relationships have motivated their exploration for applications in extreme-environment structural components, catalysis, spintronics, and thermoelectric energy conversion. Despite their technological relevance, the microscopic nature of charge redistribution and electrostatic fluctuations intertwined with complex disorder phenomena remains a long-standing problem that is still incompletely understood. To address this issue, we develop a theoretical framework that uncovers the universal statistical trend of disorder-driven charge transfer and Madelung-field fluctuations in disordered alloys. Our analytical formalism
demonstrates that local charge transfer and electrostatic potentials naturally exhibit Gaussian-like statistics and universal linear charge–potential (qV) correlations that emerge directly from the underlying disorder landscape. Our theory pinpoints the physical origin of these correlations in the interplay between electronic screening and impurity scattering, while elucidating how their statistical properties are controlled by carrier density, disorder strength, and compositional complexity. We further derive universal scaling relations governing the evolution of (qV) trends across binary and multi-component alloys, including high-entropy materials as examples. Large-supercell-based density-functional theory (DFT) calculations show good quantitative agreement with the predicted statistical behavior across representative metallic alloys. Beyond resolving longstanding questions surrounding charge-transfer statistics in disordered metals, our work establishes a computationally efficient framework for incorporating disorder-driven electrostatic fluctuations into effective-medium electronic-structure theories. Our unified statistical-physics framework provides a pathway toward the predictive design of high-performance multifunctional alloys.

\end{abstract}
\maketitle


\section{INTRODUCTION}

An in-depth understanding of the role of disorder in solids is one of the most difficult conundrums in condensed matter physics\cite{annurev:/content/journals/10.1146/annurev-conmatphys-031218-013433, DavidLPrice_2003,RevModPhys.80.1355,RevModPhys.46.465}
While idealized solids are often described by periodic crystalline lattices, real materials frequently exhibit deviations from perfect order. Such disorder may arise from random substitution of atomic species (chemical disorder), lattice distortions (structural disorder), or their coexistence. These features introduce additional complexity, including spatially fluctuating local environments that modify force constants and local bonding \cite{cain1999phase,Biswasoffdiagonal,PhysRevB.99.134203,PhysRevB.90.094208,app8122401,PhysRevB.55.4149}), breakdown of translational symmetry in chemically disordered alloys \cite{li2026unfolding}, disorder-induced modifications of electronic band structure and Fermi surface topology \cite{eibert2025effects}, enhanced scattering and transport renormalization due to disorder \cite{rossi2024disorder}, and strong coupling between lattice, charge, and spin degrees of freedom leading to localization and emergent phases \cite{shin2023structural}, all of which complicate predictive modeling of material response. These forms of randomness can dramatically alter material behavior, often giving rise to rich and unexpected physical phenomena, including Anderson localization of electronic states in disordered potentials \cite{Anderson1958}, disorder-driven metal–insulator transitions \cite{abrahams1979scaling}, emergent magnetic frustration and spin-glass behavior in chemically disordered systems \cite{binder1986spin}, and unconventional superconductivity in strongly disordered or compositionally complex materials \cite{sacepe2011localization}.

\begin{figure*}[t!]
    \centering
    \includegraphics[width=0.94\textwidth]{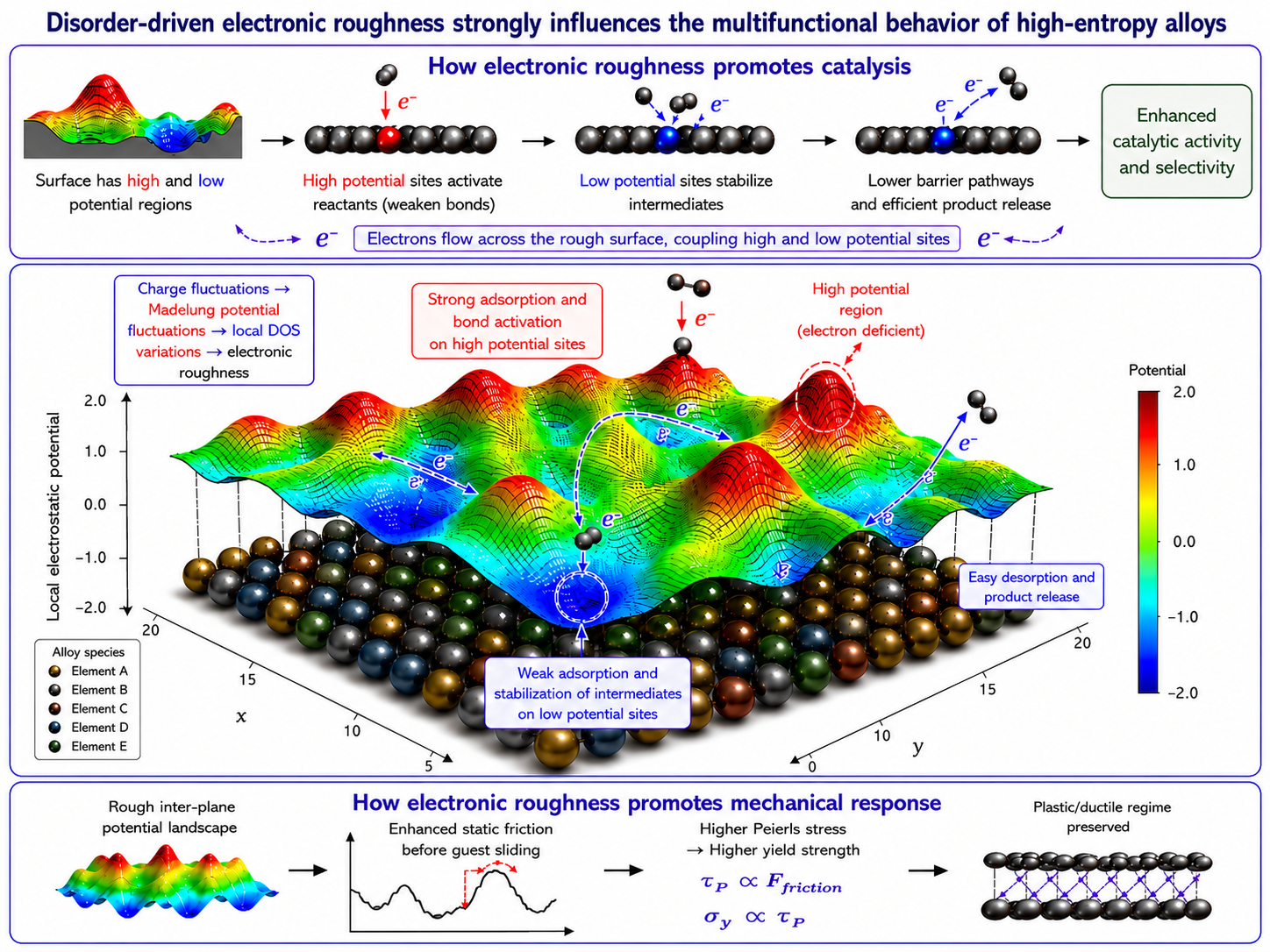}
    \caption{Schematic demonstration of how disorder-driven charge fluctuations and Madelung potential variations in high-entropy alloys (HEAs) facilitate electronically rough energy landscapes that influence both catalytic activity and mechanical response.}
\end{figure*}

High-entropy alloys (HEAs), are a class of disordered solids comprising multiple elements in near-equiatomic or high concentrations and often form single-phase solid solutions \cite{cantor2004,HEA_Yeh2004,george_high-entropy_2019,george:2020,varvenne:2017}. Their phase stability has been widely associated with the large configurational entropy of mixing, which can offset enthalpic driving forces for phase separation and thereby suppress the formation of competing ordered or multiphase states \cite{HEA_Yeh2004}. Since their emergence less than two decades ago~\cite{HEA_Yeh2004,cantor2004}, they have garnered intense interest due to their exceptional and often tunable properties, including superior mechanical strength, corrosion resistance, and unique electromagnetic or optical responses~\cite{HEA_Yeh2004,cantor2004,HEA_SC,li_mechanical_2019,george_high-entropy_2019,zhou2023}. Many conventional alloys readily undergo precipitation under stress or thermal exposure \cite{ghosh2020influence,ghosh2021precipitation}; in contrast, several HEAs exhibit remarkably stable solid-solution phases. These attributes position HEAs as promising candidates for multifunctional applications in aerospace \cite{Dada19,DADA202143,ZHUO20241097,MENON2025130091,Dixit2022,liu2019dislocation,ponga2022effects}, energy storage\cite{FAN2025103954,Ouyang2024,D3YA00319A}, and biomedical technologies\cite{RASHIDYAHMADY2023100009,Liu2022-lo,met12111940,Liu2022-qa,10.3389/fbioe.2022.977282,RASHIDYAHMADY2023100009,Chang2024-dn,LV2025114721}, fueling global efforts to optimize and commercialize these complex systems.

{Experimental design of HEAs is often impeded by the challenges involved with vast compositional search space, difficulties in compositional control and phase identification, challenges in experimentally disorder modeling, including lattice distortion and short-range ordering, and involvement of rare or expensive elements\cite{Tsai03072014, MIRACLE2017448}. Prior computational synthesis or design knowledge, along with a computationally driven, in-depth understanding of HEAs, can greatly accelerate the experimental design of new HEAs and enable industrial-scale production at lower cost.

Despite such computational demand for HEAs,
a quantitatively computational simulation-driven predictive description of their structural and electronic properties remains incomplete. Prior studies have highlighted the central role of chemical disorder and local lattice distortions in governing electronic structure and associated functionalities, yet a unified framework that consistently links composition, disorder, and emergent electronic response is still lacking \cite{HEA_Yeh2004,george_high-entropy_2019,miracle2017critical,pickering2016high,ghosh2024violation,anand2023order}. This limitation continues to constrain the development of transferable, physics-based design strategies for HEAs. Elucidating these fundamental uncertainties in disordered solids is necessary to move the field beyond empirically guided trial-and-error methodologies toward a rigorously predictive framework. This transition requires a quantitatively robust, mechanistic understanding of the governing physics across relevant length and time scales, thereby enabling the rational design, optimization, and controlled deployment of these materials in technologically relevant regimes.
Understanding charge transfer and electrostatic field fluctuations in high-entropy metallic alloys is essential for rationalizing their performance across a range of functional applications. Charge-density perturbations in crystalline solids may extend over several unit cells \cite{ghosh2022spectral}. Chemical disorder in these systems gives rise to substantial spatial variations in local charge density and electrostatic potential, which directly influence adsorption energetics and reaction pathways in catalytic environments, enabling enhanced activity and selectivity relative to conventional alloys \cite{batchelor2019high,yao2018carbothermal,BATCHELOR2019834,PEDERSEN2021100651,doi:10.1126/science.abn3103,Li2026,Wang2025}. Beyond catalysis, such local electronic heterogeneity is expected to play a central role in electrochemical energy storage and conversion\cite{FATHYUNES20263034,en15197130,10.3389/fenrg.2022.862551,Shin2023,Dong2024-lg,LI2024103468,ZHANG20205367,DANGWAL2024115774}, where fluctuations in interfacial potential landscapes modulate reaction kinetics and ion transport \cite{Li2020}, as well as in thermoelectric materials, where disorder-induced potential variations can enhance carrier scattering and reduce thermal conductivity \cite{doi:10.1126/science.abe1292,TANG20241641,Liu2025,10.1063/1.4935489,Ren_2024,Xia2024,met8100781, Karati2019,Wang2021,Fan04052017,D3MH02181E}). In magnetic systems, variations in local charge and bonding environments give rise to complex exchange interactions and emergent magnetic behavior \cite{KUMARI2022169142,Lan2022-nd,Chen2024,Kitagawa2024,Lan2022-av,PhysRevB.96.014437,LI201755,https://doi.org/10.1002/adem.201700048,PhysRevMaterials.4.014402,ZUO20181,moitzi2025inversion}, while in structural and corrosive environments, electrostatic heterogeneity influences defect energetics, diffusion, and passivation processes \cite{cryst13081231,MIRACLE2017448,QUIAMBAO2019362,met13020363,TSAI20134887,ZHANG2025113318}. Despite these broad implications, a mechanistic understanding of how compositional disorder governs charge redistribution and the resulting electrostatic landscape remains limited, hindering the development of predictive, physics-based design strategies for HEAs.


Various computational methods, including Density Functional Theory (DFT)-based {\it ab initio} methods\cite{ZHANG2022104059,HoulngSiGeSn}, CALPHAD\cite{Senkov2015,Feng2021}, molecular dynamics (MD)\cite{10.1063/5.0025310,SHARMA2017292,Singh2021}, kinetic and atomistic Monte-Carlo method\cite{YE2022130907,Liu2015}, high-throughput
method\cite{Moorehead2025,10.3389/fmats.2020.00290,LEE2022111259}, and machine learning (ML)\cite{Zhang2022,HUANG2019225,Singh2023, QIAO2021160295,https://doi.org/10.1002/adem.202402504}, have been employed to investigate high entropy alloys. These methods have their own advantages and disadvantages for accurately calculating the mechanical, electronic, and vibrational properties of HEAs at a lower computational cost. For example, Kohn-Sham DFT, by its construction, applies only to periodic solids, and its applicability becomes questionable for disordered cases. MD simulations suffer from the availability of reliable interatomic potentials. Due to the vast configurational space arising from multiple principal elements in HEAs, Monte Carlo simulations require extensive sampling to achieve convergence, especially at low temperatures, making calculations computationally costly, especially for five-element complex HEAs. CALPHAD is mainly devoted to exploring phase stability and relies on databases. The performance of the ML model also heavily depends on the availability and quality of HEAs data. Though most of these methods are mainly devoted to exploring computational synthesis, structural stability, mechanical, and thermodynamic properties, a comprehensive study using these methods to achieve a fundamental understanding of the charge-transfer mechanism in HEA remains scarce.

Due to challenges involved in the exact solution of the disordered problem, development of efficient computational methods for disordered solids has still been an active area of research for more than five decades. As a promising alternative to directly addressing the disordered problem, the Coherent Potential Approximation was first introduced in 1967\cite{CPA} as a computationally inexpensive and easy-to-implement scheme based on an effective-medium theory. It maps the complex system onto an analogous, translationally invariant single-site problem, and each atomic constituent is embedded in a "coherent" effective medium that represents the configurationally averaged environment. While the CPA offers exceptional computational efficiency and conceptual clarity, it is inherently a local mean-field theory. By construction, it neglects the stochastic fluctuations of the local chemical environment—thereby discarding the unique local ``identity'' and associated energetics of individual atoms within the alloy. Following the idea of CPA, many methods, such as first-principles combined CPA\cite{PhysRevB.5.2382, 10.1143/PTPS.53.1}, virtual crystal approximation (VCA)\cite{HUANG2021109859, WANG2021128754, LIU2019109161}, and special quasi-random structures (SQS)\cite{PhysRevB.69.214202,VANDEWALLE201313, PhysRevLett.65.353}, were developed and applied to disordered alloys.

As another alternative computational scheme, the disordered-solid problem has been reformulated in terms of multiple-scattering theory (MST), and first-principles-based Green's function methods, such as the Korringa-Kohn-Rostoker Green function (KKR-GF)\cite{KORRINGA1947392,PhysRev.94.1111}, have been developed. The central idea is to compute the single-particle Green's function rather than calculating Kohn–Sham orbitals by direct diagonalization of the Kohn–Sham Hamiltonian. However, this method still suffers from cubic scaling in terms of computational cost. A supercell approach based on the MST, called the Locally Self-Consistent Multiple Scattering (LSMS) method \cite{lsms1,lsms2}, has been developed. One of the main advantages of LSMS over CPA is that it achieves linear scaling, whereas KKR still suffers from cubic scaling in terms of computational cost. 



Enabled by modern parallel architectures, the LSMS method renders large-scale supercell calculations (in the order of $10^3-10^5$ atoms) computationally tractable. Crucially, by accounting for the unique local environment of every site during the self-consistent DFT cycle, this method preserves the intraspecies statistics and environmental fluctuations that the CPA fails to capture, providing a more rigorous description of the disordered state.

Beyond supercell-averaged electronic properties, our recent DFT calculations of HEAs~\cite{MuSTpaper2022} reveal a remarkable consistency in the stochastic distributions governing local environments across diverse alloy systems. Specifically, by analyzing two fundamental site-dependent quantities---the electronic charge transfer ($q$) and the local electrostatic "Madelung" potential ($V$)---we have shown both quantities exhibit qualitative behaviors and scaling trends consistent with earlier studies of conventional binary and ternary alloys \cite{oldAlloys1,oldAlloys2,oldAlloys3,oldAlloys4}. These include Gaussian-like intraspecies distributions for $q$ and $V$ when considered independently and most significantly a robust, statistically linear correlation between them, hereafter referred to as the $qV$ relationship.

The emergence of this $qV$ linearity across various chemical species suggests the existence of universal mechanisms underlying charge transfer and Madelung field statistics within the broader class of disordered metals. However, the governing principles behind this coupling remain elusive. Despite the success of the DFT in uncovering these trends, several open questions persist: specifically, the physical origin of the $qV$ trendline parameters (slope and intercept), their sensitivity to specific chemical descriptors, and the minimal theoretical framework required to describe these emergent statistics without relying on exhaustive, computationally demanding supercell simulations.

The primary objective of this work is to provide a rigorous foundation for the $qV$ relation. We achieve this by first constructing a minimal model that captures the essential features common to all chemically disordered alloys
\footnote{Following the scope of \cite{MuSTpaper2022} and related studies \cite{oldAlloys1,oldAlloys2,oldAlloys3,oldAlloys4}, we restrict our analysis to pure chemical disorder. While we recognize that structural defects and non-crystalline phases (e.g., amorphous or secondary phases) are critical to alloy microstructure and mechanical properties, we maintain that the single-crystal phase presents a more fundamental problem. Establishing a clear understanding of this ideal case is a necessary prerequisite for treating more complex structural deviations.}. 
Using this simplified theory, we demonstrate that the statistical features previously observed using DFT calculations can be reproduced and further analyzed to reveal their physical origins. Ultimately, this framework allows us to isolate the variables governing these trends and arrive at a comprehensive description of charge-potential correlations in disordered systems.

As a final motivation, we highlight that recent work \cite{MuSTpaper2022} demonstrated how $qV$ trends derived from DFT calculations can be integrated back into the CPA’s effective medium. This integration provides systematic corrections that significantly enhance the accuracy of CPA outputs—including total energies, effective charge transfers, and Madelung potentials, while preserving the method's inherent computational efficiency. Consequently, by clarifying the underlying physics of these $qV$ correlations, our work offers a pathway toward improved CPA-based methodologies. Specifically, our model provides a means to furnish these essential $qV$ trends without the prohibitive computational demands of large supercell DFT simulations. Potential extensions and implementations of this enhanced CPA approach are discussed in the concluding sections of this paper.



\section{THEORETICAL FRAMEWORK} \label{sec:theory}


\subsection{Disordered Hartree model for random alloys} \label{subsec:DHmodel}

For a simplified description of the chemically disordered alloy, we consider just a single-band model of spinless fermions which occupy an $N$-site lattice with moderate impurity disorder. Our full Hamiltonian is given by
\begin{equation}
    \hat{H} = \hat{H}_0 + \hat{V},
    \quad {\rm where} \quad 
    \left\{
    \begin{aligned}
        \\[-0.35cm]
        \hat{H}_0 
        & = \sum_{\vb{k} } \xi_{\vb{k}} \,  \hat{c}^{\dagger}_{\vb{k}}\hat{c}_{\vb{k}};
        \\
        \hat{V} 
        & = \sum_i (\epsilon_i + \phi_i ) \, \hat{c}^{\dagger}_i \hat{c}_i.
    \end{aligned}
    \right.
    \label{eq:H}
\end{equation}
The bare term $\hat{H}_0$, which is trivially diagonal in momentum $\vb{k}$-space, captures nearest-neighbor tight-binding kinetics
\footnote{Although our choice to use this tight-binding representation is made rather arbitrarily here, both for its working simplicity and its demonstrative usefulness, we note that much of the work described in this paper is not particularly sensitive to finer bandstructure details.}.
Taking $\vb{a} \in \{\vb{a}_1,\vb{a}_2,...,\vb{a}_d\}$ as the primitive lattice vectors of a generic $d$-dimensional crystal, our one-band dispersion relation is then given by $\xi_{\vb{k}} = -2 \, t \sum_{\vb{a}} \cos(\vb{k}\cdot\vb{a})$, with tunneling amplitude $t$.
The local perturbations in $\hat{V}$ consist of two distinct contributions. First are the random site-energies $\epsilon_i$ assigned to each site $i$ at position $\vb{r}_i$, which represent the disordered background of chemical impurities. Second are the on-site potentials $\phi_i$, which arise from the attractive ionic potential and the inter-electron Coulomb repulsion treated at the Hartree mean-field level. Specifically, $\phi_i$ corresponds to the Madelung potential exerted by all other sites $j$ on site $i$:
\begin{equation}
\phi_i = \sum_{j \neq i} V^C_{ij} (n_j - \langle n \rangle), \quad \text{where} \quad V^C_{ij} = \frac{1}{|\vb{r}_i - \vb{r}_j|}. 
\label{eq:MadPot}
\end{equation}
Here, $V^C_{ij}$ represents the intersite Coulomb potential, $\langle n \rangle$ is the average electron density, and $n_j - \langle n \rangle$ is the excess charge on site $j$. To enforce global charge neutrality, $\langle n \rangle$ serves as a compensating uniform positive background in (\ref{eq:MadPot}).


\subsection{General perturbative approach} \label{subsec:PTgen} 

Using standard perturbation theory, we may then relate the electronic profiles of the bare ($\hat{H}_0$) and the full ($\hat{H}$) problems by constructing their 
Green's functions
\begin{equation}
    \hat{G}_{(0)} = (\omega^+ - \hat{H}_{(0)} )^{-1},
    \; {\rm where} \;
    \omega^+ = \lim_{\eta \rightarrow 0^+} (\omega + i\eta)
    \label{eq:GF}
\end{equation}
where the subscript ``(0)'' denotes with or without ``0'', and using the Dyson equation
\begin{equation}
    \hat{G} = \hat{G}_0 + \hat{G}_0 \hat{V }  \hat{G}_0 + \hat{G}_0 \hat{V}  \hat{G}_0 \hat{V}  \hat{G}_0 + \cdots
\end{equation}
to express $\hat{G}$ as a series in $\hat{G}_0$ and $\hat{V}$.

Now the local densities of states can be obtained from the local Green's functions (i.e. the on-site matrix elements of $\hat{G}_{(0)}$ above) by extracting their imaginary parts. And integrating these up to the chemical potential $\mu[\langle n\rangle]$ will further provide us with the local charge. For either the bare or the full problem, we have
\begin{equation}
    n_{(0)i} = \langle \hat{c}^{\dagger}_i \hat{c}_i \rangle_{(0)}   
    = -\frac{1}{\pi} \int_{-\infty}^\mu \textrm{Im}[\hat{G}_{(0)}]_{ii} \, \dd \omega,
\end{equation}
where the angled braces around the number operator $\hat{n}_i = \hat{c}^{\dagger}_i \hat{c}_i$ are understood to represent its quantum average, and in particular, the charge of the bare model is $n_{0i} = \langle n \rangle$. 
Starting from the Dyson equation, we can derive how the local charge density of the bare model is modified by perturbations to produce the full self-consistent solution. These local charge deviations---or the excess charge $\delta n_i$---are given by a first-order expansion in the on-site perturbation energy $\widetilde{\epsilon}_i$:
\begin{equation}
    {\delta n}_{i} = n_i - \langle n \rangle = 
    A \widetilde{\epsilon}_i + \sum_{j \neq i} B_{ij} \widetilde{\epsilon}_j + \mathcal{O}[\widetilde{\epsilon}^2],
    \label{eq:DH_eqn1}
\end{equation}
with the on-site perturbation energy defined as:
\begin{equation}
\widetilde{\epsilon}_i = \epsilon_i + \phi_i, \label{eq:epsTilde}
\end{equation}
which represents the eigenvalues of the perturbation potential $\hat{V}$ and accounts for both the random site-energies and the local electrostatic potential. The expansion coefficients are given by:
\begin{align}
    A & = -\frac{1}{\pi} \int_{-\infty}^{\mu}\textrm{Im} [{G_0}^2_{ii} ] \, \dd \omega ,
    \label{eq:coefficientAquantum}
    \\
    B_{ij} & = - \frac{1}{\pi} \int_{-\infty}^{\mu}\textrm{Im}[ {G_0}_{ij}^2] \, \dd \omega 
    \quad (\textrm{for $i \neq j$}).
    \label{eq:coefficientBijquantum}
\end{align}
These coefficients depend exclusively on the electronic structure of the reference (bare) model and collectively constitute the real-space representation of the Lindhard response function. In Eq. (\ref{eq:DH_eqn1}), we have separated the response into local ($A$) and nonlocal ($B_{ij}$) contributions. 


\subsection{Linearizing the problem and achieving self-consistency} \label{subsec:linSC}

To recover the linear $qV$ trends characterizing the joint statistics between charge transfer $\delta n$ and Madelung field $\phi$, it is sufficient to truncate our expansion (\ref{eq:DH_eqn1}) to first-order in $\widetilde{\epsilon}$'s. Together with our original definition for the Madelung potential (\ref{eq:MadPot})---which we now notice contains a $\delta n_j$ nested within its sum---this provides us with a pair of equations which must be solved in a self-consistent manner. 

In our linearized theory, self-consistency is straight\-forward to achieve by lattice Fourier transforming our system of equations,(\ref{eq:MadPot}) and (\ref{eq:DH_eqn1}), together with expression (\ref{eq:epsTilde}), 
and reduce them to a more algebraic form over the momentum space,
\begin{equation}
    \left\{
    \begin{aligned}
        \delta n_{\vb{k}} & =  A \widetilde{\epsilon}_{\vb{k}} + B_{\vb{k}} \widetilde{\epsilon}_{\vb{k}},  
       \quad {\rm with} \quad \widetilde{\epsilon}_{\vb{k}} = \epsilon_{\vb{k}} + \phi_{\vb{k}};
        \\[0.2cm]
        \phi_{\vb{k}} & = V^C_{\vb{k}} \delta n_{\vb{k}}.
    \end{aligned}
    \right.
    \label{eq:FT_sysEqns}
\end{equation}
Their $\vb{k}$-space solution is then given by
\begin{equation}
    \left\{
    \begin{aligned}
    \delta n_{\vb{k}}
     & = M_{\vb{k}} \epsilon_{\vb{k}},
    & {\rm with} \quad 
    M_{\vb{k}} 
    & = \frac{A+B_{\vb{k}} }{1- (A+B_{\vb{k}}) V_{\vb{k}}^C };
    \\[0.2cm]
    \phi_{\vb{k}}
    & = \widetilde{M}_{\vb{k}} \epsilon_{\vb{k}},
    & {\rm with} \quad 
    \widetilde{M}_{\vb{k}} 
    & = V^C_{\vb{k}} M_{\vb{k}}. 
    \label{eq:Mk}
    \end{aligned}
    \right.
\end{equation}
%
In the site-space expression, (\ref{eq:Mk}) becomes
\begin{align}
    \delta n_i & = \sum_j M_{ij} \epsilon_j;
    \label{eq:δni}
    \\
    \phi_i & = \sum_j \widetilde{M}_{ij} \epsilon_j,
    \label{eq:φi}
\end{align}
where $M_{ij}$ (or $\widetilde{M}_{ij}$) is the inverse Fourier transform of $M_{\vb{k}}$ (or $\widetilde{M}_{\vb{k}}$).
By inspection, the prior result reveals that $M_{ij}=\partial \delta n_i / \partial \epsilon_j $ and $\widetilde{M}_{ij}= \partial \phi_i / \partial \epsilon_j $; these $M$ and $\widetilde{M}$ objects introduced here shall therefore be understood to represent the (linear) response of the charge fluctuation $\delta n$ and Madelung fields $\phi$, respectively, to the disordered background of impurities $\epsilon$.

At the single impurity limit, the local charge correction and the electrostatic Madelung field at site $i$ due to a single impurity placed on site $j$ are given by 
\begin{align}
\delta n_i & = M_{ij} \epsilon_j \,\,\, {\rm with} \,\,\, M_{ij} = A \delta_{ij} + B_{ij}, 
\label{eq:δni_singleimp}
\\
\phi_i & = \widetilde{M}_{ij} \epsilon_j
        \,\,\, {\rm with} \,\,\, \widetilde{M}_{ij} = \frac{1}{N} \! \sum_{ \vb{k} \in \textrm{BZ} }  
        \underbrace{ 
        \frac{A}{\frac{1}{V^C_{\vb{k}}} + A} 
        }_{ 
        \mathclap{ \hspace{1.2cm} \widetilde{M}_{\vb{k}} \textrm{ (classical)} }
        } \, e^{i \vb{k} \cdot { (\vb{r}_i - \vb{r}_j) } }.
        \label{eq:φi_classicalSingleImp}
\end{align}
as a result of equations (\ref{eq:DH_eqn1}), (\ref{eq:δni}) and (\ref{eq:φi}).
A careful analysis (for which we refer to \cref{app:FriedelOsc} for the details) reveals that
\begin{align}
         \delta n_i , M_{ij} 
         & \sim   
         \frac{ \cos{ \left[ 2 k_F |\vb{r}_{i} - \vb{r}_{j}| \right]  } }{ |\vb{r}_{i} - \vb{r}_{j}|^3 } + \mathcal{O}\left[ |\vb{r}_i - \vb{r}_j|^{-4} \right],
         \\[0.2cm]
         \widetilde{M}_{ij} & \sim e^{-|\vb{r}_{i} - \vb{r}_j|/\ell},
\end{align}
where $k_F$ is the Fermi wavevector and $\ell$ is the screen length. This behavior is analogous to the so-called Friedel oscillations produced around a single impurity scatterer embedded in an ideal Fermi gas. Since our linearized framework is an additive one, the way to construct the $M$ and $\widetilde{M}$ response functions for multiple impurities is rather straightforward, as shown in the next section.

\subsection{Disordered Hartree statistics for multiple impurities} \label{subsec:manyImps}



Consider a system of $N$ random impurities, characterized by a set of site-energies $\{\epsilon_1, \epsilon_2, \dots, \epsilon_N\}$ in the Hamiltonian (\ref{eq:H}). For each site $i$, we apply Eqs. (\ref{eq:δni}) and (\ref{eq:φi}) to compute the corresponding local charge transfer $\delta n_i$ and Madelung potential $\phi_i$. This process is repeated for $N_r$ realizations of the random site-energies to accumulate the local statistics for each disorder instance. The results for a converged sample size of $N_r = 10^6$ instances are presented in \Cref{fig:DHstats} for both quarter- and half-filled cubic lattices with equiprobable binary disorder.

At quarter-filling, screening is sufficiently weak that both the local charge transfer ($\delta n_i = \sum_j M_{ij} \epsilon_j$) and the Madelung potential ($\phi_i = \sum_j \tilde{M}_{ij} \epsilon_j$) are composed of a large number of random terms. Consequently, the Central Limit Theorem drives both quantities toward Gaussian statistics, as shown in the top-right panels of \Cref{fig:DHstats}. In contrast, the half-filled case (\Cref{fig:DHstats}, middle row) demonstrates that an increased carrier concentration enhances the screening effect. This more effectively shields local sites from distant impurities, restricting influential interactions to the nearest neighbors. While a broad Gaussian envelope remains visible, the statistics begin to develop more distinct features dependent on specific model parameters. 

The individual statistics of $\phi_i$ and $\delta n_i$ are combined in the bottom row of \Cref{fig:DHstats}. When these joint statistics are segregated by species and visualized as a scatter plot, they reveal a robust linear correlation. This verifies that our first-order treatment of the disordered Hartree model successfully recovers the linear $qV$ relationship observed in random fluctuations. These results—the Gaussian distributions in the weakly screened limit and the linear $qV$ trends—are in excellent qualitative agreement with prior DFT studies of both conventional and high-entropy alloys~\cite{MuSTpaper2022,oldAlloys1,oldAlloys2,oldAlloys3,oldAlloys4}. However, the fundamental question remains: what physical principles govern these statistics, and how can they be captured in a quantitatively accurate yet simple manner?

\begin{figure*}[t!]
    \centering
    \hspace{-0.12cm}
    \includegraphics[trim={0.3cm 0.2cm 0 0.8cm}, clip, width=1.005\linewidth]{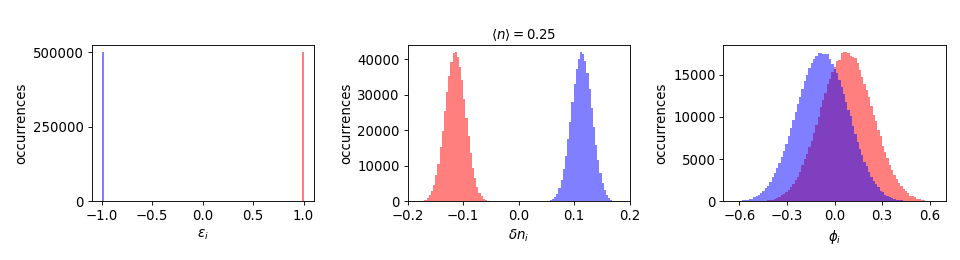}
    \\
    \includegraphics[trim={0.3cm 0.2cm 0 0.8cm}, clip, width=1.015\linewidth]{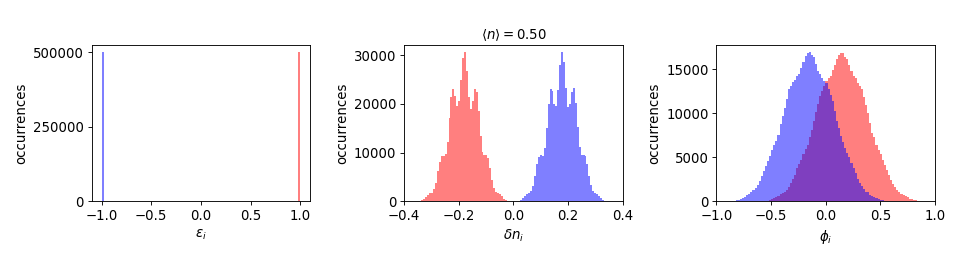}
    \\
    \begin{tabular}{c c}
        \includegraphics[trim={0.3cm 0.6cm 0 0.8cm}, clip, width=0.35\linewidth]{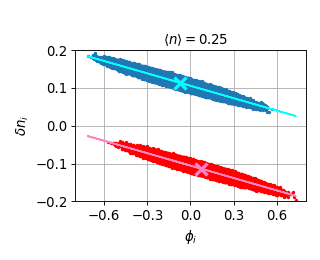}
        &
        \includegraphics[trim={0.3cm 0.6cm 0 0.8cm}, clip, width=0.35\linewidth]{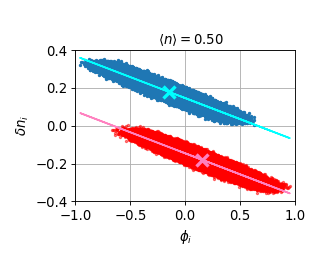}
    \end{tabular}    
    \caption{Local disordered Hartree statistics are sampled for site $i\equiv$ origin in the quarter-filled (top row) and half-filled (middle row) cubic models, both with equiprobable binary disorder of strength $w = |\epsilon_{j}| = 1$. These 100-bin histograms for the (left column) local charge correction $\delta n_i$, (middle column) impurity potential $\epsilon_i$, and (right column) Madelung field $\phi_i$ all share a converged sample size of $N_r = 10^6$ disorder instances.  In the bottom row, the joint $(\phi_i,\delta n_i)$ data points are then combined and scatterplotted for both fillings, revealing their linear $qV$ relationships, the associated means (crosses) and least squares fit lines also being displayed. All plots in this figure generally distinguish between the positive and negative impurity species by using red and blue shades of coloring, respectively.}
    \label{fig:DHstats}
\end{figure*}

Despite decades of research, fundamental questions regarding the nature of these statistical correlations remain unresolved. In \cref{sec:DHstats}, we utilize our simplified framework to investigate the physical mechanisms governing the $qV$ relationship. Specifically, we address several core questions: What physical parameters control the characteristic features of these trends---namely, their center of mass in $(\phi_i, \delta n_i)$ space, their variance, and the slope of the correlation? Furthermore, what governs the distribution of data points around the $qV$ trendline, and what are the underlying physical dependencies of these properties? To resolve these issues, we establish a formal statistical description of the impurity site-energies $\{\epsilon_i\}$. We then analyze how these input statistics propagate through our disordered Hartree calculations, allowing us to characterize and understand the emergent features observed in the resulting histograms and scatterplots.


\section{ANALYSIS OF STATISTICS IN THE LINEARIZED DISORDERED HARTREE FRAMEWORK} \label{sec:DHstats}


\subsection{Overall statistics} \label{subsec:DHstats_overall}

For an overall description of our theory's disorder-driven statistics, we first consider impurity disorder that is generally both conserving and truly random. With the ensemble average over all $N_r$ disorder instances denoted by double brackets $\llangle \rrangle$, these two conditions are captured by the following pair of equations
\begin{align}
    \left \llangle \epsilon_j \right \rrangle
    & = 0,
    \label{eq:<εj>}
    \\
     \left \llangle  \epsilon_j \epsilon_{j'}  \right \rrangle
     & = w^2 \delta_{jj'},
    \label{eq:<εjεj'>}
\end{align}
where we introduce $w^2=\llangle \epsilon_j^2 \rrangle$ to gauge the disorder strength. Note that the above encompasses the equiprobable binary case presented in \Cref{fig:DHstats}, where we additionally stipulate that $\epsilon_j = \pm w$ with 50/50 odds. Since $M_{ij}$ and $\widetilde{M}_{ij}$ are constructed using only the bare profile and lattice structure, the consequences of these for the statistics of our local charge corrections (\ref{eq:δni}) and Madelung fields (\ref{eq:φi}) become immediately apparent: $\delta n_i$ and $\phi_i$'s ensemble averages respectively given by
%
%
\begin{align}
    \left \llangle  \delta n_i  \right \rrangle
    & =  
    \sum_{j} M_{ij} \left  \llangle \epsilon_j \right  \rrangle   = 0,
    \label{eq:δni_avgall}
    \\
    \left  \llangle \phi_i \right \rrangle
    & 
    =  \sum_{j} \widetilde{M}_{ij} \left  \llangle \epsilon_j \right \rrangle = 0.
    \label{eq:φi_avgall}
\end{align}
Hence, by design, the fluctuations in the charge and Madelung field distributions are centered at zero and are thus globally conserving. Next, measuring the statistical spread about these zero-means, we compute the mean squares as follows 
\begin{align}
    \left  \llangle    \delta n_i^2  \right \rrangle     
    & = \sum_{jj' }  M_{ij} M_{ij'} \left  \llangle \epsilon_j \epsilon_{j'} \right \rrangle
    = w^2 \sum_{j} M_{ij}^2,
   \label{eq:δni^2_avgall}
    \\
    \left  \llangle   \phi_i^2  \right \rrangle
    & =  \sum_{jj' }  \widetilde{M}_{ij}  \widetilde{M}_{ij'}   \left  \llangle     \epsilon_j \epsilon_{j'} \right \rrangle
    = w^2 \sum_{j} \widetilde{M}_{ij}^2.
    \label{eq:φi^2_avgall}
\end{align}
%


\subsection{Species-resolved statistics} \label{subsec:DHstats_byspecies}

Now to distinguish finer chemical features in our disordered Hartree statistics, we define a species-resolved average $\llangle \rrangle_{i \in \alpha}$, taken over the subset of $N_\alpha$ lattice sites (with $N_\alpha < N_r$) where the local site-energy belongs to species $\epsilon_i$ belongs to species $\alpha$. 
This conditioning fixes the local site-energy while leaving the statistics of the surrounding environment ($j \neq i$) largely unchanged. The resulting species-resolved correlators are:
\begin{align}
    \left \llangle \epsilon_j \right \rrangle_{i \in \alpha}  
    & = \epsilon_{\alpha} \delta_{ij} , 
    \label{eq:<εj>α}
    \\
    \left \llangle  \epsilon_j \epsilon_{j'}  \right \rrangle_{i \in \alpha}
     & = w^2 \delta_{jj'}
     +(\epsilon_{\alpha}^2 - w^2) \delta_{ij} \delta_{jj'} .
    \label{eq:<εjεj'>α}
\end{align}
Compared to the prior defined correlator (\ref{eq:<εjεj'>}), 
the species-resolved version (\ref{eq:<εjεj'>α}) includes a correction term to ensure that for $i=j=j'$, the second moment correctly returns the constrained value $\epsilon_{\alpha}^2$.
By partitioning the disordered Hartree equations (\ref{eq:δni}) and (\ref{eq:φi}) into local and nonlocal contributions, we apply (\ref{eq:<εj>α}) to find the intraspecies averages:
%
%
%
\begin{align}
    \llangle  \delta n_i  \rrangle_{i \in \alpha} & = 
    M_{ii} \epsilon_{\alpha},
    \label{eq:δni_avgα}
    \\
    \llangle \phi_i \rrangle_{i \in \alpha} & = 
    \widetilde{M}_{ii} \epsilon_{\alpha}.
    \label{eq:φi_avgα}
\end{align}
These averages correspond to the centers of the species-specific data clusters (marked by crosshairs in the bottom panels of \Cref{fig:DHstats}).

Next, we evaluate the spread of these distributions. The species-resolved mean squares are determined using (\ref{eq:<εjεj'>α}), 
\begin{align}
    \left \llangle    \delta n_i^2  \right \rrangle_{i \in \alpha}     
    &
    = \sum_{jj' }  M_{ij} M_{ij'}  \llangle \epsilon_j \epsilon_{j'} \rrangle_{i \in \alpha}
    \nonumber
    \\
    & = w^2 \sum_{j} M_{ij}^2 + M_{ii}^2 (\epsilon_{\alpha}^2 - w^2) ,
   \label{eq:δni^2_avgα}
    \\
    \left \llangle  \phi_i^2  \right \rrangle_{i \in \alpha} 
    &
    = 
    \sum_{jj' }  \widetilde{M}_{ij}  \widetilde{M}_{ij'}    \llangle     \epsilon_j \epsilon_{j'}  \rrangle_{i \in \alpha}
    \nonumber
    \\
    & = w^2 \sum_{j} \widetilde{M}_{ij}^2 + \widetilde{M}_{ii}^2 (\epsilon_{\alpha}^2 - w^2),
   \label{eq:φi^2_avgα}
\end{align}
and they lead to the following intraspecies variances and covariance:
\begin{align}
    \left \llangle    ( \delta n_i - \llangle \delta n_i \rrangle_{i\in\alpha} )^2   \right \rrangle_{i\in\alpha}     
    & = 
    \left \llangle    \delta n_i ^2   \right \rrangle_{i\in\alpha} - \llangle    \delta n_i    \rrangle^2_{i\in\alpha},
    \nonumber \\[0.2cm]
    & = w^2 \sum_{j\neq i} M_{ij}^2,
    \label{eq:δni_varα}
    \\[0.2cm]
    \left \llangle  ( \phi_i - \llangle \phi_i \rrangle_{i \in \alpha}  )^2  \right \rrangle_{i\in\alpha}
    & = \left \llangle  \phi_i ^2  \right \rrangle_{i\in\alpha} - \llangle  \phi_i  \rrangle^2_{i\in\alpha}
    \nonumber \\[0.2cm]
    & =  w^2 \sum_{j \neq i} \widetilde{M}_{ij}^2,
    \label{eq:φi_varα}
\end{align}
as well as the co-relationship between $\delta n_i$ and $\phi_i$
\begin{align}
    \llangle \delta n_i \phi_i \rrangle_{i \in \alpha}
    & = \sum_{jj' }  M_{ij} \widetilde{M}_{ij'}  \llangle \epsilon_j \epsilon_{j'} \rrangle_{i \in \alpha}
    \nonumber 
    \\
    & = w^2 \sum_{j} M_{ij} \widetilde{M}_{ij} + M_{ii} \widetilde{M}_{ii} (\epsilon_{\alpha}^2 - w^2),
    \label{eq:δniφi_avgα}
\end{align}
\vspace{-0.4cm}
\begin{align}
    \llangle (\delta n_i - \llangle \delta n_i \rrangle_{i \in \alpha})  (\phi_i - & \llangle \phi_i 
    \rrangle_{i \in \alpha}) \rrangle_{i \in \alpha} 
    \nonumber \\[0.2cm]
    & = \llangle \delta n_i \phi_i  \rrangle_{i \in \alpha} 
    - \llangle \delta n_i \rrangle_{i \in \alpha} \llangle \phi_i  \rrangle_{i \in \alpha}
    \nonumber \\[0.2cm]
    & = w^2 \sum_{j \neq i} M_{ij} \widetilde{M}_{ij} .   
    \label{eq:δniφi_covα}
\end{align}     
%
Remarkably, despite the site-specific constraint $\epsilon_i = \epsilon_\alpha$, these second-order moments are entirely independent of species $\alpha$. In our linearized theory, every impurity species experiences identical fluctuations and "spread" in its local charge and potential, governed solely by the global disorder strength $w^2$ and the nonlocal response functions. As we show in the following sections, this species-independence has profound implications for the symmetry and universality of $qV$ trends in complex alloys.


\subsection{Least-squares fit of $qV$ trendlines} \label{subsec:leastsqs}

With the intraspecies statistics obtained in the previous \cref{subsec:DHstats_byspecies}, we may now determine the line of best fit for our species-resolved scatterplots (\Cref{fig:DHstats} bottom row). The optimal fit line parameters for species $\alpha$'s statistics are obtained by minimizing the sample's mean squared error 
\begin{align}
    \mathbb{E}_{\alpha} = \left\llangle \left( \delta n_i - (m_{\alpha} \phi_i + b_{\alpha}) \right)^2 \right\rrangle_{i \in \alpha}
\end{align}
with respect to the intraspecies slope  $m_{\alpha}$ and $y$-intercept  $b_{\alpha}$. Applying this optimization condition gives rise to a pair of equations
%
%
\begin{align}
    m_{\alpha} 
    & = 
    \frac{ 
        \left\llangle 1 \right\rrangle_{i \in \alpha}  \left\llangle \delta n_i \phi_i \right\rrangle_{i \in \alpha}     
        - 
        \left\llangle \delta n_i \right\rrangle_{i \in \alpha}  \left\llangle \phi_i \right\rrangle_{i \in \alpha}       
    }{
        \left\llangle 1 \right\rrangle_{i \in \alpha} \left\llangle \phi_i^2 \right\rrangle_{i \in \alpha}
        - 
        \left\llangle \phi_i \right\rrangle_{i \in \alpha}^2  
    },
    \label{eq:mα_moments}
    \\
    b_{\alpha} 
    & = 
    \frac{ 
         \left\llangle \delta n_i \right\rrangle_{i \in \alpha} \left\llangle \phi_i^2 \right\rrangle_{i \in \alpha}
        - \left\llangle \phi_i \right\rrangle_{i \in \alpha}  \left\llangle \delta n_i \phi_i \right\rrangle_{i \in \alpha}
    }{
        \left\llangle 1 \right\rrangle_{i \in \alpha} \left\llangle \phi_i^2 \right\rrangle_{i \in \alpha}
        - 
        \left\llangle \phi_i \right\rrangle_{i \in \alpha}^2    
    }.
    \label{eq:bα_moments}
\end{align}
Now we can identify here $\phi_i$'s species-resolved variance in the denominators of both $m_{\alpha}$ and $b_{\alpha}$ above, this having been computed previously in (\ref{eq:φi_varα}). Further, we note that the prior's numerator corresponds to the covariance provided in (\ref{eq:δniφi_covα}). 
Using the disordered Hartree moment identities derived in the previous \cref{subsec:DHstats_byspecies}, we find that the result reduces to
\begin{align}
    m_{\alpha} 
    & = 
    \frac{
        \sum\limits^{j \neq i} M_{ij} \widetilde{M}_{ij}  
    }{
        \sum\limits_{j \neq i} \widetilde{M}_{ij}^2 
    } ,
    \label{eq:mα_MM~}
    \\
    b_{\alpha} 
    & =
    \frac{ 
        \sum\limits^j \widetilde{M}_{ij} ( M_{ii} \widetilde{M}_{ij} -  M_{ij} \widetilde{M}_{ii} ) 
    }{
        \sum\limits_{j \neq i} \widetilde{M}_{ij}^2    
    } 
    \cdot \epsilon_{\alpha}
    \label{eq:bα_MM~}
\end{align}
Thus, using the linear response language with which we formulated our theory, we can completely describe the $qV$ trendlines which best fit the joint $(\phi_i,\delta n_i)$ intraspecies statistics -- the associated fit parameters primarily involving our charge and Madelung response functions $M$ and $\widetilde{M}$, respectively. Interestingly enough, we observe that the slope of these fit lines $m_{\alpha}$ are entirely independent of chemical species; as alluded to prior, this directly results from the $\alpha$-independence of our second-order central moments.

\subsection{Bivariate normal approximation for joint probabilities} \label{subsec:bivarNormApprox}

Now, for a more complete description of the joint statistical relationship between $\delta n_i$ and $\phi_i$, we revisit the scatterplots presented in \Cref{fig:DHstats}'s bottom row and consider the probability distribution which produces them. First, we observe that the scatterplotted data points all do combine to form fairly ellipsoidal contours. Furthermore, we recall that the individual statistics of either $\delta n_i$ and $\phi_i$ did present with some Gaussian qualities, these being quite apparent at quarter-filling, though perhaps more envelopic at half-filling. Together, these observations suggest that the joint probability distribution $P(\phi_i, \delta n_i)$ may obey, or otherwise be decently approximated, by a two-dimensional generalization of the usual Gaussian curve.  

\begin{figure*}[t!]
    \centering
    \begin{tabular}{c c}
        \includegraphics[trim={0 0.3cm 0 1.5cm}, clip, width=0.475\linewidth]{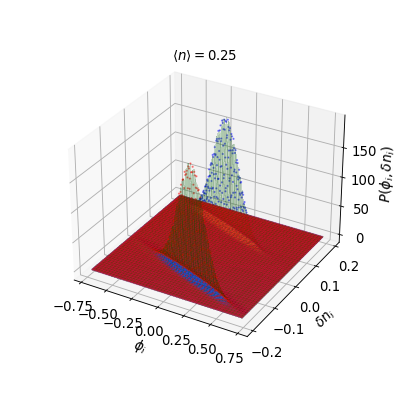}
        &
        \includegraphics[trim={0 0.3cm 0 1.5cm}, clip, width=0.475\linewidth]{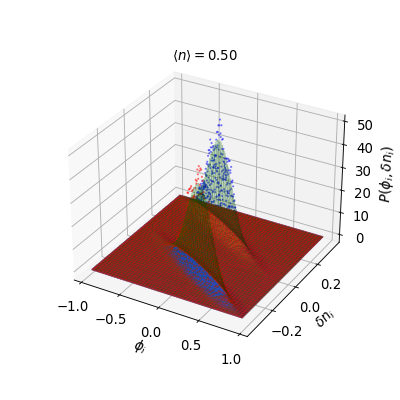}
    \end{tabular}
    \caption{Joint probability distributions $P(\phi_i, \delta n_i)$ for the local intraspecies statistics of the quarter-filled (left) and half-filled (right) cubic models with equiprobable binary disorder. The data points shown here are colored red/blue corresponding to the positive/negative impurity species, and these represent both the histogramming and subsequent normalization of the scatterplots contained in the bottom row of \Cref{fig:DHstats}. Our two-dimensional histograms span $(100\times100)$-grids each, and are compared with bivariate normal distributions (green surface plots) which are parameterized using the disordered Hartree moments derived in \cref{subsec:DHstats_byspecies}. } 
    \label{fig:bivarNorm}
\end{figure*}

We verify this correspondence in \Cref{fig:bivarNorm}, using the previously discussed binary disorder model for both quarter- and half-filled cubic lattices. By normalizing the intraspecies scatterplots from \Cref{fig:DHstats}, we obtain the joint probability distributions $P(\phi_i, \delta n_i)$ for the positive and negative binary species. We then compare these results against a bivariate normal distribution (shown as green surface plots), which takes the general form:
\begin{equation}
P(x, y) = \frac{\exp\left[ -\frac{1}{2(1-\rho^2)} \left( \frac{(x-\mu_x)^2}{\sigma_x^2} + \frac{(y-\mu_y)^2}{\sigma_y^2} - \frac{2\rho(x-\mu_x)(y-\mu_y)}{\sigma_x\sigma_y} \right) \right]}{2\pi \sigma_x \sigma_y \sqrt{1-\rho^2}}
\label{eq:Pxy}
\end{equation}
where $\rho = \sigma_{xy} / (\sigma_x \sigma_y)$ is the correlation coefficient. Much like the univariate Gaussian distribution, this standard form is parameterized entirely by moments up to second order, all of which are analytically derived from our theory in \cref{subsec:DHstats_byspecies}.

To simplify the notation in Eq. (\ref{eq:Pxy}), we map $(\delta n_i, \phi_i) \to (x, y)$ and use $\mu_{x,y}$, $\sigma_{x,y}^2$, and $\sigma_{xy}$ to denote the intraspecies averages, variances, and covariance, respectively. These parameters are captured by the following relations:

%
%
\begin{equation}
    \left\{
    \begin{aligned}
        \mu_{x} & = \widetilde{M}_{ii} \epsilon_{\alpha}, \quad \mu_{y} = M_{ii} \epsilon_{\alpha}, 
        \\[0.4cm]
        \sigma_{x}^2 & = w^2 \sum_{j \neq i} \widetilde{M}_{ij}^2, \quad \sigma_{y}^2 = w^2 \sum_{j \neq i} M_{ij}^2, \quad {\rm and}
        \\[0.1cm]
        \sigma_{xy}  & = \; w^2 \sum_{j \neq i} M_{ij} \widetilde{M}_{ij},
    \end{aligned}
    \right.
    \label{eq:DHmoms_xymap}
\end{equation}
due to (\ref{eq:δni_avgα}), (\ref{eq:φi_avgα}), (\ref{eq:δni_varα}), (\ref{eq:φi_varα}), and (\ref{eq:δniφi_covα}). The comparison in \Cref{fig:bivarNorm} shows excellent agreement, particularly for the quarter-filled case. At these reduced carrier densities, weak screening allows for the accumulation of many independent contributions, driving the statistics toward a Gaussian limit. In the half-filled case, while stronger screening introduces non-trivial features such as sharp "horns" and peaks that Eq. (\ref{eq:Pxy}) cannot capture, the bivariate normal distribution remains a satisfactory approximation of the overarching Gaussian envelope.


\subsection{Isoelectronic doping with binary disorder} \label{subsec:scanConcProcedure}

While the equiprobable binary case ($p_\pm = 0.5$) provides a clear baseline, exploring the full concentration range requires a more careful treatment of the reference potential. If we naively vary the concentration $p$ while keeping site-energies $\epsilon_\pm$ fixed, the average site-energy $\llangle \epsilon_j \rrangle = w(p_+ - p_-)$ becomes non-zero. Physically, this would shift the entire energy scale of the system, potentially altering the total electron count and making comparisons to the bare reference system difficult.


To preserve the statistical identity $\llangle \epsilon_j \rrangle = 0$ while tuning concentration $p$, we fix the interspecies site-energy splitting: 
\begin{equation}
\Delta = \epsilon_+ - \epsilon_- . \label{eq:Delta}
\end{equation}
The individual site-energies are then recomputed at each concentration step to satisfy the conservation criteria:
\begin{equation}
\left\{
\begin{aligned}
\epsilon_+ &= (1-p) \Delta,
\\[0.2cm]
\epsilon_- &= -p \Delta.
\end{aligned}
\right.
\label{eq:ε+-}
\end{equation}
By adopting this shift, we ensure that the average electron density remains invariant across all concentrations, as the response functions of the bare model are preserved. Consequently, the system is fully described by two primary degrees of freedom: the concentration $p$ and the splitting $\Delta$. Furthermore, the total disorder in the system, represented by the variance $w^2$, scales naturally with both parameters:
\begin{equation}
\llangle \epsilon_j^2 \rrangle = p(1-p)\Delta^2 = w^2 . \label{eq:w^2}
\end{equation}
In the clean limits ($p=0$ or $p=1$), the disorder $w$ correctly vanishes. This framework allows us to "dope" the alloy while ensuring the emergent statistics remain in compliance with our first-order response theory.

Using the isoelectronic procedure, we can characterize the $qV$ statistics across the full range of impurity concentrations $p$. By fixing the energy splitting $\Delta = 2$ and calculating the response for various fillings $\langle n \rangle$, we identify the universal scaling of the trendline parameters. These results, obtained using the same cubic lattice model as in previous sections, are summarized in \Cref{fig:concTrends}. To reveal the underlying $p$-dependence, the data sets for different fillings are rescaled by their half-concentration ($p=0.5$) values.

\begin{figure*}[t!]
    \centering
    \includegraphics[trim={0.3cm 0.2cm 0 0}, clip, width=0.8\linewidth]{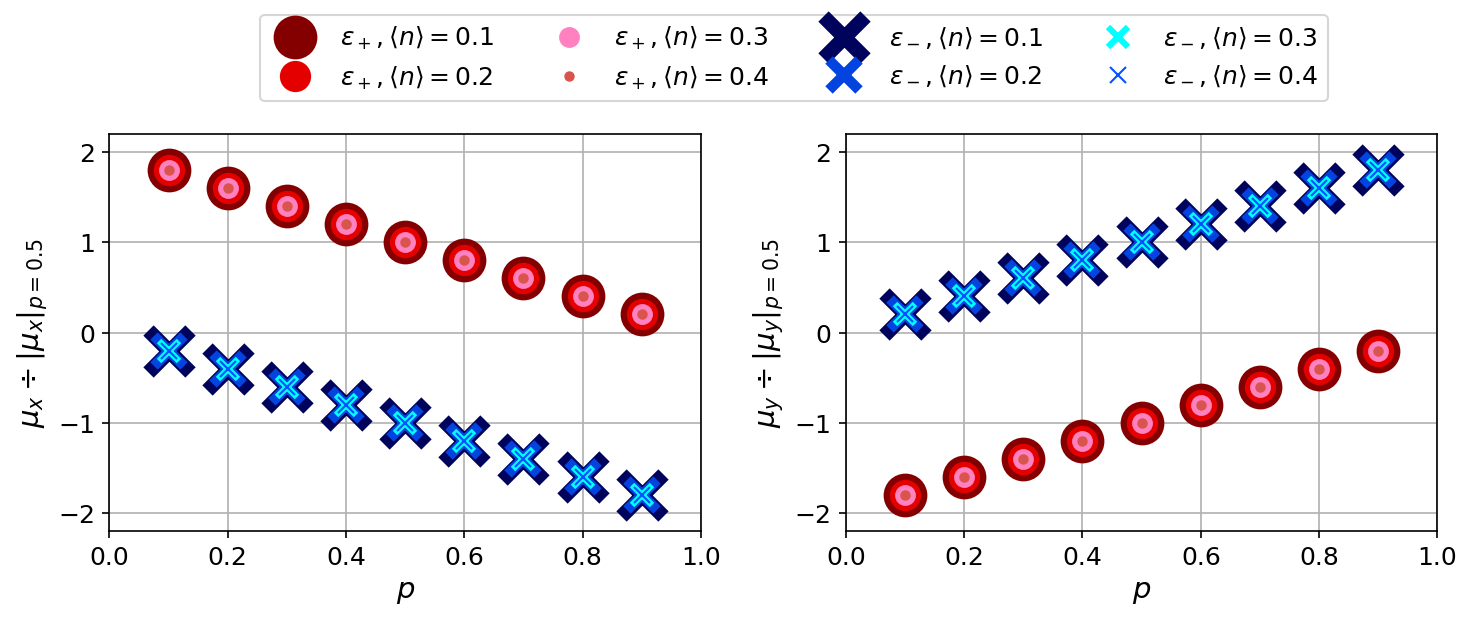}
    \\
    \includegraphics[trim={0.1cm 0.2cm 0 0}, clip, width=0.4\linewidth]{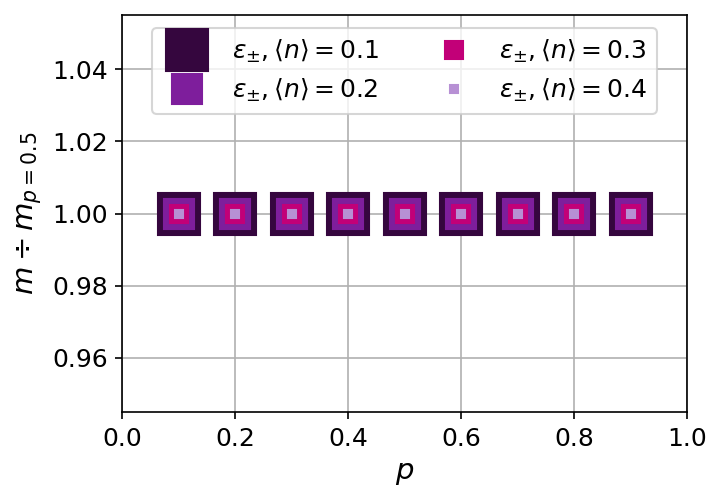}
    \\
    \includegraphics[trim={0.3cm 0.2cm 0 0}, clip, width=0.8\linewidth]{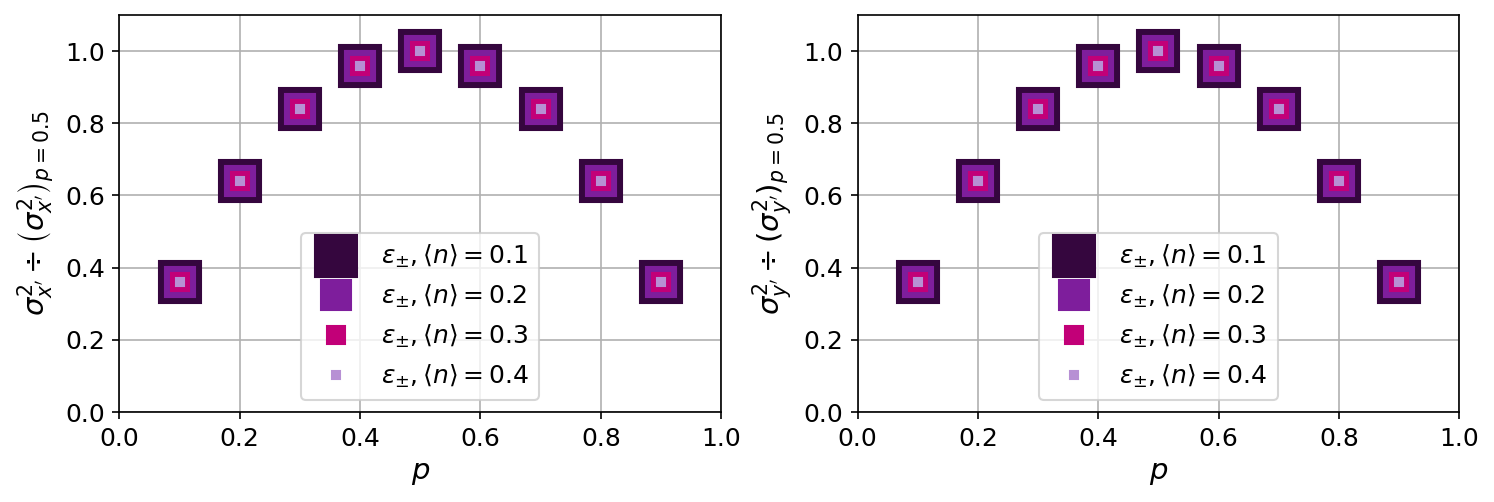}
    \caption{Disordered Hartree statistical averages (top row), slopes (middle row), and line dimensions (bottom row) characterizing the $qV$ trendlines for the isoelectronically doped binary alloy with positive impurity concentration $p$ and interspecies energy splitting $\Delta = \epsilon_+ - \epsilon_- = 2$. These have been rescaled according to their half-concentration values, revealing their universal $p$-dependences irrespective of filling $\langle n \rangle$.}
    \label{fig:concTrends}
\end{figure*}

From \Cref{fig:concTrends}, several key scaling behaviors emerge:
\begin{enumerate}[nosep,leftmargin=*]
\item \textbf{Averages ($\mu_x, \mu_y$):} The intraspecies averages for both the Madelung potential and charge transfer scale linearly with $p$. This is expected since $\mu_{x,y} \propto \epsilon_{\alpha}$ (\cref{eq:DHmoms_xymap}), and for fixed $\Delta$, the site-energies $\epsilon_{\pm}$ vary linearly with concentration (\cref{eq:ε+-}).
\item \textbf{Slopes ($m$):} The species-independent $qV$ slopes show no concentration dependence. As derived in (\ref{eq:mα_MM~}), $m$ is determined solely by the response functions $M$ and $\widetilde{M}$, which are properties of the bare model and remain invariant during isoelectronic doping.
\item \textbf{Dimensions ($\sigma_{x'}^2, \sigma_{y'}^2$):} The $qV$ line lengths and widths exhibit a quadratic dependence on $p$. Given that $m$ is constant, these dimensions are linear combinations of the variances $\sigma_x^2$ and $\sigma_y^2$. Because $\sigma_{x,y}^2 \propto w^2$ and $w^2 \propto p(1-p)$ (\cref{eq:w^2}), the quadratic scaling follows directly.
\end{enumerate}

\begin{figure*}[t!]
    \centering
    \includegraphics[trim={0.3cm 0.2cm 0 0}, clip, width=0.8\linewidth]{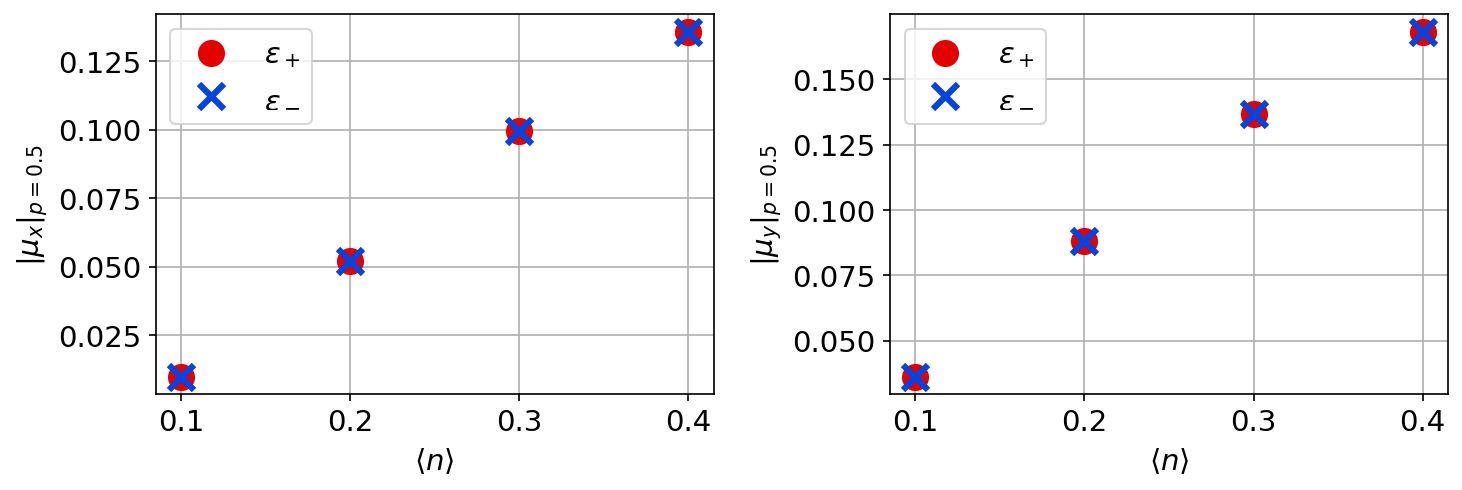}
    \\
    \includegraphics[trim={0.1cm 0.2cm 0 0}, clip, width=0.4\linewidth]{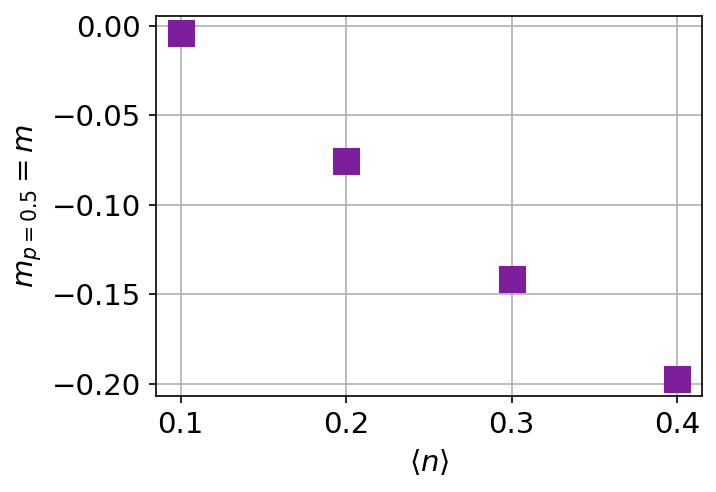}
    \\
    \includegraphics[trim={0.3cm 0.2cm 0 0}, clip, width=0.8\linewidth]{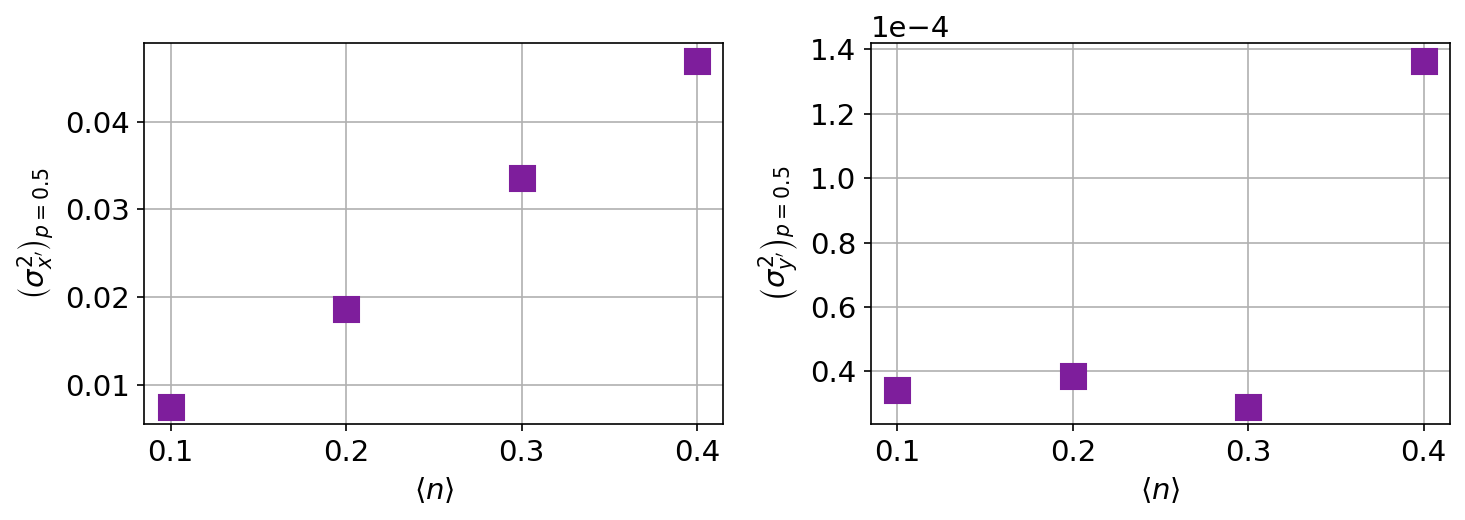}
    \caption{Disordered Hartree absolute averages (top row), slopes (middle row), and line dimensions (bottom row) for the $qV$ trendlines as a function of filling $\langle n \rangle$ in an equiprobable binary alloy with positive impurity concentration $p=0.5$ and interspecies energy splitting $\Delta=\epsilon_+ - \epsilon_- = 2$. These half-concentration values correspond to those which rescale \Cref{fig:concTrends}'s trends and reveal their universal $p$-dependence.}
    \label{fig:rescaleFactors}
\end{figure*}

The rescaling factors themselves reveal how the $qV$ features depend on the carrier concentration (filling $\langle n \rangle$), as illustrated in \Cref{fig:rescaleFactors}. We find that as $\langle n \rangle$ increases, the magnitudes of the averages $\mu_x$ and $\mu_y$ grow, reflecting an enhanced local response facilitated by the higher density of carriers. Similarly, the slopes $m$ become more sharply negative and the line lengths $\sigma_{x'}^2$ increase with filling. In contrast, the line widths $\sigma_{y'}^2$ exhibit no monotonic trend with respect to $\langle n \rangle$.

This concludes our analysis of the binary alloy, which has served as a prototypical system for resolving the physical origins and scaling laws of $qV$ statistics. In the following section, we extend this framework to the more complex case of quaternary disorder to address the multi-principal element nature of high-entropy alloys.


\subsection{\hspace{-0.1cm}Isoelectronic and bispecific doping with quaternary disorder} \label{subsec:scanConcProcedure_highent}

To generalize our framework to quaternary alloys (species $\alpha \in \{a,b,c,d\}$), we utilize a bispecific isoelectronic doping procedure. This approach simplifies the high-dimensional parameter space of multi-component alloys by treating two species as static "control" backgrounds while tuning the relative concentrations of the remaining two.

For the quaternary doping procedure, we fix the concentrations of species $a$ and $b$ ($p_a, p_b$) and define the relative site-energy splittings $\Delta_{\alpha} = \epsilon_{\alpha} - \epsilon_d$ as constants. By choosing the concentration of species $c$ ($p = p_c$) as our independent variable, the site-energies are recomputed at each step to satisfy the conservation criteria $\llangle \epsilon_j \rrangle = 0$:
\begin{equation}
\left\{
\begin{aligned}
\epsilon_d &= -p_a \Delta_a - p_b \Delta_b - p \Delta_c, \\
\epsilon_c &= \epsilon_d + \Delta_c, \\
\epsilon_b &= \epsilon_d + \Delta_b, \\
\epsilon_a &= \epsilon_d + \Delta_a.
\end{aligned}
\right.
\label{eq:eps_quat}
\end{equation}
This procedure ensures the average electron density remains invariant, allowing us to describe the quaternary system using only the relative splittings and the concentrations $p_{a,b,c}$.

The total disorder strength $w^2$ in this quaternary system is given by:
\begin{align}
    \llangle \epsilon_j^2 \rrangle 
    & = p_a \, \epsilon_a^2 + p_b \, \epsilon_b^2 + p \, \epsilon_c^2 + (1-p_a - p_b - p) \, \epsilon_d^2
    \nonumber
    \\
    & = p \, \Delta_c^2 + p_a \, \Delta_a^2 + p_b \, \Delta_b^2 - \left(  p \, \Delta_c  + p_a \, \Delta_a + p_b \, \Delta_b \right)^2
    \nonumber
    \\
    & =  w^2,        
    \label{eq:w^2_quaternary}
\end{align}
Unlike the symmetric parabola of the binary case, the presence of control species $a$ and $b$ introduces an asymmetric broadening of $w^2(p)$ and ensures that finite disorder persists even at the endpoints of the doping range ($p=0$ or $p=1-p_a-p_b$).

\begin{figure*}[t!]
    \centering
    \includegraphics[trim={0.3cm 0.2cm 0 0}, clip, width=0.8\linewidth]{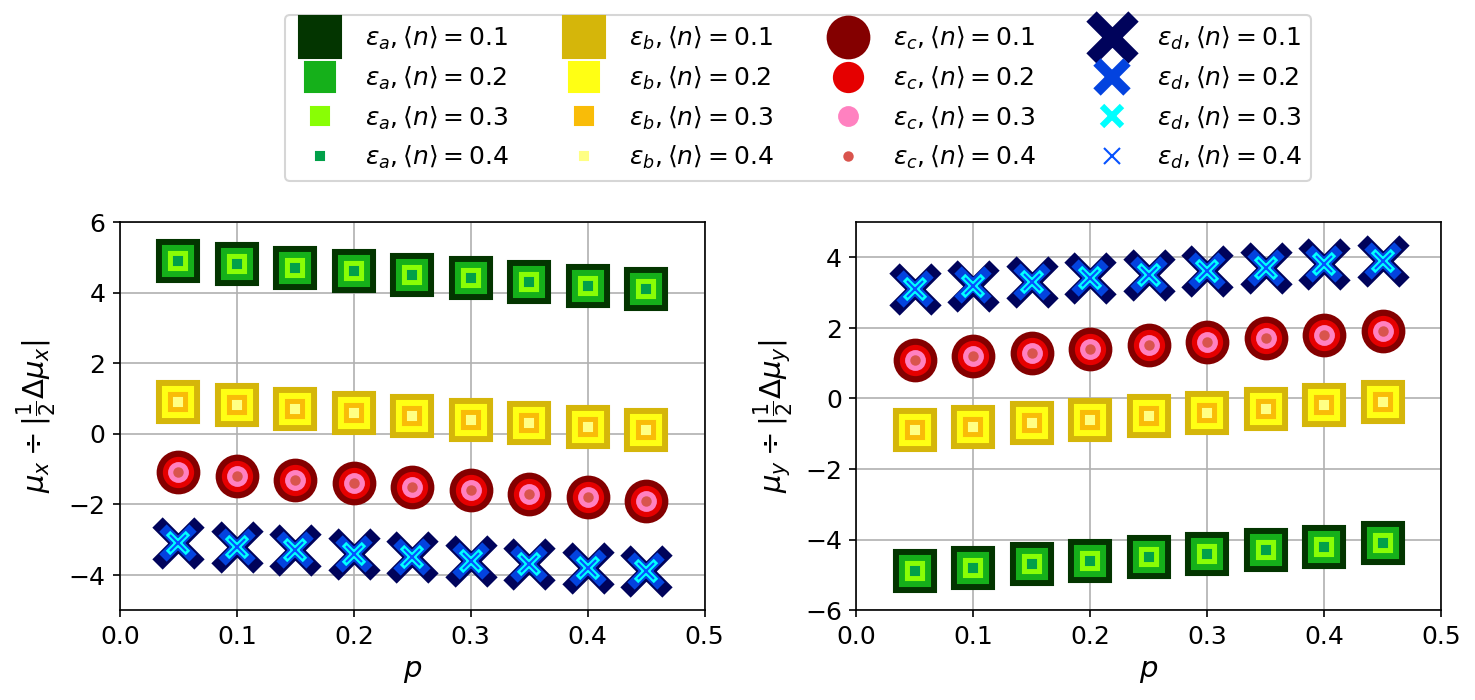}
    \\
    \includegraphics[trim={0.1cm 0.2cm 0 0}, clip, width=0.4\linewidth]{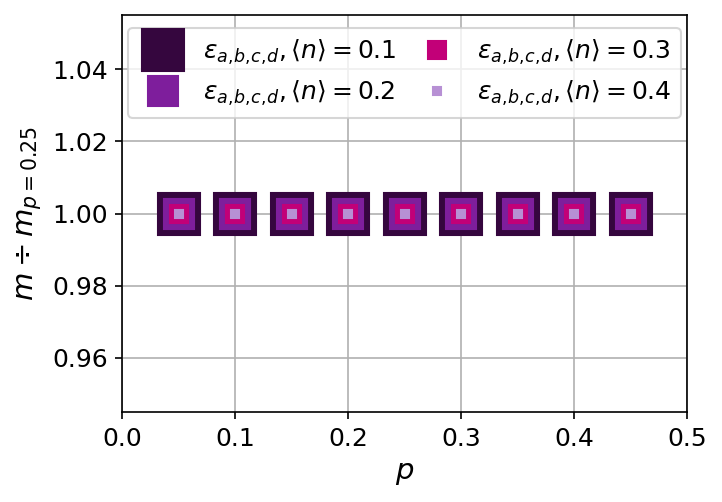}
    \\
    \includegraphics[trim={0.3cm 0.2cm 0 0}, clip, width=0.8\linewidth]{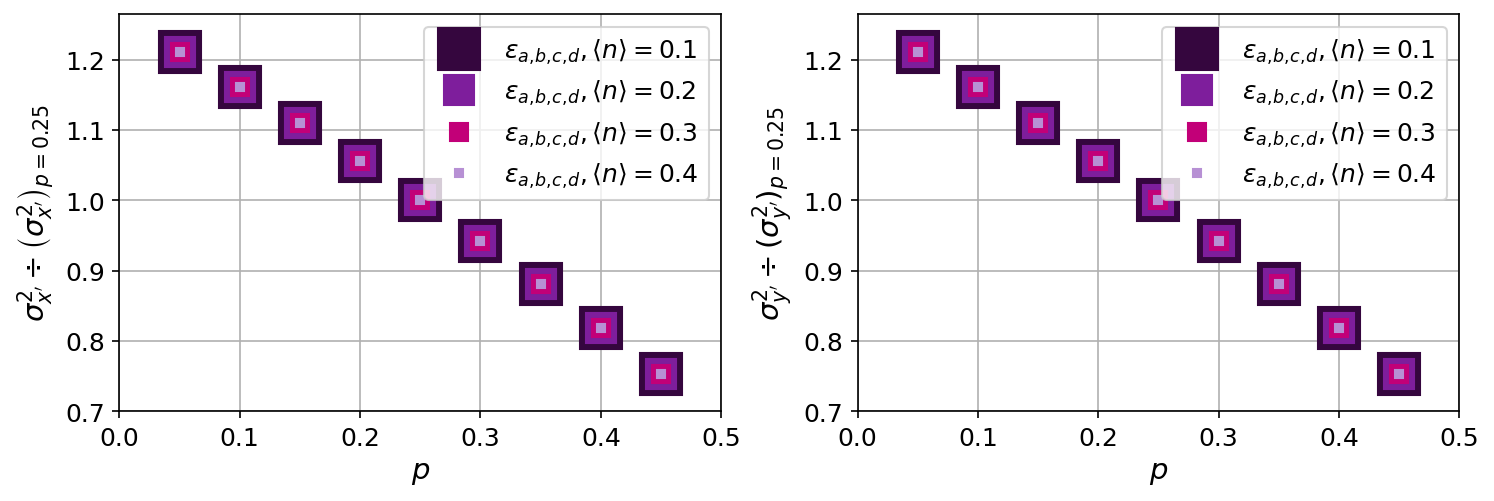}
    \vspace{-0.2cm}
    \caption{Disordered Hartree statistical averages (top row), $qV$ slopes (middle row), and qV
    line dimensions (bottom row) for the quaternary $a_{0.25} b_{0.25} c_{p} d_{0.5-p}$ alloy with variable $p$ and relative site-energies $\Delta_{\{a,b,c\}} = \epsilon_{\{a,b,c\}} - \epsilon_d = \{4,2,1\}$, fixed with respect to that of the $d$ species. Averages are rescaled by half the absolute difference between those of the $c$ and $d$ species, while remaining quantities are rescaled by their quarter-concentration ($p=0.25$) values, to collapse the various curves for different fillings $\langle n \rangle$.}
    \label{fig:concTrendsQuaternary}
\end{figure*}

We apply this model to a cubic lattice with $\{\Delta_a, \Delta_b, \Delta_c\} = \{4, 2, 1\}$ and $p_a = p_b = 0.25$. As shown in \Cref{fig:concTrendsQuaternary}, the scaling behaviors mirror those of the binary system but with secondary features driven by the control species:
\begin{enumerate}[nosep,leftmargin=*]
\item \textbf{Averages ($\mu_{x,y}$):} Remain linear in $p$, as $\mu_{x,y} \propto \epsilon_{\alpha}$ and $\epsilon_{\alpha}$ depends linearly on concentration.
\item \textbf{Slopes ($m$):} Continue to be independent of both species and concentration, as $m$ is a property of the bare response functions $M$ and $\widetilde{M}$.
\item \textbf{Dimensions ($\sigma_{x',y'}^2$):} Maintain a quadratic dependence on $p$, scaling with $w^2$. However, the curves are no longer symmetric about the midpoint due to the underlying background of the control species.
\end{enumerate}
These findings suggest that the fundamental $qV$ statistics of complex, multi-principal element alloys can be understood through the lens of linearized response theory. The key features—linearity of averages, invariance of slopes, and quadratic scaling of widths—are universal across arbitrary degrees of disorder.



\section{COMPARISON WITH DFT CALCULATIONS} \label{sec:LSMSresults}


\subsection{Statistics and concentration dependence of binary alloys} \label{subsec:LSMSresults_binary}
Reserving more details on the LSMS approach and on our calculations for the Supplemental Materials (see \cref{ssec:LSMSdetails}),  we first consider the independent/unconditional statistics of the LSMS charge transfers and Madelung fields in three binary alloys. Those corresponding to equiatomic mixtures of Cu and Au, Cu and Zn, and of Co and Ni are provided in \Cref{fig:LSMS_rawStatsBinary}'s two leftmost columns below. And it is here we find, with ultra-fine binning, that the statistics may bear sharply peaked structures which develop under broader Gaussian-shaped envelopes. Now, similarly broad features have also been reported, with more coarse-grained binning, in previous and related works \cite{oldAlloys2,MuSTpaper2022}, as mentioned briefly both in our introduction and again towards the end of \cref{subsec:manyImps}. However, we may now recall from the more recent of these two mentions -- where we had also discussed more refined details encountered in our own disordered Hartree statistics (i.e. \Cref{fig:DHstats}, and also \ref{fig:bivarNorm}) -- that we can understand these histograms' keener qualities to be related to the high screening efficiency and strong metallic character of these alloys. 

\vfill\null

\vspace{-0.3cm}
\begin{figure*}[t!]
    \centering
    \includegraphics[ width=0.9\linewidth]{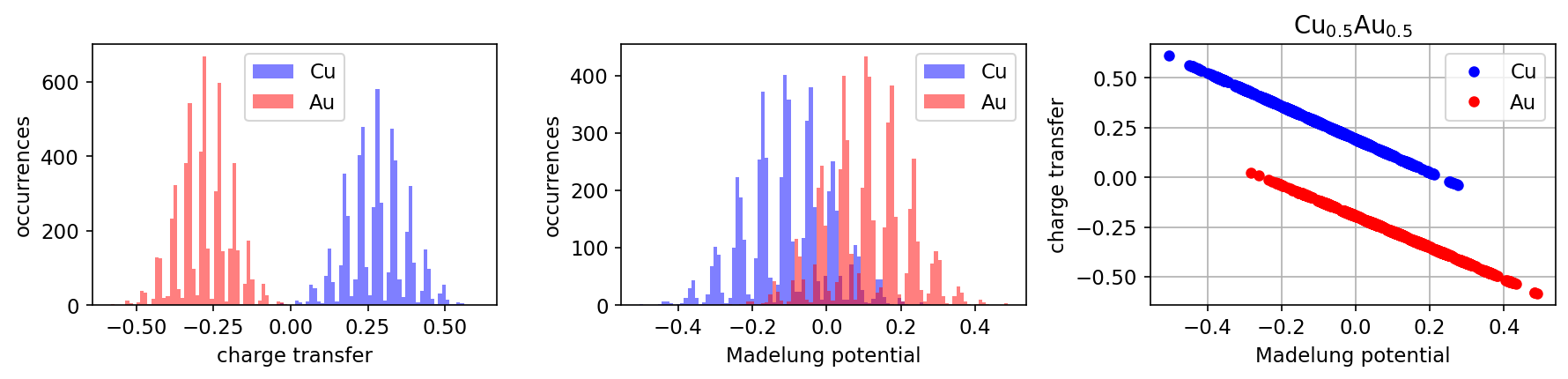}
    \includegraphics[ width=0.9\linewidth]{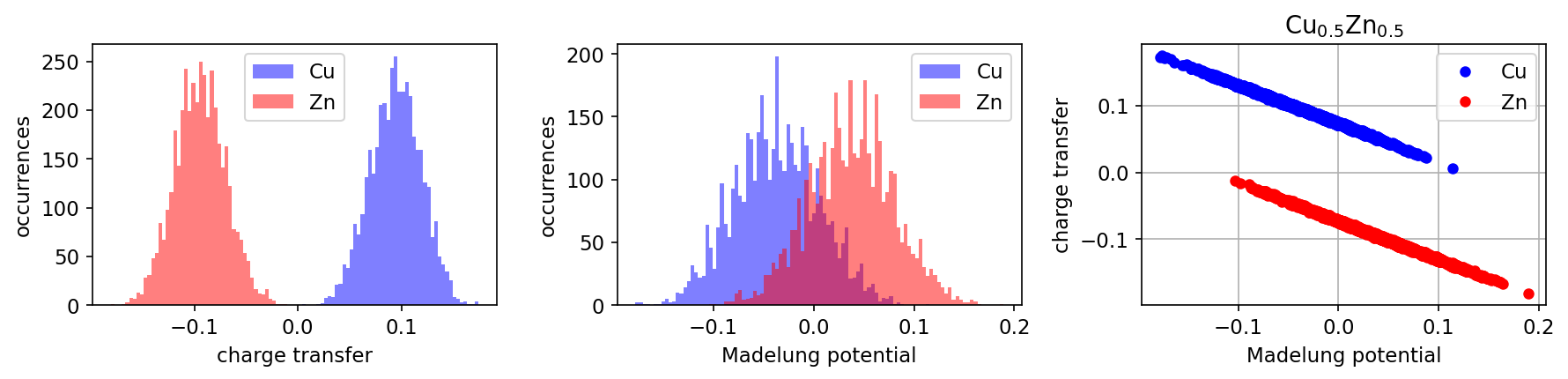}
    \includegraphics[ width=0.9\linewidth]{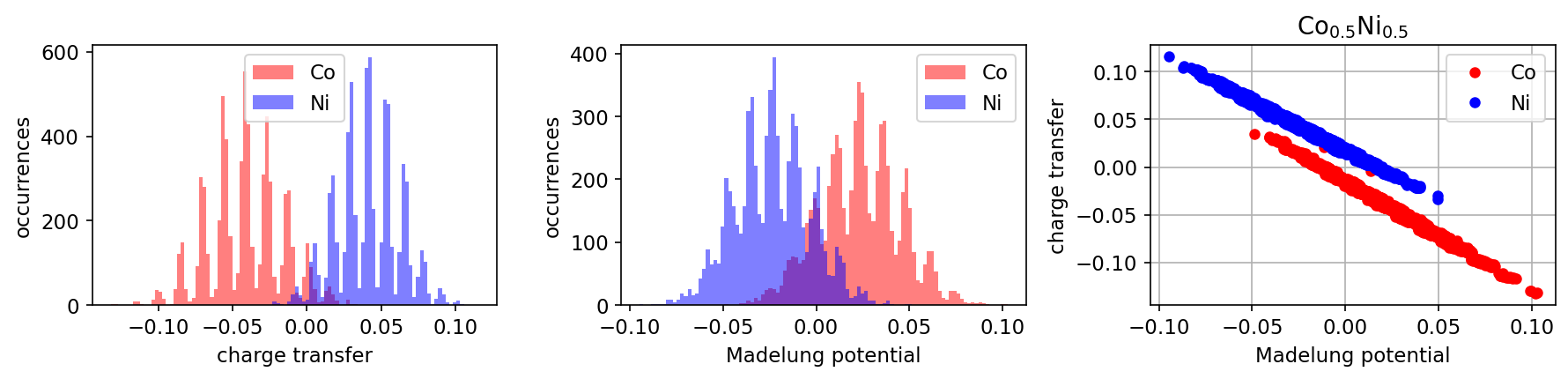}
    \caption{DFT calculated charge transfers (left column), Madelung potentials (middle column), and their joint statistical $qV$ relations (right column) are shown for the equiatomic CuAu, CuZn, and CoNi alloys in the top, middle, and bottom rows, respectively. The histograms in the two leftmost columns are finely binned (100 bins per panel) to reveal the sharper features of the associated statistics.}
    \label{fig:LSMS_rawStatsBinary}
\end{figure*}

\begin{figure*}[t!]
    \centering
    \begin{tabular}{c}
        \includegraphics[ width=0.8\linewidth]{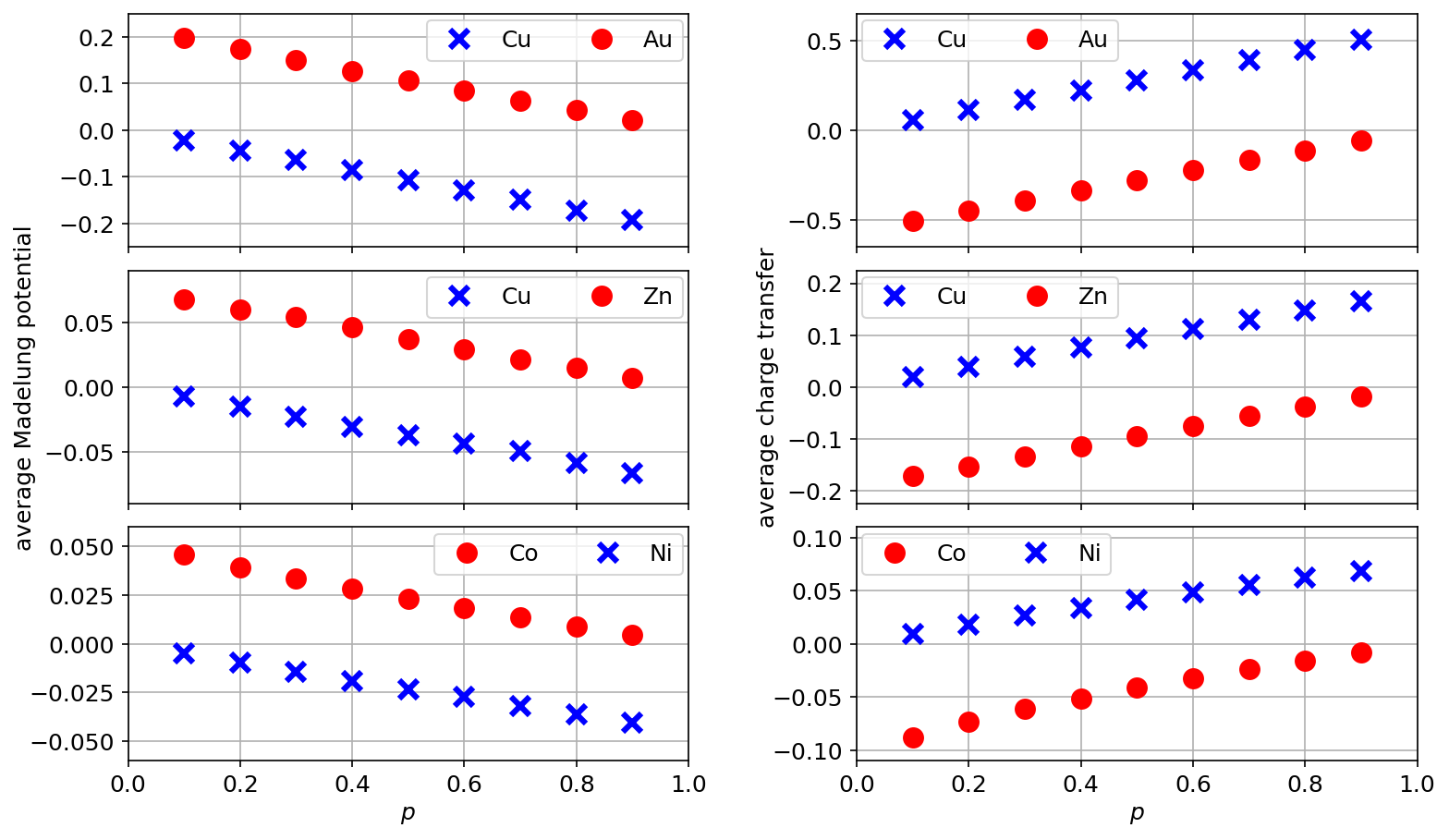}
        \\
        \includegraphics[ width=0.8\linewidth]{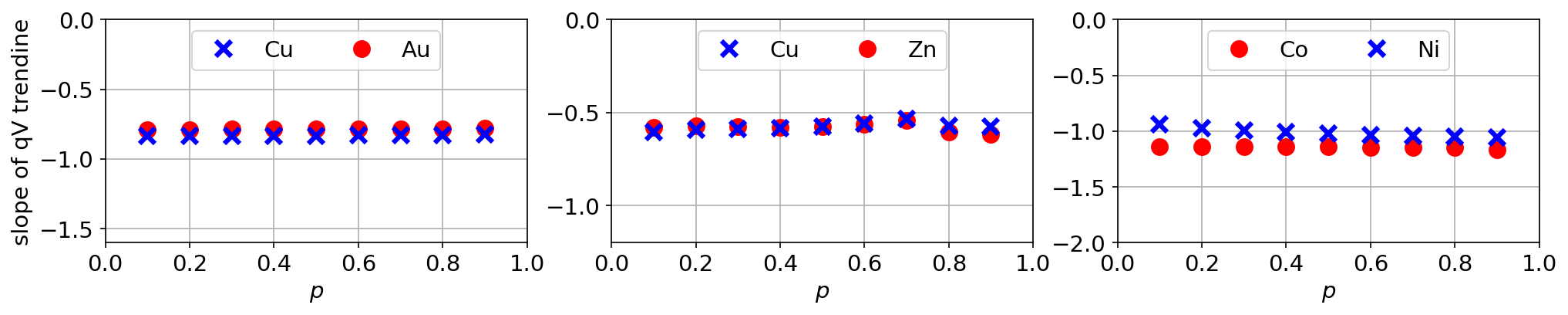}
        \\
        \includegraphics[ width=0.8\linewidth]{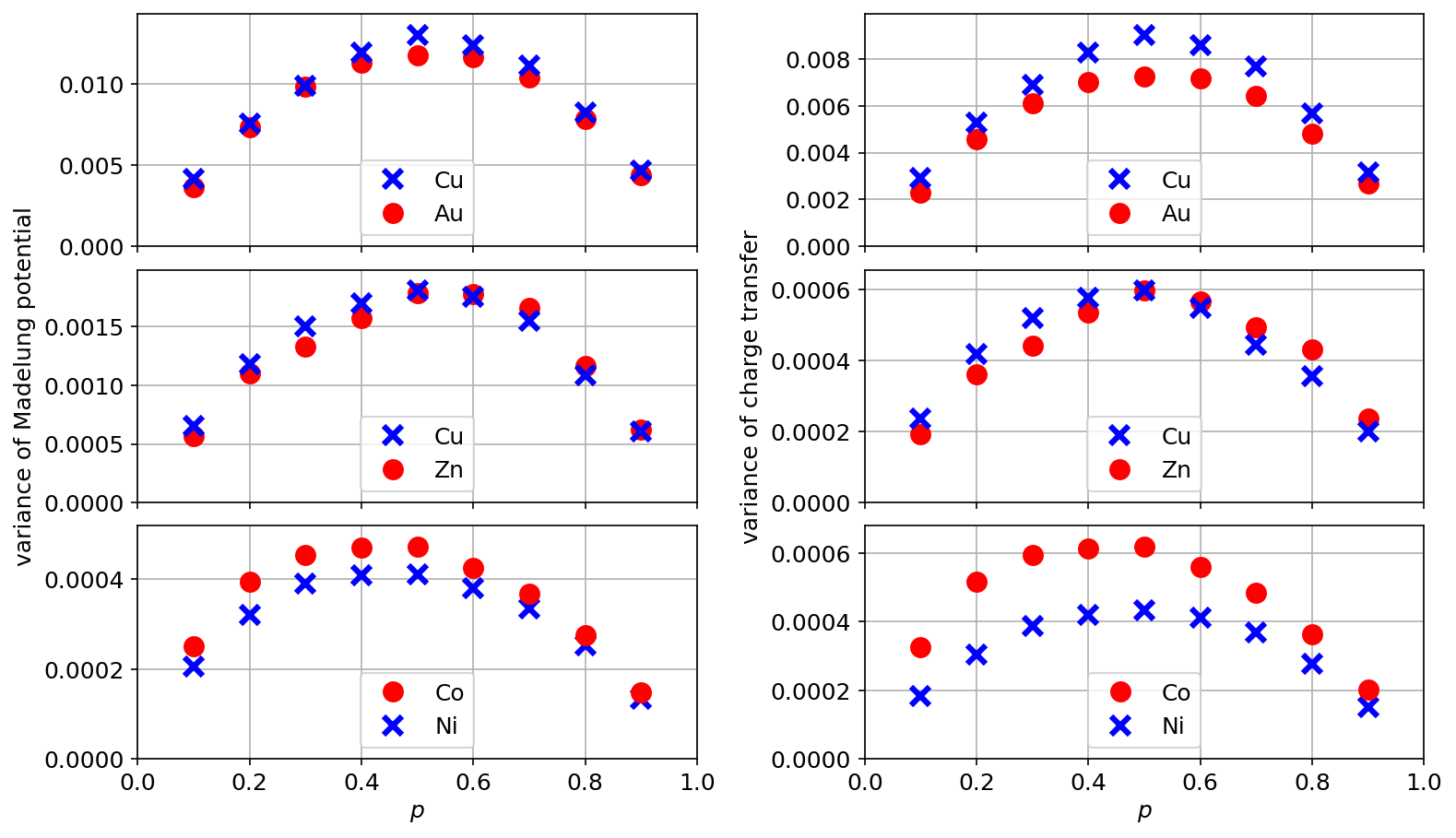}
    \end{tabular}
    \caption{ DFT calculated statistical averages (top six panels), $qV$ slopes (middle three-panel row), and variances (bottom six panels) for $\textrm{Cu}_{1-p}\textrm{Au}_{p}$, $\textrm{Cu}_{1-p}\textrm{Zn}_{p}$, and $\textrm{Co}_{p}\textrm{Ni}_{1-p}$ as functions of cation concentrations $p$ (cationic trends shown with red data points).  }
    \label{fig:LSMS_concTrendsBinary}
\end{figure*}

In addition to the unconditional statistics for the DFT calculated charge transfers and Madelung fields, we have also their joint statistical $qV$ relationships, shown in \Cref{fig:LSMS_rawStatsBinary}'s rightmost column for the equiatomic binary alloys we consider here. Now, to determine how all of these features depend on the present disorder, in \Cref{fig:LSMS_concTrendsBinary} above, we show how various relevant quantities change with atomic fractions in each alloy. Given the simpler nature of disorder in these binary solids, we may take such trends to be simple functions of atomic concentrations -- particularly of those elements serving a more cationic role%
\footnote{Our decision to track cationic concentrations ($p$) is made analogously to the choice we had made previously when studying the binary alloy's disordered Hartree statistics (\cref{subsec:scanConcProcedure}). There, we had chosen concentration $p$ of the $\epsilon_+>0$ impurities as our independent variable -- such positive defects to the one-electron potential ($\widetilde{\epsilon}$) causing a net loss of local charge.}. %
Such atoms have a higher tendency to donate their electrons and experience more negative charge transfers. And as shown in \Cref{fig:LSMS_rawStatsBinary}, it is the Zn, Au, and Co atoms which act as the cations in the CuZn, CuAu, and CoNi alloys, respectively. 

In \Cref{fig:LSMS_concTrendsBinary}, we now observe qualitative trends which are comparable to those obtained previously when investigating the statistics of the isoelectronically doped binary alloy. Note currently, however, that the CuAu alloy is the only mixture of same-group(-11) elements we consider here, with constituent atoms having valence occupations that are nominally equivalent; the doping of this alloy may therefore be regarded as truly isoelectronic. Alternatively, CuZn and CoNi consist of elements lying adjacent to one another on the periodic table, and hence the number of electrons participating in charge transfer should, in fact, change as these alloys are doped. Despite this detail, though, \Cref{fig:LSMS_concTrendsBinary} shows that the three alloys we discuss here all display: 
\begin{enumerate}[nosep,leftmargin=*]
    \item Charge transfer and Madelung potential averages that are (approximately) linear in cation concentrations.
    \item The $qV$ trendline slopes with little-to-no dependence on cation concentrations.
    \item Variances that assume a more parabolic form.
\end{enumerate}
These behaviors are all broadly consistent with what we had found previously in our simpler, model-based approach (\Cref{fig:concTrends}, \cref{subsec:scanConcProcedure}). There do remain, however, some additionally nuanced aspects of the above DFT calculated trends -- some of these perhaps being more subtly connected to this matter of (non)isoelectronicity. 


\subsection{Potential consequences of nonisoelectronicity and otherwise} \label{subsec:nonIsoDiscussion}

To elaborate on this last point, we can first note that, upon closer inspection, \Cref{fig:LSMS_concTrendsBinary}'s DFT calculated averages may not all be so perfectly linear after all. For a better perspective of this, \Cref{fig:deltas} provides, as a function of cation concentrations, the difference between average charge transfers of the two elements comprising each binary alloy.  We call this quantity "delta", in keeping with nomenclature established in previous work \cite{oldAlloys1}, where CuZn's delta was reported to be $\simeq 0.2$, as we find similarly here%
\footnote{See FIG. 3 of \cite{oldAlloys1}. Note also their CuZn slopes ($\simeq 2$), which they report in the same figure and are for $qV$ trends that have been inverted with respect to our own -- hence we must take the reciprocal of their slopes in order to find decent agreement with what we obtain here ($|m| \simeq 0.5$ in \Cref{fig:LSMS_concTrendsBinary}).}. 
Now from \Cref{fig:deltas}'s insets, we see that delta actually falls off rather nonlinearly in both the CuZn and CoNi alloys, indicating that in either alloy, at least one of the charge transfer trends, cationic or anionic, is (weakly) nonlinear as well. In particular, as the cation fraction is increased, we find that CuZn's and CoNi's deltas drop by roughly 3\% and 20\%, respectively, while CuAu's delta merely fluctuates by $\lesssim 1\%$ across all concentrations. 

\begin{figure}[hbt!]
    \centering
    \includegraphics[ width=0.8\linewidth]{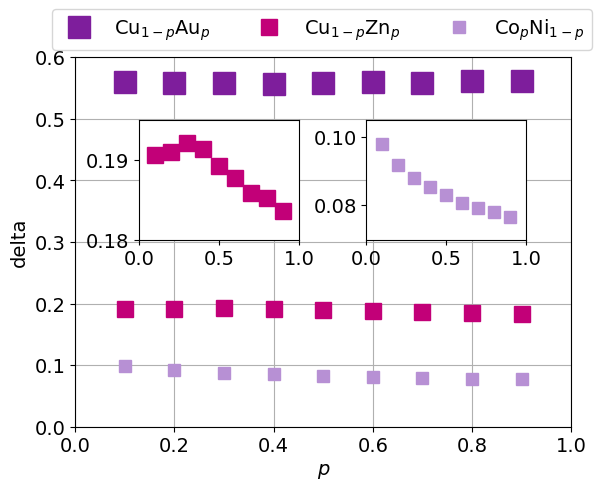}
    \caption{The difference "delta" between cation and anion charge transfer averages is plotted as a function of cation concentration $p$ for the binary CuAu, CuZn, and CoNi alloys; delta trends for the latter two are magnified within the insets.}
    \label{fig:deltas}
\end{figure}

We therefore find that CuAu's charge transfer trends may best match the expectations set by our theory, thanks likely to the fact that the doping conditions here most closely resemble those of the isoelectronically doped binary alloy that we had previously considered (\cref{subsec:scanConcProcedure}). Alternatively, this suggests that the nonisoelectronic nature of doping in the CuZn and CoNi alloys does exert some minor influence over the behavior of their DFT calculated trends. This notion can be further developed, also using our more simple model, where we may consider how variable electronic fillings affect the behavior of the doped binary alloy. As we discuss in the Supplemental Materials (see \cref{ssec:nonIso}), allowing for a small degree of nonisoelectronicity to enter our calculations can, in fact, produce some nonlinear behavior in the the concentration dependences of the charge transfer and Madelung field averages. We also find that such variable electronic fillings can induce some concentration dependence in the $qV$ trendline slopes, as well as distort/skew the parabolic variance trends -- causing some asymmetry with respect to the half-concentration levels ($p=0.5$) which correspond to the equiatomic alloy. %
Now, while there may appear to be some subtler hints of these nonisoelectronic effects in the DFT calculated trends gathered in \Cref{fig:LSMS_concTrendsBinary}, we reemphasize that such details seem also to be rather modest for the alloys we consider here, and note that further investigation is likely warranted to understand and perhaps decouple these from other possible (e.g. bandstructure) effects, or to say anything more conclusive on the matter. 
In any case, we find that the DFT calculated statistics of the doped binary alloys are in good qualitative agreement with those predicted by our much simpler theory on binary disorder.


\subsection{Bispecific doping of the high-entropy CoCrFeNi alloy}\label{eq:LSMSresults_binary}
For the final set of results that we discuss here, we revisit the topic of multi-principal alloys. Presented in \Cref{fig:LSMS_CoCrFeNi} below are trends obtained from the DFT calculated statistics of a high-entropy CoCrFeNi alloy, belonging to the Cantor class. Here, we bispecifically tune the relative concentrations of Co and Ni atoms, much in the same spirit as in our earlier model treatment of the quaternary alloy  (\cref{subsec:scanConcProcedure_highent}).\linebreak These trends may thus be taken to be simple functions of the Co concentrations, as we continue to track the bidoped species which experiences the least amount of charge transfer (on average).  Now from \Cref{fig:LSMS_CoCrFeNi}, we find concentration dependences which are in decent qualitative agreement with those obtained through our earlier model calculations (\Cref{fig:concTrendsQuaternary}, \cref{subsec:scanConcProcedure_highent}). In particular, we observe:
\begin{enumerate}[nosep,leftmargin=*]
    \item The charge transfer and Madelung field averages for all four elements trend roughly as expected, each with (approximately) linear dependence on Co concentrations.
    \item The $qV$ trendline slopes which continue to show no discernible dependence on Co concentrations.
    \item Variances which, unlike in any of the binary alloys, appear to follow a much more broadly monotonic trend. 
\end{enumerate}
Thus, despite its simplicity, our linearized disordered Hartree theory still captures the main features of the doped high-entropy alloy and reproduces much of the same DFT calculated trends.  

\begin{figure*}[t!]
    \centering
    \includegraphics[trim={0.2cm 0.1cm 0 0}, clip, width=0.70\linewidth]{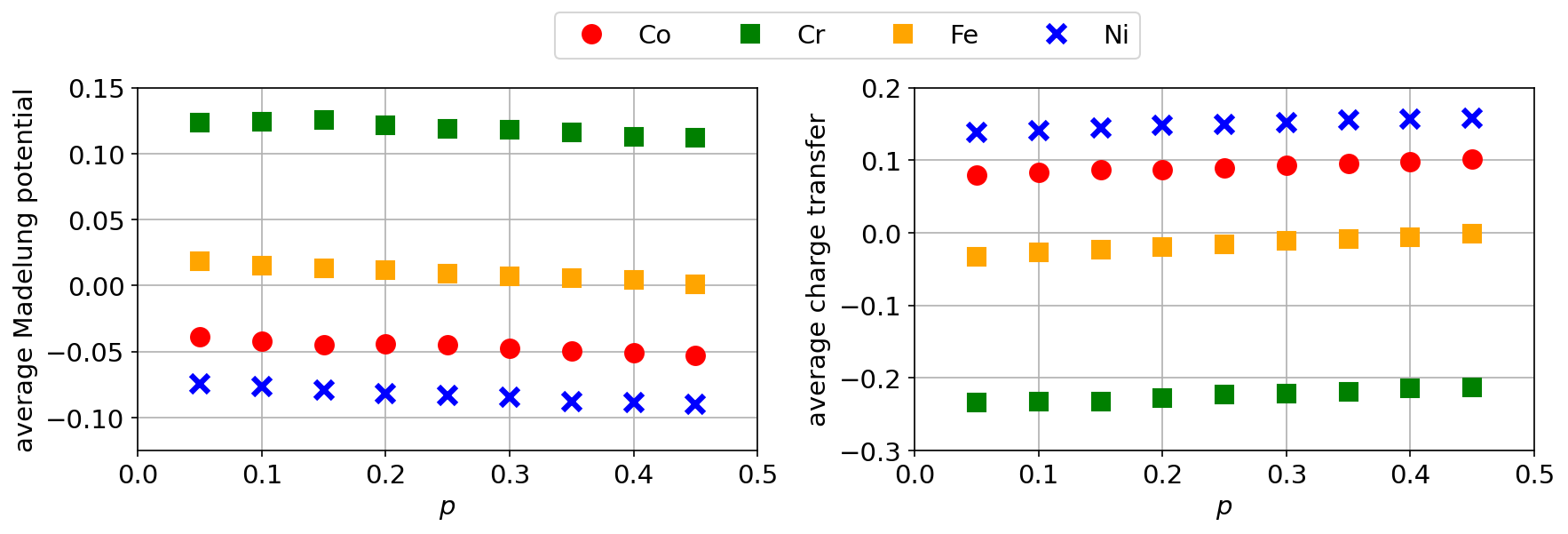}
    \\
    \includegraphics[trim={0.1cm 0.1cm 0 0}, clip, width=0.4\linewidth]{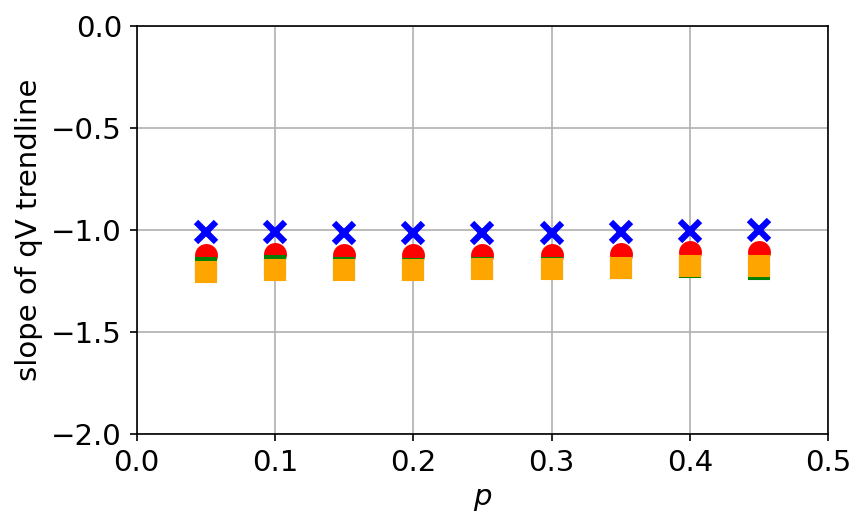}
    \\
    \includegraphics[trim={0.2cm 0.2cm 0 0}, clip, width=0.70\linewidth]{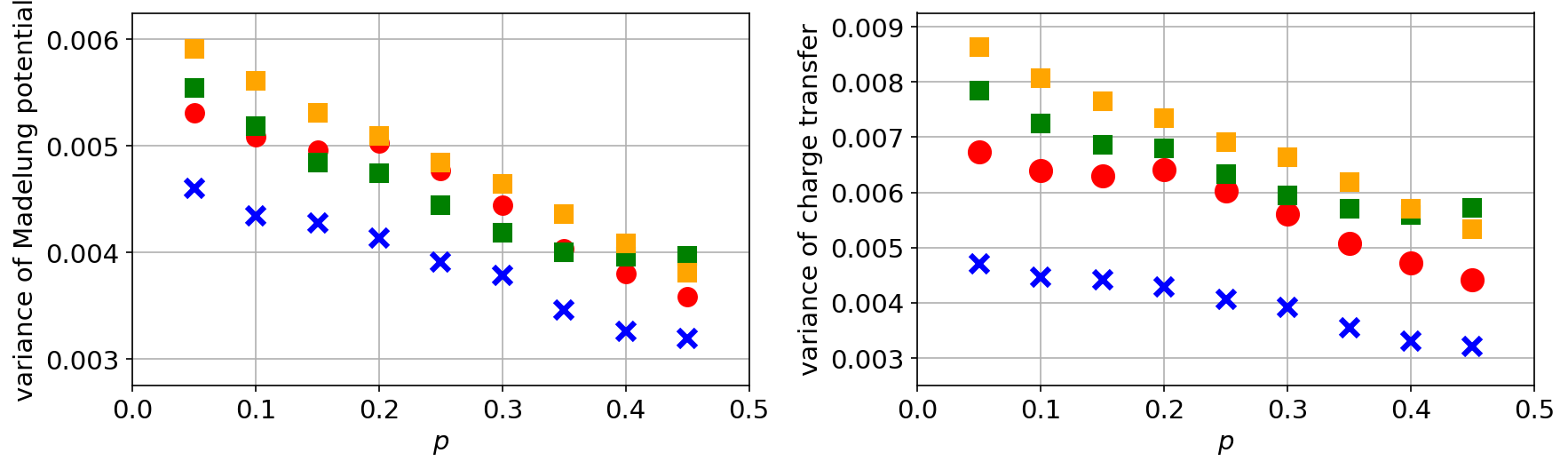}
    \caption{DFT calculated statistical averages (top row), $qV$ trendline slopes (middle panel), and variances (bottom row) for the bispecifically doped $\textrm{Co}_p\textrm{Cr}_{0.25}\textrm{Fe}_{0.25}\textrm{Ni}_{0.5-p}$ alloy as a function of Co concentration $p$.}
    \label{fig:LSMS_CoCrFeNi}
\end{figure*}


\section{FURTHER DISCUSSIONS}


\subsection{Vanishing $qV$ line widths versus Friedel oscillations} \label{subsec:narrowQV_vs_Friedel}
A notable feature of the DFT calculated statistics presented in \cref{sec:LSMSresults} is the extreme narrowness of the $qV$ trendlines. In many binary cases, the statistical widths are so negligible that identifying a conditional spread of charge transfer $\delta n_i$ for a given Madelung potential $\phi_i$ becomes numerically challenging.

Within our perturbative framework, the width of the $qV$ trend is governed by the nonlocal expansion coefficients $B_{ij}$. Physically, these coefficients represent the Friedel oscillations—the redistribution of charge at site $i$ caused by an impurity at a distant site $j$. If we neglect these nonlocal contributions, the charge transfer simplifies to:
\begin{equation}
\delta n_i = A \epsilon_i + A \phi_i.
\label{eq:delta_n_noB}
\end{equation}
In this limit, once the species ($\epsilon_i$) and the local electrostatic environment ($\phi_i$) are specified, $\delta n_i$ becomes purely deterministic.

While the random distribution of surrounding impurities still creates a spread in the values of $\phi_i$ itself, the absence of $B_{ij}$ collapses the joint distribution into a perfect, zero-width line. This suggests that the "tightness" of $qV$ trends in realistic alloys is a direct indicator of the suppression of Friedel oscillations by efficient screening.

\subsection{Semiclassical approximation and comparison to full theory} \label{subsec:semiclassicalCompare}

To determine whether these nonlocal effects are essential to the alloy's description, we can adopt a semiclassical approximation by setting $B \rightarrow 0$ throughout our theory. This modification leaves the general structure of our linear response equations (\ref{eq:Mk}--\ref{eq:φi}) intact but simplifies the response functions $M$ and $\widetilde{M}$.

We find that the results of this semiclassical approach are in excellent qualitative and quantitative agreement with our full quantum theory. As shown in \Cref{fig:rescaleFactorsCompare}, the averages ($\mu_{x,y}$), slopes ($m$), and trendline lengths ($\sigma_{x'}^2$) calculated under the $B \rightarrow 0$ limit match the full theory with only minor discrepancies in the strength of filling dependence. The only statistical quantity that fundamentally changes is the line width $\sigma_{y'}^2$, which vanishes by construction.

\begin{figure*}[t!]
    \centering
    \includegraphics[trim={0.3cm 0.2cm 0 0}, clip, width=0.65\linewidth]{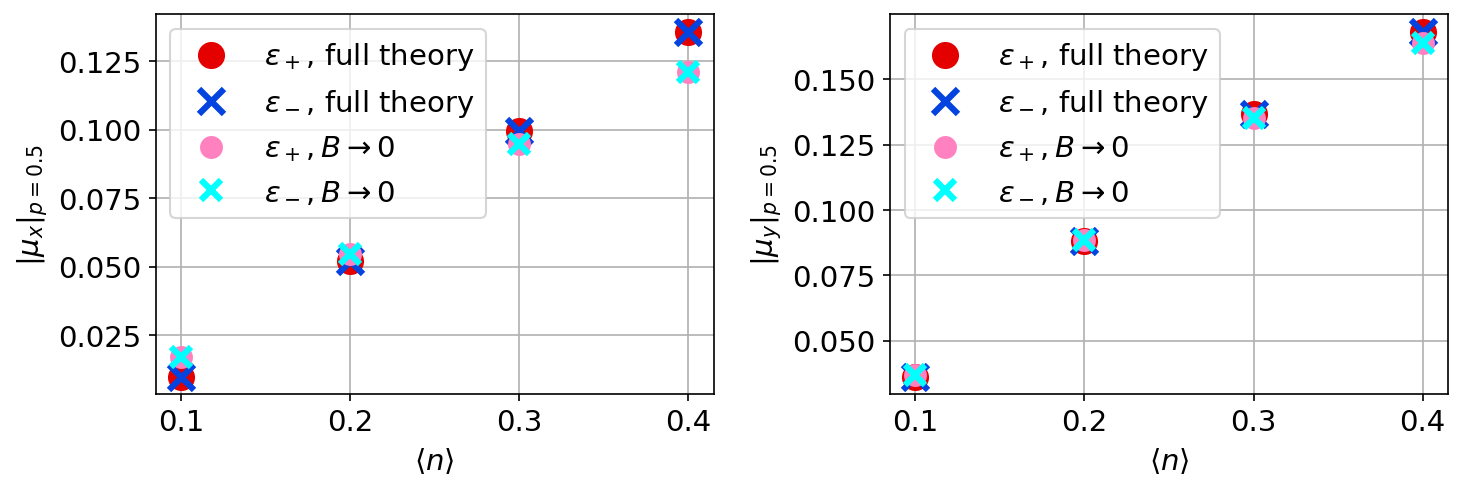}
    \\
    \includegraphics[trim={0.1cm 0.2cm 0 0}, clip, width=0.4\linewidth]{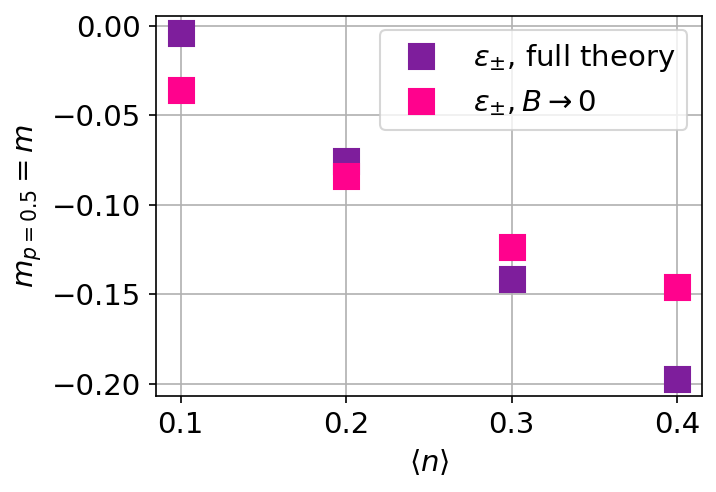}
    \\
    \includegraphics[trim={0.3cm 0.2cm 0 0}, clip, width=0.65\linewidth]{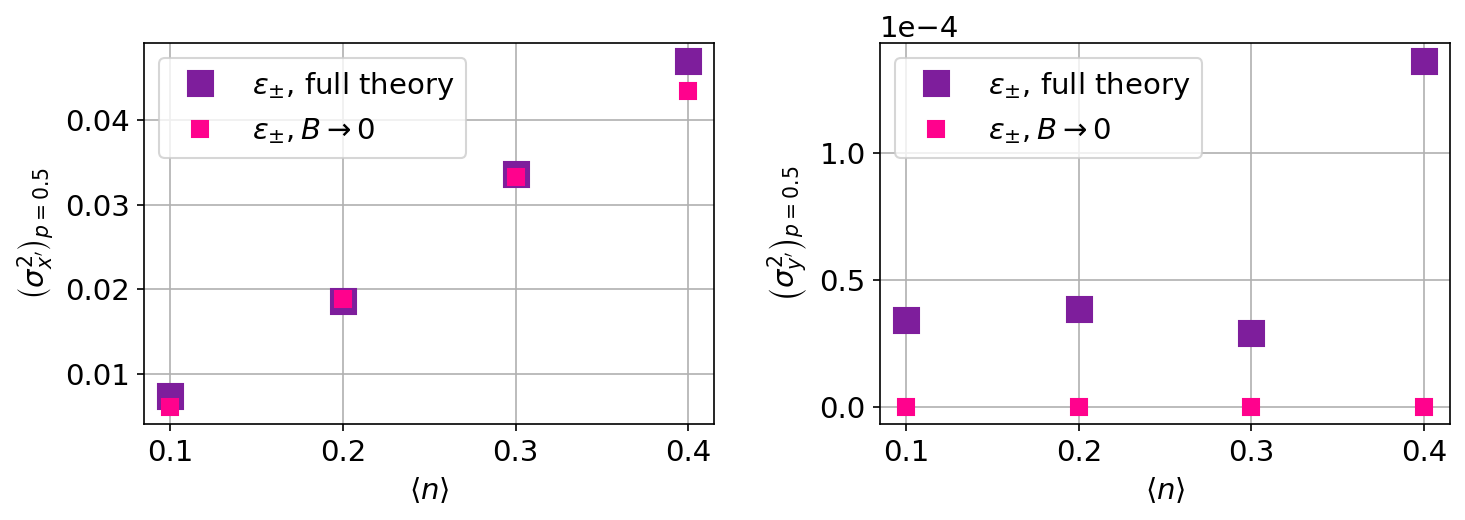}
    \caption{Disordered Hartree absolute averages (top row), $qV$ slopes (middle row), and line dimensions (bottom row) as functions of filling $\langle n \rangle$ in the equiprobable binary alloy with positive impurity concentration $p=0.5$ and interspecies energy splitting $\Delta=\epsilon_+ - \epsilon_- = 2$. Compared here are results obtained using a semiclassical version ($B \rightarrow 0$) of our theory and those produced with our full theory, the latter having been copied over from the earlier \Cref{fig:rescaleFactors}. }
    \label{fig:rescaleFactorsCompare}
\end{figure*}


This comparison suggests that the semiclassical approximation is highly effective for metallic alloys. Physically, the mobility of the electron gas in these systems allows for rapid screening of defects, which effectively truncates the long-ranged oscillations and justifies a local, semiclassical treatment of the charge-potential relationship.

\subsection{$qV$-corrected-CPA and virtual crystal approach for perturbative first principles modeling of disorder} \label{subsec:CPAextensions}
The success of our linearized disordered Hartree model offers a clear roadmap for enhancing modern first-principles techniques, most notably the Coherent Potential Approximation (CPA). In its standard formulation, CPA is a locally embedded mean-field theory that effectively replaces a disordered environment with a uniform, configurationally averaged medium. While this approach is computationally efficient, it inherently fails to capture the local electrostatic fluctuations—such as the Madelung potential variations—that are ubiquitous in real random alloys. By assuming that every atom of a given species perceives the exact same environment, conventional CPA effectively zeroes out the rich statistical landscape of charge transfer and potential shifts that our model has shown to be physically significant.

To address these shortcomings, our work proposes a more efficient ``$qV$-corrected CPA'' framework that leverages perturbative strategies to recapture these missing fluctuations without the prohibitive cost of large-scale supercell calculations. One such strategy involves the use of the Virtual Crystal Approximation (VCA) to establish a homogeneous bare reference state. By treating the VCA medium as the unperturbed system, one can analytically or numerically determine its linear response to specific atomic substitutions. This response directly provides the $qV$ slopes ($m$) and intercepts ($b$) necessary to apply electrostatic corrections within the CPA cycle, circumventing the need for computationally intensive large supercell DFT calculations.  Furthermore, while our current derivations utilize a simplified single-band representation for clarity, the underlying physics is robust and readily generalizable to complex, multiband electronic structures. As detailed in the Supplemental Materials, the fundamental linear relationship between charge transfer and the Madelung potential remains invariant even when the model is extended to accommodate the multiple orbitals and intricate symmetries characteristic of realistic crystalline materials. This generalizability ensures that $qV$-based corrections can be integrated into high-fidelity density functional theory (DFT) workflows, providing a more accurate description of the local chemical environment in high-entropy and conventional alloys alike.

\subsection{Nonperturbative extensions of CPA with semiclassical extended dynamical mean-field methods}

For regimes where perturbations are not sufficient, our theory points toward a unified CPA+EDMFT (Extended Dynamical Mean-Field Theory) approach. While standard DMFT focuses on local correlations, EDMFT introduces a self-consistent bosonic bath that can capture nonlocal interactions, such as the fluctuating Madelung fields discussed here.

A semiclassical rendering of EDMFT—where the time-dependence of the bosonic bath is ignored—effectively maps the problem onto a static yet spatially randomized potential profile. This reduction makes the problem significantly more manageable while still capturing the crucial $qV$ correlations. Combining these methods with CPA would provide a powerful, nonperturbative tool for modeling the electronic properties of high-entropy alloys and other complex disordered materials.


\section{CONCLUSIONS}

The electrostatic field within a solid, arising from the spatial distribution of electrons and atomic nuclei, defines a complex potential that governs bonding, charge density, and local dipole moments. In disordered alloys, the response of local charge transfer to fluctuations in the electrostatic environment is found to follow a linear $qV$ relationship discovered in prior first-principles studies. In this work,
we have developed a minimal theoretical framework that identifies impurity scattering and electrostatic screening as the fundamental mechanisms driving charge transfer and Madelung field fluctuations. By constructing a linearized disordered Hartree theory, we have provided the first rigorous explanation for the linear $qV$ relationship.

This $qV$ relation is fundamentally an expression of electronegativity equalization. When dissimilar metals form an alloy, electrons redistribute to balance the electrochemical potential, establishing a contact potential that dictates the local electronic landscape. This has profound implications for catalyst design. In Pt-based or Pd-based alloys, for example, the $qV$ relation allows for the quantitative prediction of how alloying modulates the charge state of surface active sites, which in turn tunes the binding energies of key intermediates such as $\text{*CO}$, $\text{*H}$, or $\text{*OH}$~\cite{Hammer1995,annurev:/content/journals/10.1146/annurev.physchem.53.100301.131630}. By establishing the electrostatic potential as a descriptor for catalytic activity, researchers can leverage these linear scaling relationships to rationally optimize adsorption energetics and lower activation barriers for the oxygen reduction or hydrogen evolution reactions. 

Beyond surface chemistry, the $qV$ relation offers a mechanistic window into the mechanical behavior of alloys. The restoring forces against atomic displacement--and thus the fundamental origins of elasticity and hardness--are deeply sensitive to charge redistribution under strain. Our framework captures this reciprocal coupling: mechanical stress alters the local potential, driving an internal charge flow that modifies the cohesive energy and the system's resistance to further deformation. In complex systems like HEAs, these fluctuations can lead to inhomogeneous charge landscapes that either pin dislocations to enhance strength or facilitate localized brittleness~\cite{CHEN2021116638,MENG2021114104,LEE2022142293,Tandoc2023,AIDHY2024112912}. Understanding the electronic origin of such redistribution reveals the underlying physics of mechanical anomalies, from the "superelasticity" in Ti-alloys~\cite{KIM2006423,KIM20062419,ALZAIN20111464} to the exceptional ductility of noble-metal solutions~\cite{HUO2018208,FU2023145733}.

Our analysis reveals that the $qV$ statistics follow universal scaling laws governed by impurity concentration and carrier filling. We have verified these predictions against large supercell DFT calculations for both binary and quaternary (Cantor) alloys, resolving longstanding questions regarding the center, slope, and spread of these statistical trends. Our investigation highlights several promising directions for future research, notably the development of $qV$-corrected CPA and semiclassical EDMFT methods, which will significantly enhance our ability to predict and engineer the properties of complex, disordered materials. Furthermore, extending the $qV$ relation to incorporate non-spherical charge and potential analysis, and applying this framework to non-metallic systems, such as oxides and two-dimensional heterostructures, may uncover novel phenomena where charge transfer and electrostatic potential couple directly to magnetic or thermoelectric responses. Addressing these frontiers will solidify the $qV$ relation as a cornerstone of next-generation materials design.

\section*{Acknowledgements}
Work in Florida (WDH and VD) was supported by the US NSF grant no. DMR-2409911, and
the National High Magnetic Field Laboratory through the NSF
Cooperative Agreement No. DMR-2128556 and the State of Florida. HT  was supported by the US NSF grant no. OAC-1931367 and NSF DMR-1944974 grant. WM was supported by the US NSF grant no. DMR-1944974. KMT was partially supported by NSF DMR-1728457 and NSF OAC-1931445. YW was partially supported by NSF OAC-1931525. 
An award of computer time was provided by the INCITE program. This research also used resources of the Oak Ridge Leadership Computing Facility, which is supported by the Office of Science of the U.S. Department of Energy under Contract No. DE-AC05-00OR22725.
%


\begin{appendices}


\section{FRIEDEL OSCILLATIONS IN THE DISORDERED HARTREE FRAMEWORK} \label{app:FriedelOsc}


To recover the Friedel oscillations in our linearized disordered Hartree theory, we consider the standard model of an ideal Fermi gas, and embed within it just a single impurity defect. In our perturbative language, this just means that the band-dispersion of our bare reference system assumes the usual quadratic form, while the perturbation is caused by a lone impurity potential. No interactions are present in this picture, and, placing the impurity of strength $\epsilon_j$ on site $j$, the full Hamiltonian is then given by (compare with (\ref{eq:H})) 
\begin{equation}
    \hat{H} = \hat{H}_0 + \hat{V}
    \qquad ; \qquad 
    \left\{
    \begin{aligned}
        \\[-0.35cm]
        \hat{H}_0 
        & = \sum_{\vb{k} } \frac{1}{2}|\vb{k}|^2 \,  \hat{c}^{\dagger}_{\vb{k}}\hat{c}_{\vb{k}}
        \\
        \hat{V} 
        & = \epsilon_j \, \hat{c}^{\dagger}_j \hat{c}_j 
    \end{aligned} .
    \right.
    \label{eq:H_FriedelApp}
\end{equation}
Of course, in the standard textbook approach, we often benefit in taking additional idealizations, though we will postpone these until they become necessary to achieve the expected analytical result. 

For now, we exploit the fact that the above has the same form as our original model (\ref{eq:H}), where comparing with this, we simply replace $\xi_{\vb{k}} \rightarrow |\vb{k}|^2/2$ and $\widetilde{\epsilon}_{i} \rightarrow \epsilon_j \, \delta_{ij}$. Thus, the formalism we had developed in \cref{subsec:DHmodel,subsec:PTgen,subsec:linSC} may still apply, and we can then immediately write down
\begin{align}
        \delta n_i = M_{ij} \epsilon_j \qquad \qquad ; \qquad \qquad M_{ij} = A \delta_{ij} + B_{ij},
        \label{eq:δni_FriedelApp}
\end{align}
as we had in equation (\ref{eq:δni_singleimp}), to treat this non-interacting, single-impurity problem. The (first-order) density response function $M_{ij}$ is given now in this non-interacting limit by our perturbative expansion coefficients, or alternatively, the Lindhard function, these being defined in (\ref{eq:coefficientAquantum}) and (\ref{eq:coefficientBijquantum}) of the main text. Thus, to compute
\begin{equation}
    M_{ij} = -\frac{1}{\pi} \int_{-\infty}^{\mu}\textrm{Im} [{G_0}^2_{ij} ] \, \dd \omega 
    \label{eq:Mij_FriedelApp}
\end{equation}
we must first obtain the position-space matrix elements of the bare (free-particle) Green's function. 

From (\ref{eq:H_FriedelApp}), we know $\hat{H}_0$ is diagonal in $\vb{k}$-space. And with 
\begin{align}
    \hat{G}_{0} 
    & \overset{(\ref{eq:GF})}{=} (\omega^+ - \hat{H}_{0} )^{-1}
    \qquad ; \qquad
    \omega^+ = \lim_{\eta \rightarrow 0^+} (\omega + i\eta),
    \label{eq:G0_FriedelApp}
\end{align}
we similarly obtain the Green's function's diagonal entries in the momentum eigenspace
\begin{align}
    {G_0}_{\vb{k}}
    = (\omega^+ - \hat{H}_{0} )^{-1}_{\vb{k}}
    = \frac{1}{\omega^+ - \frac{1}{2}|\vb{k}|^2}.
\end{align}
We will now proceed to Fourier transforming this back over to a position-space representation. 

Note that, in practice, such transformations implicitly involve finite sums over discrete $\vb{k}$-grids which span the first Brillouin zone. However, as alluded to prior, we now facilitate analytical/integration methods by taking the thermodynamic limit where $\vb{k}$ becomes a continuous variable. Additionally, we'll now treat real-space as continuous, moving away from the lattice model that we have developed and used outside of this appendix; this further allows for $\vb{k}$ to span over all $\mathbb{R}^3$. The appropriate modifications are taken below where we are now explicit with summation/integration bounds. We add further that, as ${G_0}_{\vb{k}}$ is spherically symmetric, we work in spherical coordinates for added convenience.
\begin{align}
    {G_0}_{ij}  
    &  = \frac{1}{N} \sum_{\vb{k} \in \textrm{BZ} } {G_0}_{\vb{k}} \; e^{i \vb{k} \bigcdot (\vb{r}_i - \vb{r}_j)}
    \\
    & \, \, \Big\downarrow - \textrm{ \scriptsize  continuum, thermodynamic limit}
    \nonumber
    \\[0.1cm]
    {G_0}_{ij}  
    & = \frac{ \Omega_0 }{ (2\pi)^3 } \int_{\mathbb{R}^3}  {G_0}_{\vb{k}} \;  e^{i \vb{k}\bigcdot (\vb{r}_i-\vb{r}_j ) } \, \dd^3\vb{k}
    \nonumber \\
    & = \frac{ \Omega_0 }{ (2\pi)^3 } \int_{\mathbb{R}^3}  \frac{1}{\omega^+ - \frac{1}{2} |\vb{k}|^2 } \; e^{i \vb{k}\bigcdot (\vb{r}_i-\vb{r}_j ) } \, \dd^3\vb{k}
    \nonumber \\
    & = \frac{ \Omega_0 }{ (2\pi)^3} \int_0^{\infty} \!\! \int_0^{\pi} \!\! \int_0^{2\pi}  \frac{e^{i k |\vb{r}_i-\vb{r}_j| \cos{\theta} }}{\omega^+ - \frac{1}{2} k^2 } \, k^2 \sin{\theta}  \, \dd k \, \dd\theta \, \dd\phi
    \nonumber \\ 
    \hspace{-1.5cm}  & = \frac{ \Omega_0}{(2\pi)^2} \int_0^{\infty} \frac{k^2}{\omega^+ - \frac{1}{2} k^2} \, \dd k \int_0^{\pi} \sin{\theta}  \, e^{ik|\vb{r}_i-\vb{r}_j|\cos{\theta}} \dd \theta    \hspace{1.5cm} 
    \nonumber  \\ 
    & =  \frac{\Omega_0}{2\pi^2} \frac{1}{|\vb{r}_i-\vb{r}_j|}  \int_0^{\infty} \frac{k \sin{(k|\vb{r}_i-\vb{r}_j|})}{\omega^+ - \frac{1}{2} k^2}   \, \dd k.
    \intertext{Between the final three lines above, we integrate out the azimuthal and polar angles, $\phi$ and $\theta$, respectively -- the prior being trivial and the latter manageable by standard $u$-substitution. Expressing this last result as}
    \hspace{-2cm} 
    {G_0}_{ij}  
    & =\frac{\Omega_0}{2\pi^2} \frac{1}{|\vb{r}_i-\vb{r}_j|} \frac{d}{d|\vb{r}_i-\vb{r}_j|}  \left[  \int_0^{\infty}  \frac{\cos{(k|\vb{r}_i-\vb{r}_j|)}}{\frac{1}{2}k^2 - \omega^+} \, \dd k \right], 
    \hspace{-2cm} 
\end{align}
we next note that the square-bracketed integral can be evaluated by first restoring $\omega^+$ from (\ref{eq:G0_FriedelApp}) and relabeling its real part using an auxiliary momentum variable $k_{\omega} = \sqrt{2\omega}$. This leaves the task of contour integration to achieve what is essentially just the free particle Green's function. By applying the residue theorem, we ultimately find  
\begin{align}
    {G_0}_{ij}  
    & = \frac{\Omega_0}{\pi^2} \frac{1}{|\vb{r}_i-\vb{r}_j|} \frac{d}{d|\vb{r}_i-\vb{r}_j|}  \left[ \lim_{\eta \rightarrow 0^+} \int_0^{\infty}  \frac{\cos{(k|\vb{r}_i-\vb{r}_j|)}}{k^2 - k_{\omega}^2+ i\eta  } \dd k \right],
    \nonumber
    \\
    & = \frac{\Omega_0}{\pi^2} \frac{1}{|\vb{r}_i-\vb{r}_j|} \frac{d}{d|\vb{r}_i-\vb{r}_j|}  \left[ \frac{i \pi}{2} \frac{ e^{i k_{\omega}|\vb{r}_{i} - \vb{r}_{j}| }}{ k_{\omega}} \right],
\end{align}
or more simply,
\begin{equation}
        {G_0}_{ij} = - \frac{\Omega_0}{2\pi} \frac{1}{|\vb{r}_i - \vb{r}_j|}   e^{i k_{\omega} |\vb{r}_i - \vb{r}_j|}.  
\end{equation}

Now we can finally compute the (first-order) density response function
\begin{align}
    M_{ij} 
    & \overset{(\ref{eq:Mij_FriedelApp})}{=} -\frac{1}{\pi} \int_{-\infty}^{\mu}\textrm{Im} [{G_0}^2_{ij} ] \, \dd \omega 
    \nonumber
    \\
    & \overset{\hphantom{(\ref{eq:Mij_FriedelApp}) } }{=} - \frac{\Omega_0}{4\pi^3} \frac{1}{|\vb{r}_i - \vb{r}_j|^2}  \int_{-\infty}^{\mu} \textrm{Im} [e^{2 i k_{\omega} |\vb{r}_i - \vb{r}_j|} ] \, \dd \omega ,
\end{align}
to recover the Friedel oscillations. Note however that, for $\omega < 0$, $k_{\omega} = \sqrt{2\omega}$ becomes imaginary, which results in either unphysical or exponentially suppressed contributions to the result. Indeed the bare band dispersion $|\vb{k}|^2/2$ is positive-definite, which obviates any need to consider negative $\omega$s. We thus take the opportunity to modify the lower cutoff of our integral such that 
\begin{align}
    M_{ij} 
    & =  - \frac{\Omega_0}{4\pi^3} \frac{1}{|\vb{r}_i - \vb{r}_j|^2}  \int_{0}^{\mu} \sin[2 k_{\omega} |\vb{r}_i - \vb{r}_j| ] \, \dd \omega
    \nonumber
    \\
    & =  - \frac{\Omega_0}{4\pi^3} \frac{1}{|\vb{r}_i - \vb{r}_j|^2}  \int_{0}^{k_F} \sin[2 k_{\omega} |\vb{r}_i - \vb{r}_j| ] \, k_{\omega} \,  \dd k_{\omega} 
\end{align}
where we exchange integration variables, $\omega=k_{\omega}^2/2$ for $k_{\omega}$, and we further introduce the Fermi wavevector $k_F = \sqrt{2\mu}$ as the new upper cutoff. This last line is straightforwardly evaluated using integration-by-parts, the result being
\begin{align}
    \hspace{-0.5cm}
    M_{ij}   
     =  \frac{\Omega_0 k_F}{(2\pi)^3}    
    \left[ \;  \frac{ \cos{(2k_F |\vb{r}_i - \vb{r}_j| ) } }{ |\vb{r}_i - \vb{r}_j|^3 } - \frac{ \sin{ (2k_{F} |\vb{r}_i - \vb{r}_j| ) }} {2k_F |\vb{r}_i - \vb{r}_j|^4} 
    \right].
    \hspace{-0.5cm}
\end{align}
This furnishes for us the Friedel oscillations which are generated around a single impurity defect in the ideal Fermi gas. To leading-order in impurity displacement $|\vb{r}_i - \vb{r}_j|$, we recover the standard result where $M_{ij}$ and, through (\ref{eq:δni_FriedelApp}), $\delta n_i$ both oscillate with twice the Fermi wavevector and decay as an inverse cube.


\section{CLASSICAL DISORDERED HARTREE THEORY} \label{app:classicalDH}


We construct here the classical, high-temperature analogue of our disordered Hartree model, as well as its perturbative solution relating fluctuating charge transfer and Madelung field distributions in chemically disordered metals. What follows shall correspond to the theory we had developed for the quantum (zero-temperature) limit in the main body of this work, particularly in \cref{sec:theory}.

We start with the quantum Hamiltonian we had originally presented in (\ref{eq:H}), and suppress now the hopping/tunneling term ($\{ t,\hat{H}_0 \} \rightarrow 0$), as the associated quantum fluctuations become washed out at higher temperatures. Exchanging also each quantum operator (hatted: $\hat{ }$ ) with their classical counterparts (checked: $\check{ }$ ), what remains is a purely local model that is diagonal in the site basis
\begin{align}
    \check{H} 
    = \check{V}
     = \sum_i \widetilde{\epsilon}_i \,  \check{n}_i
    \qquad ; \qquad 
    \widetilde{\epsilon}_i = \epsilon_i + \phi_i.
    \label{eq:classicalH}
\end{align}
Note here that the quantum number operator $(\hat{n}_i = \hat{c}^{\dagger}_i \hat{c}_i) $  has been replaced by its classical complement $(\check{n}_i = 0,1)$, the latter only assuming integer occupation values subject to Pauli exclusion. Now by including an explicit chemical potential term, we can construct the grand canonical Hamiltonian
\begin{align}
    \check{K} 
    & = \check{H} -\mu \sum_{i} \check{n}_i  = \sum_{i} h_i \, \check{n}_i
    \qquad  ;  \qquad
    h_i = \widetilde{\epsilon}_i - \mu,
    \label{eq:classicalK}
\end{align}
the eigenvalues ($h_i$) of which appear in various relevant and useful thermodynamic quantities. 

Of particular and immediate interest to us is the local charge distribution. And in the current high-temperature regime, these are governed by thermal fluctuations in $\check{n}_i$ which are well-described by classical statistical mechanics. Thus, with local site indices ($i$) offering a good basis to work in, we can immediately determine that the average on-site occupancy is captured by the Fermi-Dirac distribution
\begin{align}
    n_i = \langle \check{n}_i \rangle = n_F(h_i) 
    = \frac{1}{e^{\beta h_i} + 1}
    \overset{ (\textrm{\ref{eq:classicalK}}) }{=} \frac{1}{e^{\beta (\widetilde{\epsilon}_i - \mu ) } + 1},
    \label{eq:niFermi}
\end{align}
where $\beta = T^{-1}$ is the inverse temperature, and the angled braces around the classical number operator $\hat{n}_i = 0,1$ are understood to represent its classical/thermal average.

Now in our perturbative language, this above is analogous to the "full" charge profile which includes the effects of both disorder ($\epsilon$) and inter-electron Coulomb repulsion ($\phi$). However, to develop things further, we must separate these perturbative contributions from those which remain in their absence. We therefore Taylor expand the above in powers of the total perturbation ($\widetilde{\epsilon}$) 
\begin{align}
    n_i 
    &  =
    \underbrace{ n_i |_{\widetilde{\epsilon}_i \rightarrow 0}  }_{ {n_0}_i = \langle n \rangle }  
    + \underbrace{ \left. \frac{ \partial n_i }{ \partial \widetilde{\epsilon}_i } \right|_{\widetilde{\epsilon}_i \rightarrow 0} \widetilde{\epsilon}_i }_{\delta n_i}
    + \mathcal{O}[\widetilde{\epsilon}^2].
    \label{eq:niClassicalTaylor}
\end{align}
Thus we find that, to linear-order in $\widetilde{\epsilon}$s, the local charge correction is given by 
\begin{align}
     \delta n_i  & = {n}_{i} - \langle n \rangle = A \widetilde{\epsilon}_i,
     \label{eq:DH_eqn_classicalApp}
\end{align}
where $n_i$ is given in (\ref{eq:niFermi}), while
\begin{align}
    \langle n \rangle 
    = {n_0}_i 
    = n_i |_{\widetilde{\epsilon}_i \rightarrow 0} 
    \overset{ (\textrm{ \ref{eq:niFermi}}) }{=}
    n_F(-\mu)
    = \frac{1}{e^{- \beta  \mu  } + 1}.
\end{align}
is now our classical version of the bare charge profile, and
\begin{align}
    A 
    & =   
    \left. \frac{ \partial n_i }{ \partial \widetilde{\epsilon}_i } \right|_{\widetilde{\epsilon}_i \rightarrow 0} 
    =  \left[ \frac{ \partial}{ \partial \widetilde{\epsilon}_i } \frac{1}{e^{\beta h_i} + 1} \right]_{\widetilde{\epsilon}_i \rightarrow 0}
    \nonumber \\
    & =
    \beta \, \langle n \rangle \, ( \langle n \rangle - 1 ).
    \label{eq:coefficientAclassical}
\end{align}
is the only perturbative expansion coefficient we have left to consider in the classical limit. This, again, reflects the fact that the classical model is purely local, one added consequence of this being the absence of any nonlocal $B_{ij}$ terms in (\ref{eq:DH_eqn1}) as compared to the full quantum analogue (\ref{eq:DH_eqn_classicalApp}). 



\section{INCLUSION OF EXCHANGE-CORRELATION EFFECTS} \label{app:XCeffects}


We take  $\hat{H} \rightarrow \hat{H} + \hat{V}^{\textrm{XC}} $, where
\begin{equation}
    \hat{V}^{\textrm{XC}} = \sum_i  \epsilon^{\textrm{XC}}_i \, \hat{c}^{\dagger}_i \hat{c}_i
    \qquad ; \qquad 
    \epsilon_i^{\textrm{XC}} = \epsilon_i^{\textrm{XC}}[n_i]
\end{equation}
within the local-density approximation made popular by standard density-functional methods. Noting further that,  with $\delta n_i \ll $ for all $i \in N$ as assumed in our perturbative scheme,
\begin{align}
    \epsilon_i^{\textrm{XC}}[n_i] 
    & = 
    \epsilon_i^{\textrm{XC}}[\langle n \rangle] 
    + 
    \underbrace{ 
    \left. \frac{ \partial \epsilon_i^{\textrm{XC}}[n_i] }{\partial n_i}\right|_{n_i = \langle n \rangle } }_{v^{\textrm{XC}}[\langle n \rangle]} 
    \delta n_i
    + \mathcal{O}[\delta n_i^2]
    \\
    & = \epsilon^{\textrm{XC}} + v^{\textrm{XC}} \delta n_i  + \mathcal{O}[\delta n_i^2]
\end{align}
can be truncated to first-order in $\delta n_i$. The zeroth-order term is evaluated with respect to the uniform solution (hence the absence of $\epsilon^{\textrm{XC}}$'s subscript in the second line, and similarly for $v^{\textrm{XC}}$) which provide a uniform shift to every site. It may therefore be absorbed into our chemical potential $\mu$. The first-order term, on the other hand, captures the exchange-correlation effects due to charge fluctuations $\delta n_i$ about the uniform result $\langle n \rangle$, providing our model with an additional source of perturbations. We thus revise our Hamiltonian to include these exchange-correlation effects  (compare with (\ref{eq:H}))
\begin{align}
    \hat{H} = \hat{H}_0 + \hat{V}
    \qquad ; \qquad 
    \left\{
    \begin{aligned}
        \\[-0.35cm]
        \hat{H}_0 
        & = \sum_{\vb{k} } \xi_{\vb{k}} \,  \hat{c}^{\dagger}_{\vb{k}}\hat{c}_{\vb{k}}
        \\
        \hat{V} 
        & = \sum_i (\epsilon_i + \phi_i + v^{\textrm{XC}} \delta n_i ) \, \hat{c}^{\dagger}_i \hat{c}_i 
    \end{aligned}
    \right.
    \label{eq:bareH}
\end{align}
Following the procedure outlined in \cref{subsec:PTgen} of the main text, we then obtain a linearized system of self-consistent equations which are Fourier transformed as follows (compare with (\ref{eq:FT_sysEqns}))
\begin{align}
    \left.
    \begin{aligned}
        {\delta n}_{i}   & = A \widetilde{\epsilon}_i + \sum_{j\neq i} B_{ij} \widetilde{\epsilon}_j 
        \\[0.1cm]
        \phi_i  & = \sum_{j \neq i} V^C_{ij} \delta n_j  
        \\[0.1cm]
        \widetilde{\epsilon}_i & = \epsilon_i + \phi_i + v^{\textrm{XC}} \delta n_i 
    \end{aligned}
    \right\}
    \quad
    \longrightarrow 
    \quad
    \left\{
    \begin{aligned}
        \delta n_{\vb{k}} & =  A \widetilde{\epsilon}_{\vb{k}} + B_{\vb{k}} \widetilde{\epsilon}_{\vb{k}}   \vphantom{ \sum_{j\neq i}  }
        \\[0.1cm]
        \phi_{\vb{k}} & = V^C_{\vb{k}} \delta n_{\vb{k}}
        \vphantom{ \sum_{j \neq i} V^C_{ij} \delta n_j  }
        \\[0.1cm]
        \widetilde{\epsilon}_{\vb{k}} & = \epsilon_{\vb{k}} + \phi_{\vb{k}} + 
        v^{\textrm{XC}} \delta n_{\vb{k}}     
    \end{aligned}
    \right.
    .
    \hspace{-0.1cm}
\end{align}
The $\vb{k}$-space solution is then given by (compare with (\ref{eq:Mk})) 
\begin{align}
    \delta n_{\vb{k}}
     & = M_{\vb{k}} \epsilon_{\vb{k}}
     \quad ; \quad 
      M_{\vb{k}} = \frac{A + B_{\vb{k}}}
      {1 - (A + B_{\vb{k}}) \left(V^C_{\vb{k}} + v^{\textrm{XC} } \right)}
    \\
    \phi_{\vb{k}}
    & = \widetilde{M}_{\vb{k}} \epsilon_{\vb{k}} 
     \quad ; \quad 
        \widetilde{M}_{\vb{k}} = V^C_{\vb{k}} M_{\vb{k}} 
\end{align}

\section{ADDITIONAL DETAILS ON DISORDERED HARTREE CALCULATIONS} \label{ssec:DHdetails}

\subsection{Procedure for calculation of Green's functions} \label{ssubsec:GFcalc}

The (bare) Green's function in the momentum ($\vb{k}$) eigenbasis is mutually diagonal with the nearest-neighbor tight-binding Hamiltonian $\hat{H}_0$ that is chosen as our bare reference system; in this space, its elements are given by 

\begin{equation}
\begin{aligned}
G_{0\vb{k}}
&= \left(\omega^+ - \hat{H}_0\right)^{-1}_{\vb{k}}
 = \frac{1}{\omega^+ - \xi_{\vb{k}}}, \\[3pt]
\xi_{\vb{k}}
&= -2t \sum_{\vb{a}}
   \cos\!\left(\vb{k}\bigcdot\vb{a}\right),
\qquad
\omega^+
= \lim_{\eta\rightarrow0^+}(\omega+i\eta).
\end{aligned}
\label{seq:G0k1}
\end{equation}
Formally, the positive $\eta \rightarrow 0^+$ convergence factor enforces causal dynamics on this retarded Green's function, lending to its meaning and usefulness, e.g. in problems involving linear response. In the limit of vanishing $\eta$, however, ${G_0}_{\vb{k}}$ approaches the real $\omega$-axis with a singular form, making (\ref{seq:G0k1}) impractical to implement directly in our work. For example, its lattice transform,

\begin{equation}
    {G_0}_{ij}
    =
    \frac{1}{N} \! \sum_{ \vb{k} \in \textrm{BZ} }
  {G_0}_{\vb{k}} \, e^{i \vb{k} \bigcdot { (\vb{r}_i - \vb{r}_j) } },
  \label{seq:G0ij}
\end{equation}
which is a major component of our theory, may present with significant noise and sharp, unphysical features -- even if this sum, which we perform using standard fast Fourier transform (FFT) libraries, is taken over a large and demanding number of $\vb{k}$-points.

Now we note that, following major concerted efforts mainly occurring in the 1970s, it is possible to exactly compute such Green's functions for the nearest-neighbor tight-binding model  \cite{economou,ref72economou,ref73economou,ref74economou,ref75economou,ref76economou}. These methods have been developed for a number of different crystal lattices, and they operate in the thermodynamic limit, where $\vb{k}$ is continuous variable and (\ref{seq:G0ij}) becomes an integral over $\vb{k}$. Such procedures are rather involved, however, and lack generality -- often requiring substantial effort and consideration of the lattice geometry to reformulate several of these Green's functions (i.e. taken along particular cuts or planes of the lattice) in terms of elliptical integrals, followed by the derivation of recurrence relations that establish which of these Green's functions are necessary to compute those remaining for arbitrary $(i,j)$. %
%
%
And, although they are extremely precise, such accurate Green's function methods are not particularly necessary for our own purposes of investigating disorder-driven statistics.

To avoid such nontrivial steps/specifications, we instead opt for a simpler, more general procedure where $\eta$ is kept small but finite, allowing the position Green's function (\ref{seq:G0ij}) to be approximated by a contour which sits just above the real $\omega$-axis. We therefore take as our momentum Green's function

\begin{equation}
    G_{0 \vb{k}}
    = \frac{1}{\omega + i\eta  - \xi_{\vb{k}}}
    =
    \frac{ \omega - \xi_{\vb{k}} }{(\omega - \xi_{\vb{k}})^2 + \eta^2 } - i \frac{\eta}{(\omega - \xi_{\vb{k}})^2 + \eta^2 },
    \label{seq:G0k2}
\end{equation}
this having the effect of smearing out the unphysical noise to a degree which is set by our choice of $\eta$. Note that striking balance between excessive noise and loss of features due to smearing requires we set $\eta$ to be the order of nearest $\vb{k}$-point spacing, $\sqrt[3]{\Delta^3 \vb{k}}$. With further trial tuning, we eventually settle on $\eta=1.5 \sqrt[3]{\Delta^3 \vb{k}}$ since -- for a sufficiently dense $\vb{k}$-grid of $N=15\times15\times15$ points which recovers the main local features of the thermodynamic limit (see also \cref{ssubsec:convergence}) -- this choice appears to provide adequate noise suppression without sacrificing too much of the Green's function's physical details, as is shown in \Cref{sfig:G0iis_scanN_scan_eta}. We thus make this choice of $\eta$ and $N$ for all position Green's functions and use these to generate any dependent results that are presented in the main body of this manuscript. A selection of those used in our work are provided in the \Cref{sfig:GFs} as well.

Note lastly that the choice of $\eta$ should also be made under the consideration of working energy scales. Though this was not particularly discussed in the main body of our work (hopping $t=1$ taken for all tight-binding calculations presented there), we briefly mention here that when independently tuning bare energy scales -- i.e. by rescaling of hopping amplitude $t$ or, equivalently, the half-bandwidth $6t$ -- we must also rescale $\eta$ commensurately for appropriate renormalization of our Green's functions, as $\{\hat{G}_0,\hat{H}_0,\omega,\eta\}$ all share common (natural) units. See, for example, \Cref{sfig:G0iis_tunet}.

\begin{figure*}[htbp]
    \centering
    \includegraphics[width=\linewidth]{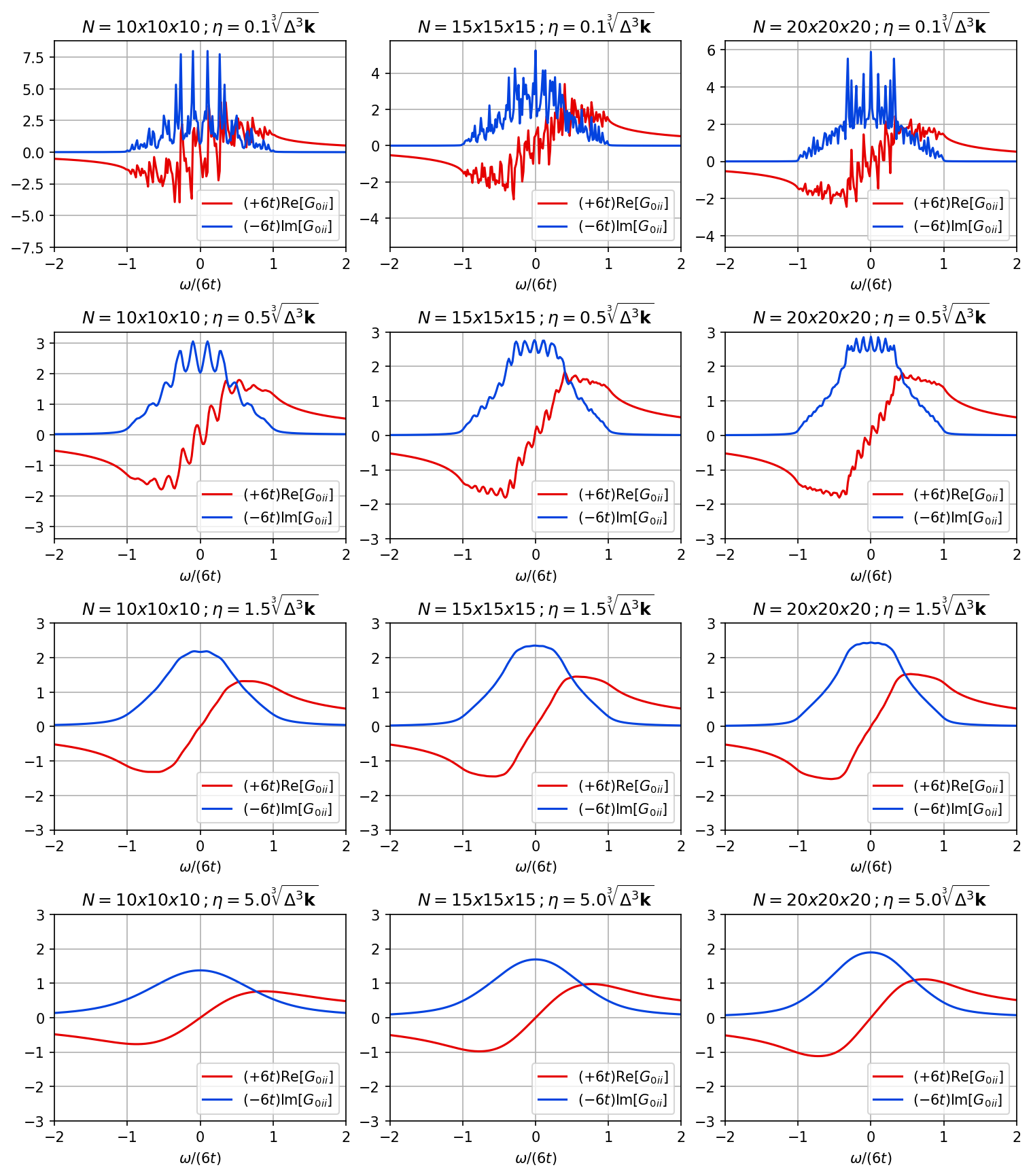}
    \caption{%
    Local Green's function ${G_0}_{ii}$, computed for the nearest-neighbor
    tight-binding model on a cubic lattice with unit hopping amplitude and
    unit lattice spacing, $t = |\vb{a}| = 1$. The red and blue curves
    correspond to the real and imaginary parts, respectively, and are
    collected for comparison after scanning through different $\vb{k}$-grid
    discretizations $N$ and smearing parameters $\eta$. For small $\eta$,
    the local Green's function spectrum is plagued by spurious and
    unphysically spiky features. When $\eta$ is on the order of the
    $\vb{k}$-grid spacing,
    $\eta \sim \mathcal{O}[\sqrt[3]{\Delta^{3}\vb{k}}]$, the spikes have
    largely been smeared out. For $\eta$ taken much larger than this,
    however, the spectral features grow increasingly compromised and
    overwhelmed by the smearing. Note that, in each panel, the frequency
    $\omega$ axis has been reduced by a factor of the half-bandwidth $6t$,
    while the $\textrm{Re}[{G_0}_{ii}]$ and $\textrm{Im}[{G_0}_{ii}]$ axes
    have been scaled by a factor of $\pm 6t$ for ease of comparison with
    the standard result (see, for example, Fig.~5.9 of Ref.~\cite{economou}).
    }
    \label{sfig:G0iis_scanN_scan_eta}
\end{figure*}

\begin{figure*}[hbt!]
    \centering
    \begin{tabular}{c c c}
        \includegraphics[trim={0 0.3cm 0 0cm}, clip, width=0.425\linewidth]{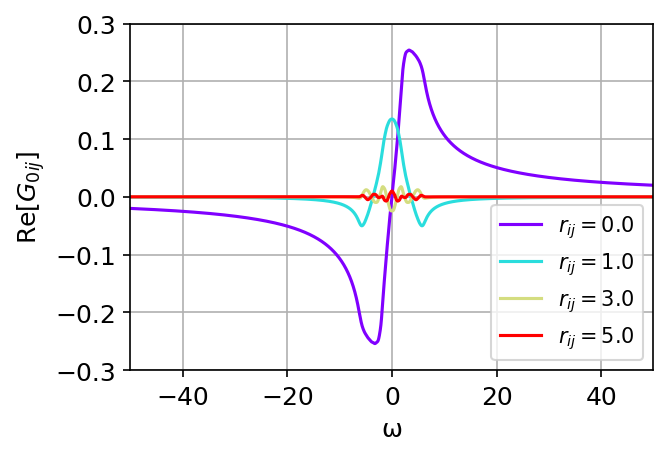}
        & \hspace{0.3cm} &
        \includegraphics[trim={0 0.3cm 0 0cm}, clip, width=0.425\linewidth]{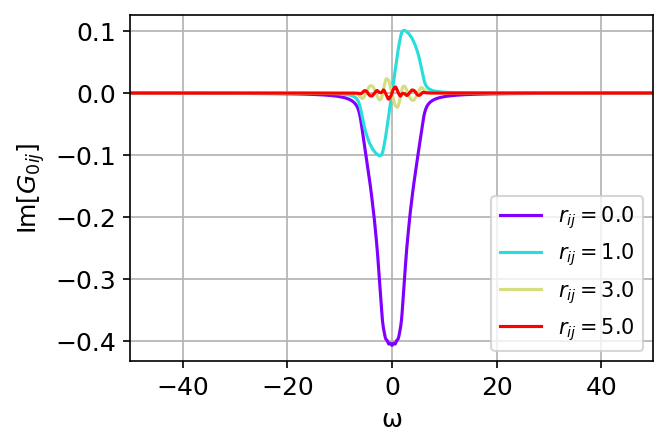}
    \end{tabular}
     \caption{Bare position Green's functions ${G_0}_{ij}$ for a nearest-neighbor tight-binding model on the cubic lattice with unit hopping and unit lattice spacing, $t = |\vb{a}| = 1$. They are obtained using a $\vb{k}$-grid discretization of $N=15\times15\times15$ points with smearing set to $\eta=1.5*\sqrt[3]{\Delta^3\vb{k}}$. They are then plotted as a function of frequency $\omega$ for a few selected site separations $r_{ij} = |\vb{r}_i - \vb{r}_j| = \{0,1,3,5 \}$, and are decomposed into their real (left panel) and imaginary (right panel) parts.}
     \label{sfig:GFs}
\end{figure*}

\begin{figure*}[t]
    \centering
    \includegraphics[
        trim={0 0.3cm 0 0cm},
        clip,
        width=0.85\textwidth
    ]{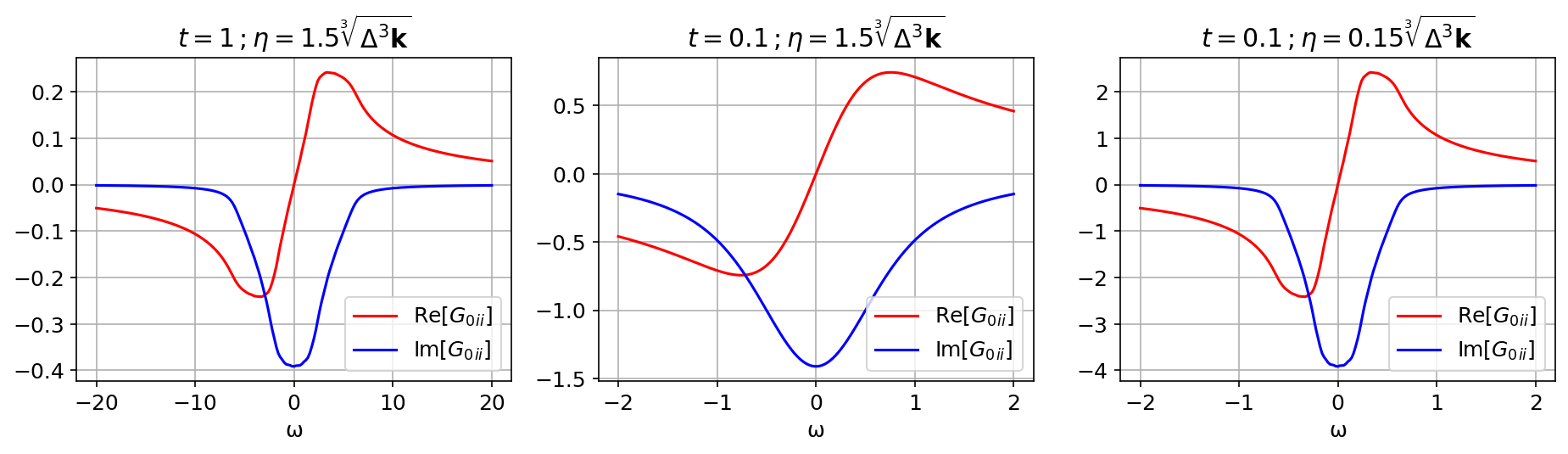}
    \caption{Local Green's functions ${G_0}_{ii}$ for the cubic nearest-neighbor
    tight-binding model, computed on a $\vb{k}$-grid of
    $N=15\times15\times15$ points. In the leftmost panel, we set the hopping
    amplitude $t=1$ and smearing
    $\eta=1.5\sqrt[3]{\Delta^3\vb{k}}$, as is taken throughout the main body
    of our work. In the two rightmost panels, hopping is scaled down to
    $t=0.1$, and the Green's functions are recomputed using both
    $\eta=1.5\sqrt[3]{\Delta^3\vb{k}}$ and
    $\eta=0.15\sqrt[3]{\Delta^3\vb{k}}$, where only the latter retains the
    same form and proportions as those of the $t=1$ curves.}
    \label{sfig:G0iis_tunet}
\end{figure*}

\subsection{Perturbative expansion coefficients} \label{ssubsec:expCoefficients}

With the bare lattice Green's functions which are obtained using the procedure described in the previous \cref{ssubsec:GFcalc}, we may construct the expansion coefficients

\begin{align}
A
&= -\frac{1}{\pi}
\int_{-\infty}^{\mu}
\textrm{Im}\!\left[{G_0}_{ii}^{\,2}\right]
\,\dd\omega ,
\label{seq:coefficientAquantum}
\\
B_{ij}
&= -\frac{1}{\pi}
\int_{-\infty}^{\mu}
\textrm{Im}\!\left[{G_0}_{ij}^{\,2}\right]
\,\dd\omega ,
\qquad (i\neq j).
\label{seq:coefficientBijquantum}
\end{align}


These integrals can be performed numerically using the efficient method of adaptive quadratures. For the cubic model with unit hopping and lattice spacing, and with an associated selection of Green's functions presented in \Cref{sfig:GFs} above, we may take, for instance, the local ($r_{ij}=0$) component and evaluate (\ref{seq:coefficientAquantum}), obtaining $A \approx -0.16...$ when the lattice is set to half-filling ($\mu=0 \, \leftrightarrow \, \langle n \rangle = 0.5$). Similarly, the remaining nonlocal ($r_{ij} \neq 0$) elements of $G_0$ may be entered into (\ref{seq:coefficientBijquantum}) to compute the $B_{ij}$ coefficients; for the same half-filled cubic model, these are collected and scatterplotted in Figure \ref{sfig:Bij_and_Bk}'s left panel. In the right panel, we have $B_{ij}$'s lattice transform $B_{\vb{k}}$ scatterplotted; this data is again obtained using standard FFT libraries and then interpolated using radial basis functions of cubic order. Note also that, because our expansion coefficients must preserve the cubic symmetry we have chosen for the associated bare model, $B_{\vb{k}}$ is plotted just along the $k_x$-axis, as plots along the $k_y$ or $k_z$ axes will be identical in form.

\begin{figure*}[hbt!]
    \centering
    \centering
    \begin{tabular}{c c c}
        \includegraphics[trim={0 0.6cm 0 0cm}, clip, width=0.425\linewidth]{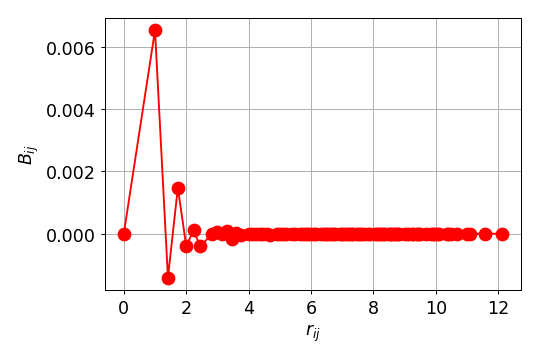}
        & \hspace{0.3cm} &
        \includegraphics[trim={0 0.6cm 0 0cm}, clip, width=0.425\linewidth]{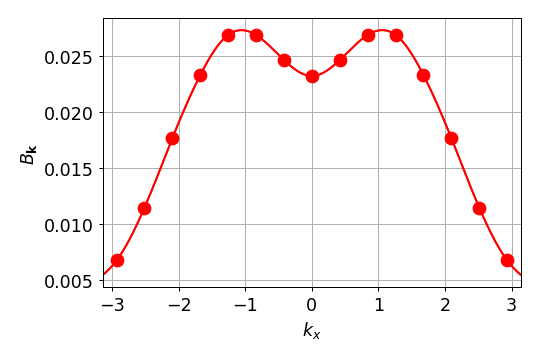}
    \end{tabular}
    \caption{Nonlocal expansion coefficients $B_{ij}$ for the half-filled cubic model are computed using the Green's functions in \Cref{sfig:GFs} and then scatterplotted as a function of scalar site separation $r_{ij}$ in the left panel. They are subsequently Fourier transformed to produce $B_{\vb{k}}$ which is interpolated using a radial cubic spline and plotted along the $k_x$-axis in the right panel.}
    \label{sfig:Bij_and_Bk}
\end{figure*}

\subsection{Note on k-grid discretization and convergence towards the thermodynamic limit} \label{ssubsec:convergence}

By Fourier duality, the aforementioned choice (\cref{ssubsec:GFcalc}) of $N=15\times15\times15$ points for $\vb{k}$-grid discretization implies that position-space is organized into periodically repeated supercells, each consisting of a commensurately numbered $N$ sites. Thus, with the cubic geometry we choose currently for our lattice, the largest site-separation $r_{ij}$ (prior to periodic redundancy) within a supercell acting as our system is $\sqrt{7^2 + 7^2 + 7^2} = 12.124...$ lattice spacings (between a corner of the supercell and the origin). Correspondingly, there lie 7 sites between the origin and any supercell face which sits on a cartesian axis.
                                                                                                                                                                                                                 Now, in principal, we wish to simulate the bulk properties of a model solid where $N$ is taken to be thermodynamically large. However, as can be inferred from \Cref{sfig:GFs} and \Cref{sfig:Bij_and_Bk}('s left panel), the amplitudes of the (nonlocal) Green's functions and corresponding $B_{ij}$ coefficients all become quickly suppressed prior to this 7-site cutoff. This suggests that the relevant nonlocal processes which have any considerable effect on the origin and are sourced from the same supercell have been properly accounted for. We may therefore take this as one indication that the $\vb{k}$-grid discretization (or system size/periodization) has been sufficiently chosen to reproduce the main local features of the thermodynamic limit -- in avoidance of finite-size effects.

This is true, at least, for the current choice of (bare/non-interacting) model parameters which have been made at the outset%
\footnote{I.e. $t = |\vb{a}| = 1$ and $\mu = 0$ for the current working example of a half-filled cubic model with unit hopping and lattice spacing.}.
Nonetheless, for different model parameterizations, it is simple enough to perform this basic check of $\vb{k}$-grid density (system size) by plotting these quantities in real-space and ensuring they become negligible as $r_{ij}$ approaches the boundary of the supercell. See also the bottom two panels of \Cref{sfig:ResponseFunx}, where we go on to incorporate the effects of electrostatic interactions and complete our linear response theory -- here we may similarly verify that the real-space profiles of the response functions $M_{ij}$ and $\widetilde{M}_{ij}$ vanish prior to the supercell boundary.

\subsection{Ewald sums for Fourierization of Coulomb lattice potentials} \label{ssubsec:Ewald}

In addition to the perturbative expansion coefficients which are discussed in the earlier \cref{ssubsec:expCoefficients}, what is lastly required to compute the desired response functions is the lattice transform $V^C_{\vb{k}}$ of the Coulomb potential
\begin{align}
    V^C_{ij} = \frac{1}{|\vb{r}_{ij}|}
    \qquad \qquad ; \qquad \qquad
    \vb{r}_{ij} = \vb{r}_{i} - \vb{r}_j.
    \label{seq:VCij}
\end{align}

The lattice sums (over $\vb{r}_{ij}$s) that are involved -- and that carry these long-ranged inverse power-law contributions, of which there are, in principal, infinitely/thermodynamically many -- are ill-posed operations, however, as attempts to directly perform this sort of arithmetic produces a $V^C_{\vb{k}}$ that converges both slowly with the number of terms, as well as conditionally -- depending even on the order in which the terms are being summed \cite{AshcroftMermin,toukmaji_ewald_1996,Ewald_LeeAndCai2009}.

To evaluate $V^C_{\vb{k}}$ in a more efficient and appropriate manner, we use an approach which is based on the original work by Peter Ewald \cite{ewald_1921} and belongs to a standard class of methods that are designed to treat such poorly convergent sums. Such Ewald summation techniques have found applications in various problems, allowing for the accurate simulation of long-ranged interactions in both model and realistic systems alike, and enabling the study of a number of important phenomena -- cohesiveness of solids \cite{AshcroftMermin,lanata_2019}, ordering, melting, and glassification of electrons \cite{Pankov_2007,nearlyFrozenCoulLiq_yohanes2011,yohanes_2012,yohanesThesis2013}, as well as the molecular dynamics of biologically relevant structures and solutions \cite{toukmaji_ewald_1996}, to name a few.

With such wide uses and applications, there are a number of ways to formulate and implement these Ewald sums, most of which are not-at-all unique nor unrelated to one another. Common to all of these schemes, for instance, are the separation of the long-ranged power law into two main parts -- one being a ``short-ranged" contribution whose sum properly converges in real-space, and another ``long-ranged" term converging in reciprocal-space. In particular, if we work explicitly under periodic boundary conditions that are applied to a supercell%

\footnote{Earlier chosen in the working example to be a cubic cell consisting of $N=15\times15\times15$ sites.} %
of volume $\Omega$, then the interaction potential between $\vb{r}_i$ and all periodic repetitions of $\vb{r}_{j}$ that are generated by an infinite set of supercell translations $\vb{nL}$ may be rewritten from as (\ref{seq:VCij}) as
\begin{align}
    V_{ij}^C(\vb{r}_{ij} \rightarrow \vb{r})
    & =
    \sideset{}{'} \sum_{ \vb{nL} } \frac{1}{|\vb{r} + \vb{nL} |}
    \nonumber
    \\
    & =
    \sideset{}{'} \sum_{ \vb{nL} }
    \frac{\textrm{erfc} ( \alpha |\vb{r}+\vb{nL}|) }{ |\vb{r} + \vb{nL} | }
    +
    \frac{ 4 \pi }{ \Omega } \sum_{ \vb{k} } \frac{ e^{-\frac{ |\vb{k}|^2 }{ 4 \alpha^2 }} }{|\vb{k}|^2} \,
    e^{- i \vb{k} \bigcdot{ \vb{r} }}
    -
    \frac{ 2 \alpha }{ \sqrt{\pi} }  \,  \delta_{\vb{r}=0}
    \label{seq:VCEwald_realspace}
\end{align}

where $V^C_{ij}$'s translational symmetry makes it convenient for us to treat $\vb{r}_{ij} \rightarrow \vb{r}$ as a single variable for displacements, and:
\begin{itemize}[nosep,leftmargin=*]
    \item the primed sum over $\vb{nL}$ excludes the home cell ($\vb{nL}=0$) whenever $\vb{r}=0$, avoiding a self-interaction contribution.
    \item the unrestricted $\vb{k}$-sum spans \textit{all} momenta satisfying $\vb{k}\bigcdot\vb{nL} = 2\pi*$integer (including $\vb{k}$ lying beyond the first Brillouin zone).
\end{itemize}
Both sums in (\ref{seq:VCEwald_realspace})'s second line converge appropriately with respect to increasing number of terms -- the first real-space sum being ``short-ranged", therefore, and the second, ``long-ranged" -- and $\alpha$ is an externally specified parameter controlling their relative rate of convergence.

This last result is valid in three dimensions, and may be obtained following a decomposition of (\ref{seq:VCEwald_realspace})'s first line and a series of transformations which are facilitated by various special/named functions (e.g. gamma, error, Misra functions, etc.) and identities. We will not discuss specifics here, but note that much of these general steps are rather standard and taken more explicitly in a number of sources \cite{de_leeuw_1980,smith_1981,Pankov_2007,yohanesThesis2013,stamm_2018}, though some details may vary. For instance, Appendix B of \cite{yohanesThesis2013} presents a periodic supercell construction akin to our own, while Appendix B of \cite{Pankov_2007} offers an alternative that is derived entirely in the thermodynamic limit. In any case, the lattice transform of this last line produces the Fourierized Coulomb potential that we seek

\begin{align}
    V^C_{\vb{k}}
    & = \sum_{\vb{r} \neq 0 }
    \frac{   \, \textrm{erfc} (\alpha |\vb{r}| )  }{ |\vb{r}| }  \, e^{i\vb{k}\bigcdot\vb{r} }
    +
    \frac{  4 \pi }{ \Omega_0 }  \sum_{ \vb{K} }
    \frac{ e^{ - \frac{ |\vb{k}+\vb{K}|^2 }{ 4 \alpha^2 } }  }{|\vb{k}+\vb{K}|^2}
    -
    \frac{  2 \alpha  }{ \sqrt{\pi} }
    \label{seq:VCEwald_momspace}
\end{align}
where we introduce $\Omega_0(=|\vb{a}|^3$ in our cubic model) as the primitive unit cell and also note that:
\begin{itemize}[nosep,leftmargin=*]
    \item the first short-ranged $\vb{r}$-sum is now unrestricted, spanning all sites throughout all (home + periodically repeated) supercells.
    \item the second long-ranged sum is now taken over all reciprocal lattice vectors $\vb{K}$ satisfying $\vb{K} \bigcdot \vb{r} = 2\pi*$integer.
\end{itemize}
With (\ref{seq:VCEwald_momspace}) above, we compute the lattice transform of the interaction potential (\ref{seq:VCij}) by iteratively increasing the number of terms in each sum until both converge to a root mean-squared error < $10^{-8}$. Following the conventional wisdom \cite{toukmaji_ewald_1996}, we select $\alpha = \sqrt{\pi}/L$, where $L(=\sqrt[3]{N}|\vb{a}|)$ is the sidelength of our cubic supercell, so that the two sums converge at the same rate.

\subsection{Linear response functions}
With the lattice transformed expansion coefficients (nonlocal) and Coulomb potentials from the earlier two \cref{ssubsec:expCoefficients,ssubsec:Ewald}, the charge transfer and Madelung response functions, $M$ and $\widetilde{M}$ respectively, may now be computed. Their momentum representations are copied below from (\ref{eq:Mk})
of the main text

\vspace{-0.5cm}
\begin{minipage}[t]{0.45\textwidth}
\begin{align}
    M_{\vb{k}}
    & = \frac{A+B_{\vb{k}} }{1- (A+B_{\vb{k}}) V_{\vb{k}}^C },
    \label{seq:Mk}
\end{align}
\end{minipage}
\hfill
\begin{minipage}[t]{0.496\textwidth}
\begin{align}
    \widetilde{M}_{\vb{k}}
    & = V^C_{\vb{k}} M_{\vb{k}} =  \frac{A + B_{\vb{k}}}{\frac{1}{V^C_{\vb{k}}}- (A+B_{\vb{k}}) } .
    \label{seq:Mtildek}
\end{align}
\end{minipage} \\

We use these definitions to generate our response functions, the momentum profiles of which are given in the top row of \Cref{sfig:ResponseFunx} below and their (fast) Fourier transforms scatterplotted in the bottom row.

\begin{figure*}[t]
    \centering
    \includegraphics[
        width=0.85\textwidth
    ]{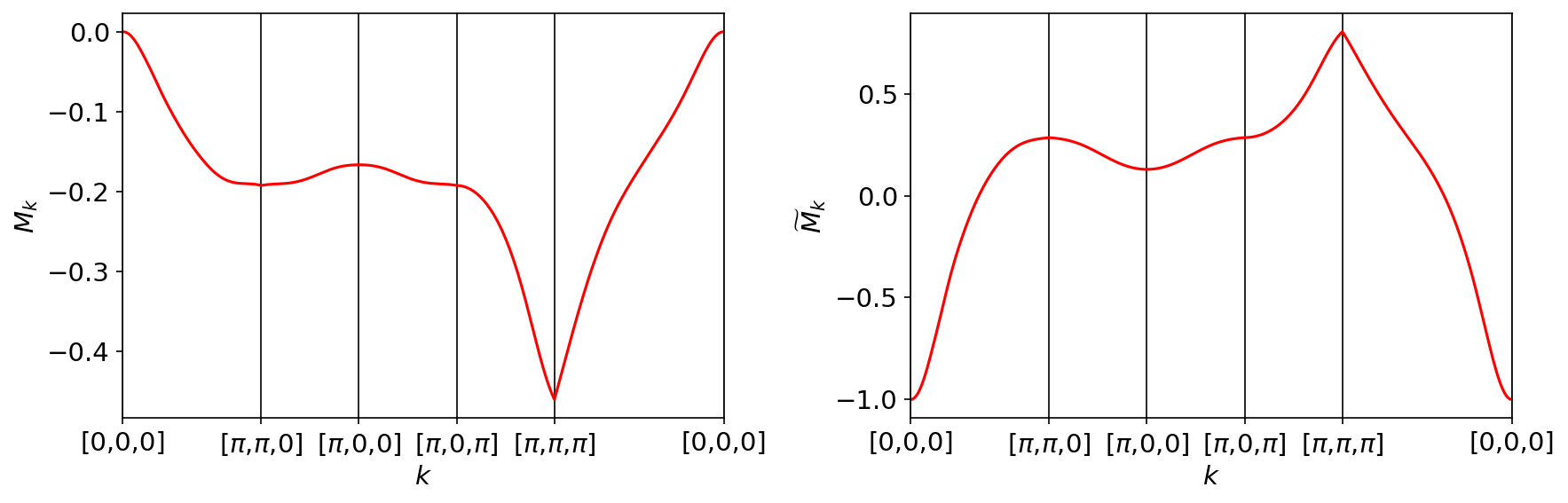}

    \vspace{0.3cm}

    \includegraphics[
        width=0.85\textwidth,
        trim={0 0 -1.5cm 0},
        clip
    ]{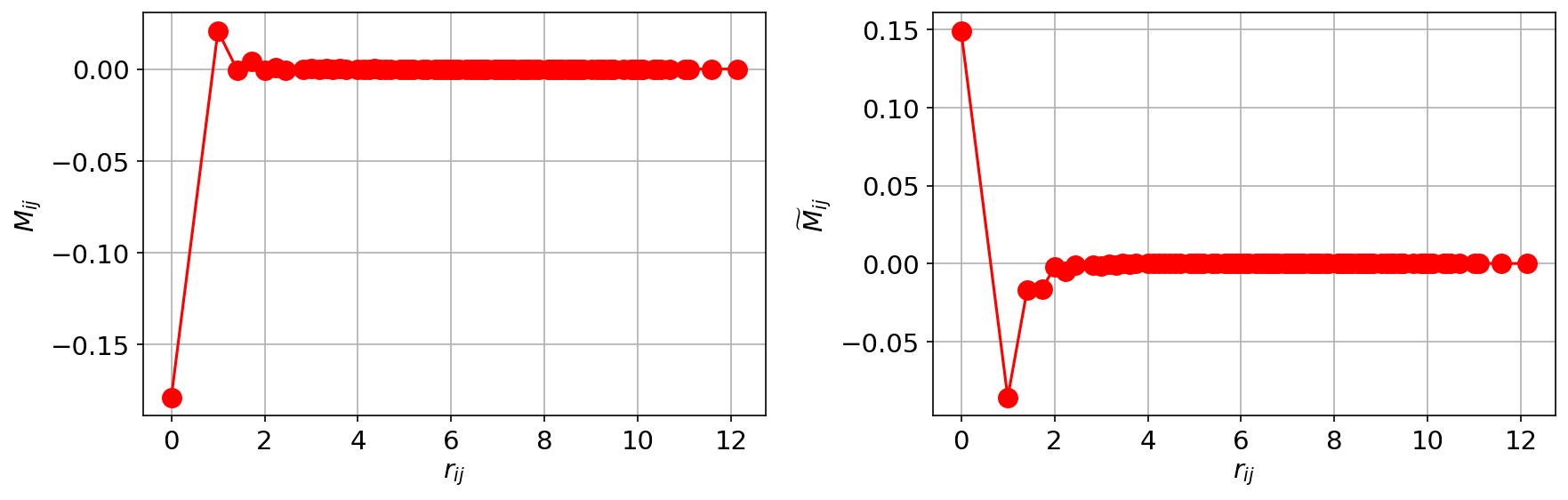}

    \caption{Momentum response functions $M_{\vb{k}}$ and
    $\widetilde{M}_{\vb{k}}$ for the charge-transfer and Madelung fields,
    respectively, are given in the top two panels for the half-filled cubic
    lattice. They are scatter-plotted as a function of scalar site separation
    $r_{ij}$ in the left panel. They are then Fourier transformed to produce
    $B_{\vb{k}}$, which is subsequently interpolated using a radial cubic
    spline and plotted along the $k_x$ axis in the right panel.}
    \label{sfig:ResponseFunx}
\end{figure*}

\section{LSMS APPROACH AND CALCULATION DETAILS} \label{ssec:LSMSdetails}


\subsection{General framework and approach}
We provide here further context and details concerning the LSMS approach implemented in \cref{sec:LSMSresults} of the main text.

In the recent decades, the {\em ab initio} methods based on density functional theory (DFT) have become a widely used tool in computational materials science which allows theoretical prediction of physical properties of materials from the first principles and theoretical interpretation of new physical phenomena found in experiment. In the framework of DFT, an electronic structure calculation for a material generally involves a self-consistent process that iterates between two computational tasks: 1) Solving the Kohn-Sham equation, to obtain the electron density and, if needed, the magnetic moment density; and 2) Solving the Poisson equation to obtain the electrostatic potential corresponding to the electron density and constructing the effective potential by adding the exchange-correlation potential to the electrostatic potential. The self-consistent process proceeds until a convergence criteria is reached. Because of the difference in the basis functions used in the representation of Kohn-Sham orbitals, {\em ab initio} methods differ in the technique used in solving the Kohn-Sham equation and they usually fall into two broad classes: local atomic-like orbital (e.g. Gaussian-type functions, linear-muffin-tin orbitals) based and plane-wave based. Of course, it is possible to avoid using a basis function representation altogether by numerically solving the Kohn-Sham equation on grids or finite elements. The choice of the technique and the basis functions is made to minimize the computational costs and ease the programming efforts, while maintaining a sufficient numerical accuracy.

Despite their methodology differences in the way of solving the Kohn-Sham equation, {\em ab initio} approaches to materials simulation usually take the same computational procedure, starting with constructing a unit cell that repeats itself along $x$, $y$, and $z$ directions to fill the entire space. The unit cell consists of the constituent atoms in a predetermined proportion and in a real space distribution to mimic the atomic composition and spatial arrangement in the actual material. A fundamental problem arises, however, with conventional {\em ab initio} methods when applied to unit cells containing a large number of atoms ($\sim{100}$) that the amount of computational work, or more precisely, the number of floating point operations, increases as $O(N^3_a)$, the third power of the number of atoms ($N_a$) in the unit cell. In plane-wave based {\em ab initio} methods, for example, the Kohn-Sham orbital wave function orthogonalization step scales as $O(N^2N_p)$, where $N$ is the number of Kohn-Sham orbitals, $N_p$ is the number of plane waves, and both $N$ and $N_p$ are proportional to $N_a$, and the electron density calculation step using FFT scales as $O(N{\cdot} N_p\log{N_p})$. For a moderate system size ($N_a \sim 100$ or less), the prefactor for the $O(N{\cdot}N_p\log{N_p})$ scaling processes dominates so that the electronic structure calculation scales approximately as $O(N^2_a\log{N_a})$. For larger system sizes, however, the computing time spent on the orthogonalization step, which essentially scales as $O(N^3_a)$, will become dominant. As for the multiple scattering theory based methods, the $O(N_a^3)$ scaling problem arises from inverting the KKR matrix for the the calculation of the $\tau$-matrix. Since the size of the KKR matrix is proportional to $N_a$, the computational cost for taking KKR matrix inverse for each energy and $\bf{k}$ point scales cubically with respect to $N_a$. Another problem with conventional {\em ab initio} methods is the lack of efficient schemes for parallel implementation, mainly due to the fact that the dominating computational tasks are global in nature. Especially, when the number of atoms is large, very few $\bf{k}$-points are actually needed for the Brillouin zone integration so that the parallelization over the $\bf{k}$-points for the band structure calculation and the Brillouin zone integration no longer shows any advantage.
                                                                                                                                                                                                                 Because of the computational difficulties mentioned above, applying conventional {\em ab initio} methods to the electronic structure calculation for materials with complex structures (e.g., nanostructures, interfaces, defects, etc) that require large unit cell sizes is obviously prohibitive. To overcome this computational bottleneck, much efforts have been made since early 1990s to develop approximate methods to solve the electronic structure for systems involving large unit cell with an acceptable computational cost. As a result of these efforts, several order-$N$ {\em ab initio} methods have appeared, for which the computational effort of the methods scale linearly, i.e. $O(N_a)$, with respect to the number of atoms in the unit cell, rather than cubically like the conventional {\em ab initio} methods. For the rest of this chapter, we present an order-$N$ {\em ab initio} method based on multiple scattering theory that shows clear linear scaling property, and we demonstrate its petascale computing capability in a recent effort in the {\em ab initio} calculation of magnetic phase transition temperatures.  

In the framework of DFT, a major task in an {\em ab initio} electronic structure calculation is to solve the Kohn-Sham equation (in atomic units $\hbar=1$ and $m_e=\frac{1}{2}$) as follows,
\begin{equation}
\left[-\nabla^2+\underline{V}_{\rm eff}({\bf r})\right]\underline{\Psi}_\alpha({\bf r})=\epsilon_\alpha\underline{\Psi}_\alpha({\bf r}),\label{eq:kse}
\end{equation}
where the effective potential $\underline{V}_{\rm eff}({\bf r})$ is a functional of electron density in local density approximation or generalized gradient approximation, and in general, $\underline{V}_{\rm eff}({\bf r})$ is written as a $2\times2$ matrix in spin space since in the case of non-collinear magnetism spin is no longer a good quantum number that the wave function needs to be treated as a $2\times1$ vector in spin space and it is not necessarily an eigenstate of the spin "up" or "down" measurement. Subscript $\alpha$ is a label identifying the quantum state of the Kohn-Sham orbital wave function $\underline{\Psi}_\alpha({\bf r})$, associated with energy eigenvalue $\epsilon_\alpha$. The valence electron density is a summation of the probability density for both spin up ($\sigma=1$) and spin down ($\sigma=2$) states of the valence band:
\begin{equation}
\rho({\bf r})=\sum_{ \substack{ \alpha \\ \epsilon_\alpha\leq\epsilon_{\rm F} } }\sum_{\sigma=1}^{2}\left|\Psi_{\sigma,\alpha}({\bf r})\right|^2,\label{eq:den0}
\end{equation}
provided that the Kohn-Sham orbital wave functions are orthonormal and the Fermi energy $\epsilon_{\rm F}$ is known. In solving the Kohn-Sham equation (\ref{eq:kse}), multiple scattering theory offers a powerful tool. In the multiple scattering theory approach, a crystal is considered as made up of non-overlapping and space filling atomic cells (or voronoi polyhedra), $\left\{\Omega_n,n=1,2,\ldots\right\}$, each of which is centered at an atomic site described by a position vector ${\bf R}_n$ in the crystal and each atomic cell is considered as an electron scattering center with scattering potential $v^n({\bf r}_n)$ identical to the effective potential inside the cell and zero outside, i.e.,
\begin{equation}
\underline{v}^n({\bf r}_n)=\left\{
\begin{array}{ll}
\underline{V}_{\rm eff}({\bf r}), & \mbox{if ${\bf r}\in\Omega_n$;} \\[10pt]
0, & \mbox{otherwise,}
\end{array}
\right. \label{eq:ssp}
\end{equation}
where ${\bf r}_n={\bf r}-{\bf R}_n$ is a coordinates vector centered at ${\bf R}_n$. Solving the one-electron Schr\"{o}dinger (\ref{eq:kse}) can thus be cast into a multiple scattering problem, in which the Bloch wave is essentially a standing wave solution of the multiple scattering processes.

A multiple scattering process in a crystal can be described in terms of so-called scattering path matrix $\underline{\tau}^{nm}(\epsilon)$, which depicts all possible scattering processes of an electron traveling from site $n$ to site $m$. In real space formalism, the scattering path matrix is given by
\begin{equation}
\begin{aligned}
\underline{\tau}^{nm}(\epsilon)
={}&\underline{t}^{n}(\epsilon)\delta_{nm}
\\
&+\underline{t}^{n}(\epsilon)
\sum_{k\neq n}
\underline{g}^{nk}(\epsilon)
\underline{t}^{k}(\epsilon)
\\
&+\underline{t}^{n}(\epsilon)
\sum_{k\neq n}
\underline{g}^{nk}(\epsilon)
\underline{t}^{k}(\epsilon)
\sum_{j\neq k}
\underline{g}^{kj}(\epsilon)
\underline{t}^{j}(\epsilon)
\\
&+\underline{t}^{n}(\epsilon)
\sum_{k\neq n}
\underline{g}^{nk}(\epsilon)
\underline{t}^{k}(\epsilon)
\sum_{j\neq k}
\underline{g}^{kj}(\epsilon)
\underline{t}^{j}(\epsilon)
\\
&\qquad\times
\sum_{i\neq j}
\underline{g}^{ji}(\epsilon)
\underline{t}^{i}(\epsilon)
+\ldots
\\
={}&\underline{t}^{n}(\epsilon)\delta_{nm}
+\underline{t}^{n}(\epsilon)
\sum_{k\neq n}
\underline{g}^{nk}(\epsilon)
\underline{\tau}^{km}(\epsilon).
\end{aligned}
\label{eq:tau_exp}
\end{equation}
where the subscript $nm$ on the right hand side implies the sub-block at the $n$th row and $m$th column of the big matrix after the inverse is taken. For a crystal with infinite number of atoms, the number sub-blocks of the big matrix is of course infinite. However, if the crystal has a periodic structure, the matrix inverse in () can be reduced to the inverse of a matrix with $N_a\times N_a$ sub-blocks.

One important advantage of the MST method is its convenient access to the Green function for one-electron Schr\"{o}dinger (\ref{eq:kse}). In the vicinity of atomic site $n$, the Green function can be expressed in terms of the single site regular solutions $Z^n_L({\bf r}_n;\epsilon)$ and irregular solutions $J^n_L({\bf r}_n;\epsilon)$ due to single site potential $v^n({\bf r}_n)$ as follows,
\begin{equation}
\begin{aligned}
G_{\sigma\sigma^\prime}
({\bf r}_n,{\bf r}^\prime_n;\epsilon)
={}&
\sum_{LL^\prime}
Z^{n}_{L\sigma}({\bf r}_n;\epsilon)
\tau^{nn}_{L\sigma L^\prime\sigma^\prime}(\epsilon)
\\
&\times
Z^{n\ast}_{L\sigma^\prime}
({\bf r}_n;\epsilon)
\\
&-
\sum_{L}
Z^{n}_{L\sigma}({\bf r}_n;\epsilon)
J^{n\ast}_{L\sigma}({\bf r}_n;\epsilon)
\delta_{\sigma\sigma^\prime}.
\end{aligned}
\label{eq:MSTG}
\end{equation}
where matrix index $L$ simply represents a combination of angular momentum quantum number $l$ and magnetic quantum number $m$. This allows us to calculate the electron density and magnetic moment density associated with the valence states in atomic cell $\Omega_n$ by taking the imaginary part of Green function trace integrated in valence energy band,
\begin{equation}
\begin{split}
&\rho^n({\bf r}_n)=-\frac{1}{\pi}\Im{\rm Tr}\int_{\epsilon_{\rm B}}^{\epsilon_{\rm F}}\underline{G}({\bf r}_n,{\bf r}_n;\epsilon)d\epsilon,
\\
&{\bf m}^n({\bf r}_n)=-\frac{1}{\pi}\Im{\rm Tr}\int_{\epsilon_{\rm B}}^{\epsilon_{\rm F}}\underline{\bm{\sigma}}\cdot\underline{G}({\bf r}_n,{\bf r}_n;\epsilon)d\epsilon.
\end{split}
\label{eq:mmd}
\end{equation}
In expression (\ref{eq:mmd}), $\underline{\bm{\sigma}}=(\underline{\sigma}_x,\underline{\sigma}_y,\underline{\sigma}_z)$ is the Pauli matrix vector, $\epsilon_{\rm B}$ is the bottom energy of the valence band, and the energy integration can be conveniently carried along a contour in the complex energy plane so that only few tens of complex energy points are necessary. Obviously, this Green function approach makes the calculation of the band structures and Kohn-Sham orbital wave functions unnecessary, and consequently the time-consuming procedure for orthogonalizing and normalizing the wave functions can be entirely avoided. It is necessary to note that the MST method is an all-electron approach to the {\em ab initio} electronic structure calculation. By all-electron, it means that electronic states for both valence and core electrons are treated on equal footing – in contrast to pseudopotential methods where core electrons are not explicitly treated. Another important feature of the MST method is that the only global operation required for obtaining the Green function is the calculation of a multiple scattering matrix for each atom. It is this step that accounts for the major portion of the floating point operations of the entire electronic structure calculation. Specifically, like other conventional {\em ab initio} methods, the MST method still suffers cubic scaling limitation since the calculation of multiple scattering path matrices in () requires inverse of a matrix whose size is proportional to the number of atoms in the unit cell, and therefore its application is essentially restricted to problems whose size is within less than a thousand atoms.

Moreover, the MST method becomes linear scaling if one approximates the calculation of $\tau^{nn}_{L\sigma L^\prime\sigma^\prime}(\epsilon),~n=1,2,\ldots N_a,$ by neglecting the multiple scattering processes that involve atoms at a distance greater than a cut-off radius $R_{\rm LIZ}$ from atom $n$. This is the essence of the locally self-consistent multiple scattering (LSMS) method. The idea behind this approximation is based on the observation that the scattering processes involving far away atoms influence the local electronic states less and less as the distance from the scatter under study is increased, an example of nearsightedness proposed by W. Kohn. In the LSMS method, the space within $R_{\rm LIZ}$ centered at an atom is called local interaction zone (LIZ) of the atom. If there are $M_a$ atoms in the LIZ centered at atom $n$, the time cost for calculating $\tau^{nn}_{L\sigma L^\prime\sigma^\prime}(\epsilon)$, and thus the Green function, for atom $n$ does not depend on $N_a$, rather it depends on $M_a$. Since we only have to repeat the Green function calculation for each atom, the total time cost for the entire electronic structure calculation will scale linearly with respect to $N_a$, the number of atoms in the unit cell. In addition, parallelism is intrinsic to the method since the Green function calculation for each atom and each energy point along the complex contour is essentially independent. Consequently, there are no global operations involved in the process of calculating the Green function other than few trivial global sum operations, e.g., the summation of the net charge in each atomic cell for the determination of the electron chemical potential, and therefore the code is highly parallel, as demonstrated in Figure. The LSMS method has proved to be a very useful tool for performing the {\em ab initio} calculation for complex structures involving tens of thousands of atoms. It was the first scientific application to pass the teraflop computing speed barrier while investigating the magnetic properties of a non-collinear magnetic structure of 1458 iron atoms, and was one of the very few scientific applications that demonstrated petascale computing capability.

\subsection{LSMS calculation details}

\begin{figure}[htbp]
    \centering
    \includegraphics[
        width=\columnwidth
    ]{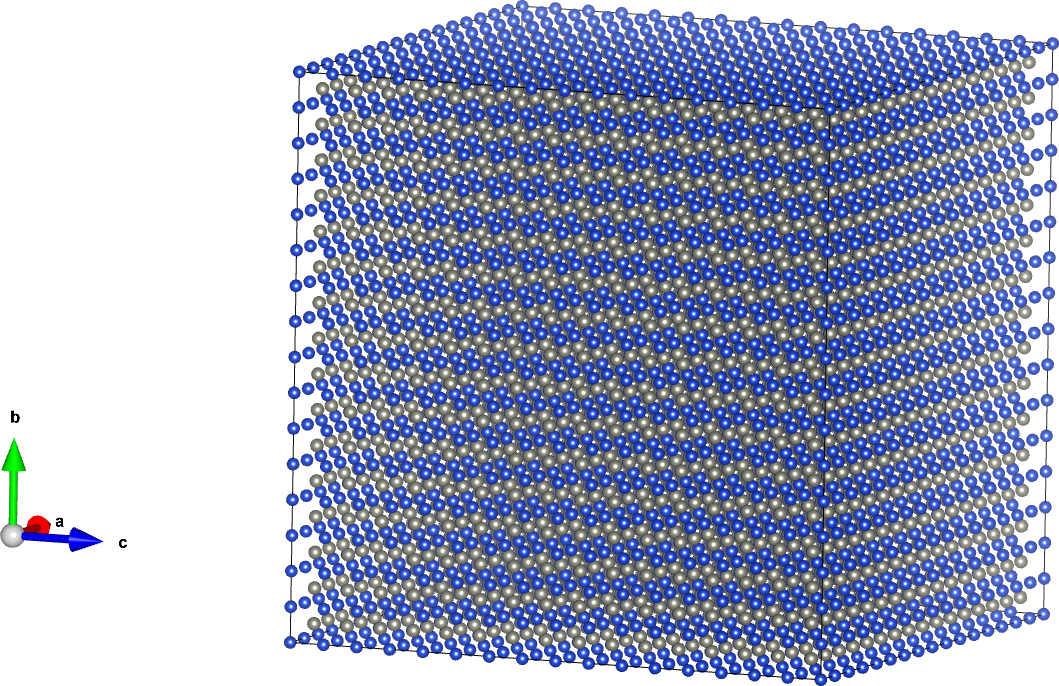}
    \caption{The supercell of CuZn containing 8192 atoms.}
    \label{fig:superCuZn}
\end{figure}

\begin{figure}[htbp]
    \centering
    \includegraphics[
        width=\columnwidth
    ]{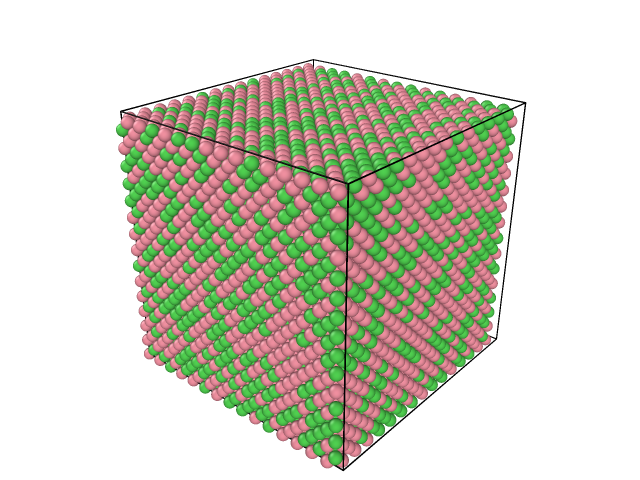}

    \vspace{0.2cm}

    \includegraphics[
        width=\columnwidth
    ]{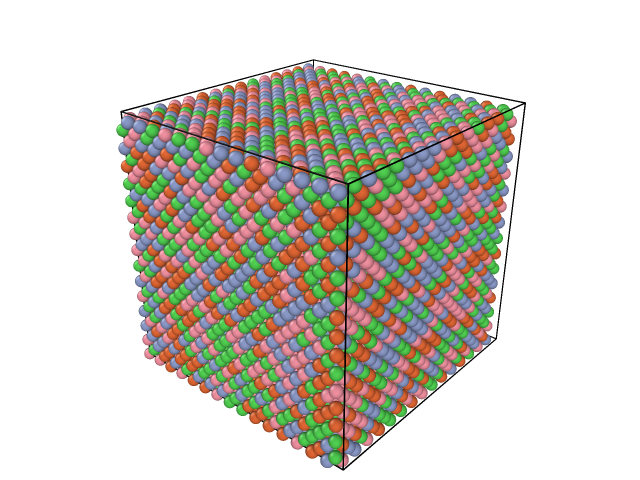}

    \caption{Supercells of CoNi (top) and CoCrFeNi (bottom) with disorder, containing 13500 atoms.}
    \label{fig:FCCHEA}
\end{figure}

Our first-principles calculations were performed using the locally self-consistent multiple scattering (LSMS) method implemented in the MuST package\cite{lsms1,lsms2}. We considered 16$\times$ 16 $\times$ 16 supercells comprising 8192 atoms for the body-centered cubic (BCC) CuZn crystal structure, 14$\times$ 14 $\times$ 14 supercells containing 10,976 atoms for the face-centered cubic (FCC) CuAu structure, 15$\times$15 $\times$ 15 supercells consists of 13500 atoms for both the face-centered cubic (FCC) CoNi and CoCrFeNi structure. We did not perform structure optimization adopting the experimental structures in our calculations, and all these supercells were generated based on the experimental structure. The lattice constants used for CoNi and CoCrFeNi was 3.5325 Angstrom and 3.5767 Angstro, respectively. As an example, the supercell of CuZn used in our calculations is shown in Fig\ref{fig:superCuZn}, and the supercells of FCC CoNi and CoCrFeNi is shown in Fig\ref{fig:FCCHEA}. For CoNi and CoCrFeNi, we use a random distribution of atoms. In our LSMS calculations, we set the LIZ cut-off radius to 18.5 Bohr, with lmax as 4. The convergence with respect to the LIZ cut-off radius was carefully checked for each calculation. We used the generalized gradient approximation (GAA) for the exchange-correlation functional as proposed by Pedrew, Burke, and Ernzerhof\cite{PhysRevLett.77.3865}. The reciprocal space integration was performed by the Brillouin zone sampling with the 2$\times$2$\times$2 Monkhorst-Pack net\cite{PhysRevB.13.5188}.

\subsection{Muffin-tin potential zero and recentering of LSMS Madelung field statistics} \label{ssubsec:MadZero}

The muffin-tin approximation (MTA) is a popular method that allows for the efficient calculation and representation of single-particle wave-/Green's functions in solid bandstructure calculations. It sees regular usage in various implementations of DFT, including those which are suitable for random alloys, such as the KKR-CPA method as well as the LSMS method employed here in our work. In general terms, the MTA typically involves compartmentalizing the solid into (1) nonoverlapping, atom-centered, spherical ``muffin tins" and (2) interstitial regions where the one-electron potential is understood to be (approximately) constant. The Schrodinger equation is then solved in both of these regions with solutions matched at their boundaries in real-space.

There is some arbitrariness, though, associated with how one may construct these muffin-tins, and thus, select the accompanying ``muffin-tin zero" potential found within the interstitial zones -- arbitrariness that should, in principle, be of little consequence in ideally converged DFT results for systems where the MTA is valid\footnote{reference on poor behavior in semiconductors? keller reference...}. This ambiguity, therefore, reflects a kind of gauge freedom that's available in specifying a zero-point offset to the one-electron DFT potential.  In the MuST implementation of LSMS, the muffin-tin zero is taken to be the average electrostatic potential per unit cell. The overall/interspecies averages of the Madelung fields contributing to the DFT potential may thus be found at an arbitrarily offset value, which then varies with the compositional mixture of the alloy (see \Cref{sfig:LSMSMadelungAvgs} below).
                                                                                                                                                                                                                 In contrast to this, we note that our linearized disordered Hartree theory involves introducing perturbatively the electrostatic Madelung (or Hartree) fields, which are taken relative to an empty potential, as well as impurity defects whose statistics are conserving (of global charge and electrostatic Madelung fields). As a result, this constrains our model calculations to produce Madelung field statistics which are globally centered on zero (as stated in (\ref{eq:φi_avgall}) of the main text). For proper comparison between the results of our simplified theory and those of the supporting LSMS calculations, we therefore shift the latter's total/interspecies Madelung field statistics in our post-processing, such that these are recentered to the same zero-point at each level of doping.

                                                                                                                                                                                                                 \begin{figure*}[hbt!] 
																											 \centering
    \begin{tabular}{c c}                                                                                                                                                                                                 \includegraphics[trim={0 0.1cm 0 0}, clip, width=0.45\linewidth]{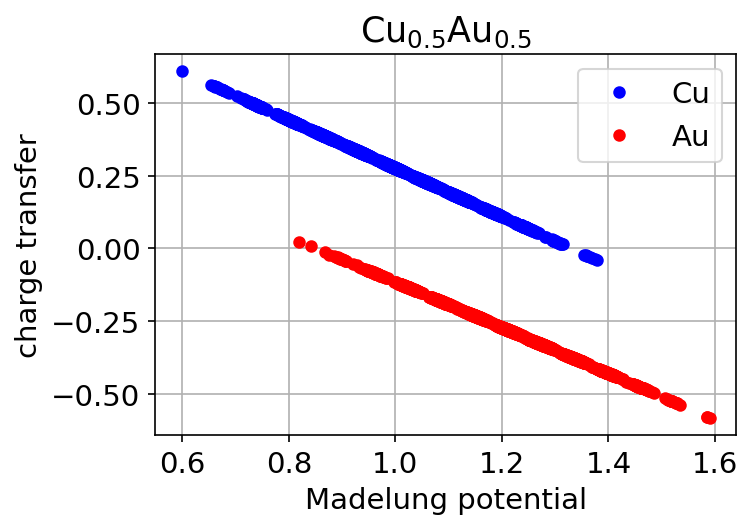}                                                                                              &                                                                                                                                                                                                                \includegraphics[trim={0 0 0.4cm -0.1cm}, clip, width=0.46\linewidth]{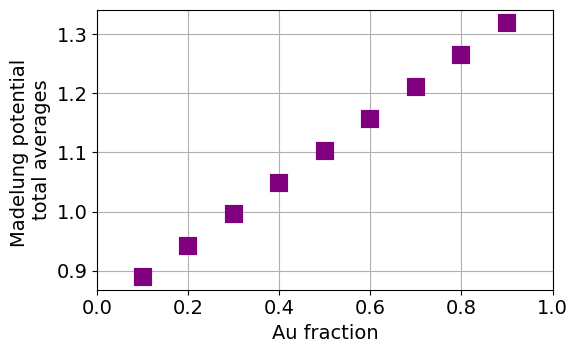}
    \end{tabular}
    \caption{(Left panel) Raw LSMS qV trendlines for the equiatomic CuAu alloy produced using the MuST implementation of LSMS. The marginal statistics of the Madelung potentials on the $y$-axis has a total/interspecies average of $\sim 1.1$. (Right panel) Au concentration dependence of the Madelung field total averages. These are subtracted out of the raw LSMS statistics in our post-processing (i.e. CuAu plots in \Cref{fig:LSMS_concTrendsBinary} of the main text) for appropriate comparison with our model calculations (\Cref{fig:concTrends}).}
    \label{sfig:LSMSMadelungAvgs}
\end{figure*}

\section{NONISOELECTRONIC DOPING AND CONCENTRATION DEPENDENCE OF BINARY STATISTICS } \label{ssec:nonIso}


In the latter parts of \cref{sec:DHstats} in the main text, we discuss how the disordered Hartree statistics vary with relative impurity concentrations as the number of electrons are held fixed across all dopings. However, we may further consider how the statistics are affected once we allow for some degree of nonisoelectronicity to enter our calculations. For the binary alloy treated in \cref{subsec:scanConcProcedure} of the main text, the simplest approach may be to assume some weak linear dependence of chemical potentials $\mu$ on (positive) impurity concentrations $p$.

\begin{figure*}[t]
    \centering
    \includegraphics[
        trim={0 0.8cm 0 0},
        clip,
        width=0.95\textwidth
    ]{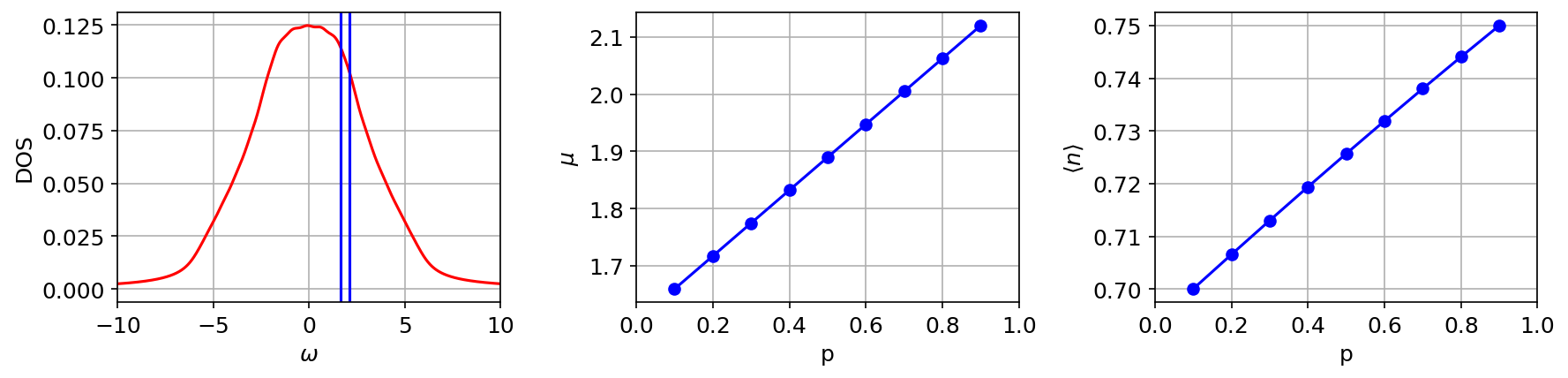}
    \caption{In the middle panel, the chemical potential $\mu$ is assumed to increase linearly with (positive) impurity concentration $p$. Correspondingly, the filling $\langle n \rangle$ is shown in the right panel to increase from $0.7 \leq \langle n \rangle \leq 0.75$, similarly in an approximately linear manner. In the left panel, the (bare) density of states or DOS is plotted in red with the range of chemical potentials bounded by the blue vertical lines.}
    \label{sfig:nonIsoProfiles}
\end{figure*}

                                                                                                                                                                                                                 Assuming, as we do in \Cref{sfig:scanConc_nonIso} above, such weak $p$-dependent profiles for the chemical potentials and fillings, we may reimplement almost exactly the same procedure we had developed in \cref{subsec:scanConcProcedure} to isoelectronically dope through the binary alloy -- the single modification being that our linear response functions $M$ and $\widetilde{M}$, which wholly depend on bare model parameters including $\mu$, are now recomputed at each level of doping $p$. As in \cref{subsec:scanConcProcedure}, we continue to enforce conserving statistics such that the binary site-energies are given as they were in (\ref{eq:ε+-}) of the main text,

\begin{equation}
\left.
\begin{aligned}
\epsilon_- &= -p\,\Delta,\\
\epsilon_+ &= (1-p)\Delta
\end{aligned}
\right\}
\longrightarrow
\begin{aligned}
\llangle \epsilon_j \rrangle
&= p\,\epsilon_+ +(1-p)\epsilon_-\\
&=0.
\end{aligned}
\label{seq:ε=+-}
\end{equation}

																											 \noindent and variances

\vspace{-0.5cm}
\begin{minipage}[t]{0.45\textwidth}
\begin{align}
    \sigma_x^2 = w^2 \sum_{j\neq i} \widetilde{M}_{ii}
\end{align}
\end{minipage}
\hfill
\begin{minipage}[t]{0.496\textwidth}
\begin{align}
    \sigma_y^2 = w^2 \sum_{j\neq i} M_{ii}
\end{align}
\end{minipage} \\

\noindent  of the Madelung field ($x\equiv \phi_i$) and charge transfer ($y\equiv \delta n_i$) statistics, . We can also compute the slope of ther joint statistical qV trendlines
\begin{align}
    m
    & =
    \frac{
        \sum\limits^{j \neq i} M_{ij} \widetilde{M}_{ij}
    }{
        \sum\limits_{j \neq i} \widetilde{M}_{ij}^2
    }
\end{align}
as defined in (\ref{eq:mα_MM~}) as well (dropping species label $\alpha$, of which the slopes are independent in our model). These quantities have been plotted as a function of positive impurity concentration $p$ in \Cref{sfig:scanConc_nonIso} below for the binary alloy with $\Delta=2$. Compared to the trends we have obtained for the binary alloy in \ref{fig:concTrends} of the main text, what differs now -- due to $M$ and $\widetilde{M}$'s newfound $p$-dependence -- is that
\begin{enumerate}[nosep,leftmargin=*]
     \item the Madelung field and charge transfer averages are no longer strictly linear.
     \item the qV trendline slopes are now (weakly) concentration dependent.
     \item the parabolic variances become skewed and are no longer symmetric with respect to half-concentrations ($p=0.5$) values.
 \end{enumerate}

\begin{figure*}[hbt!]
    \centering
    \includegraphics[trim={0.3cm 0.2cm 0 0}, clip, width=0.95\textwidth]{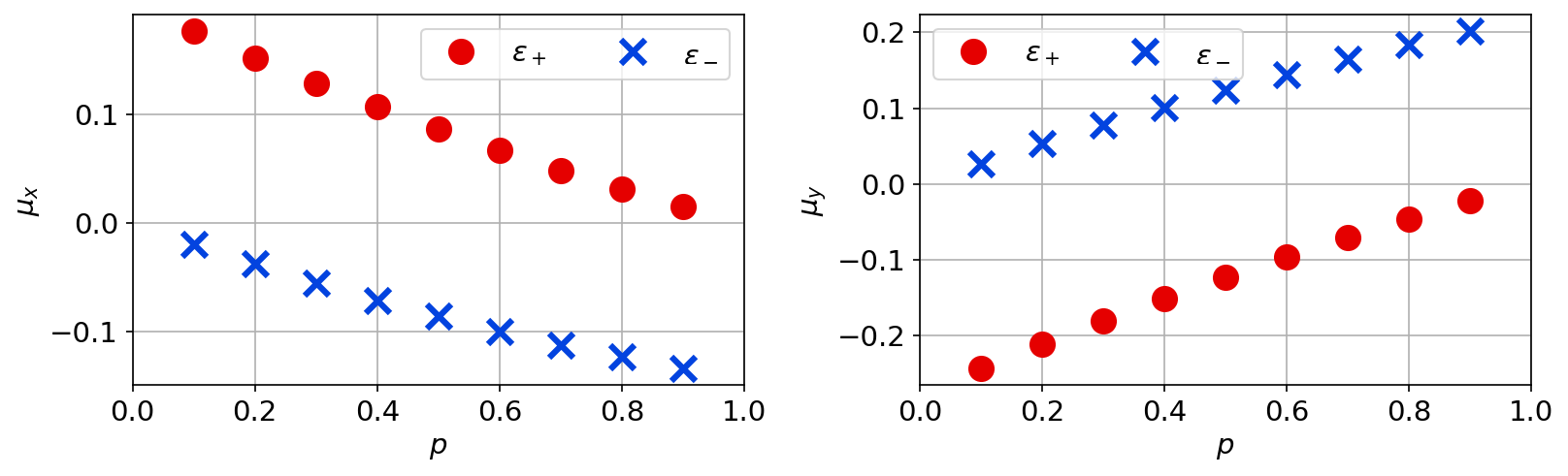}
    \\
    \includegraphics[trim={0.2cm 0.1cm 0 0}, clip, width=0.90\textwidth]{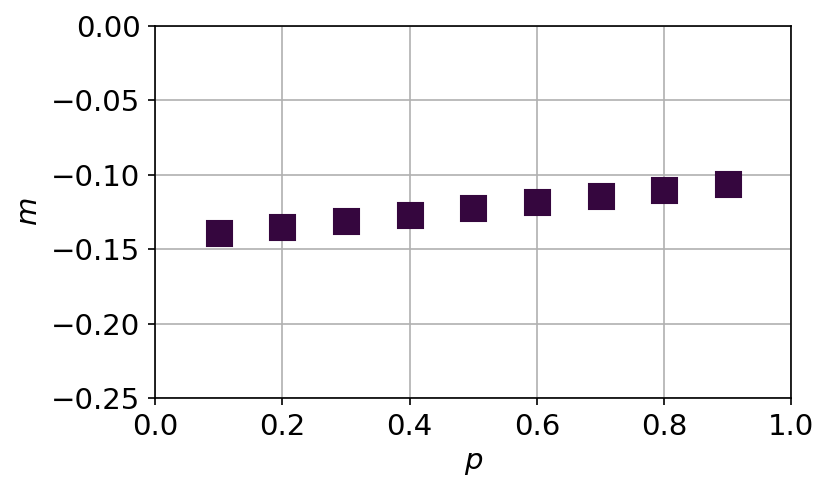}
    \\
    \includegraphics[trim={0.3cm 0.2cm 0 0}, clip, width=0.95\textwidth]{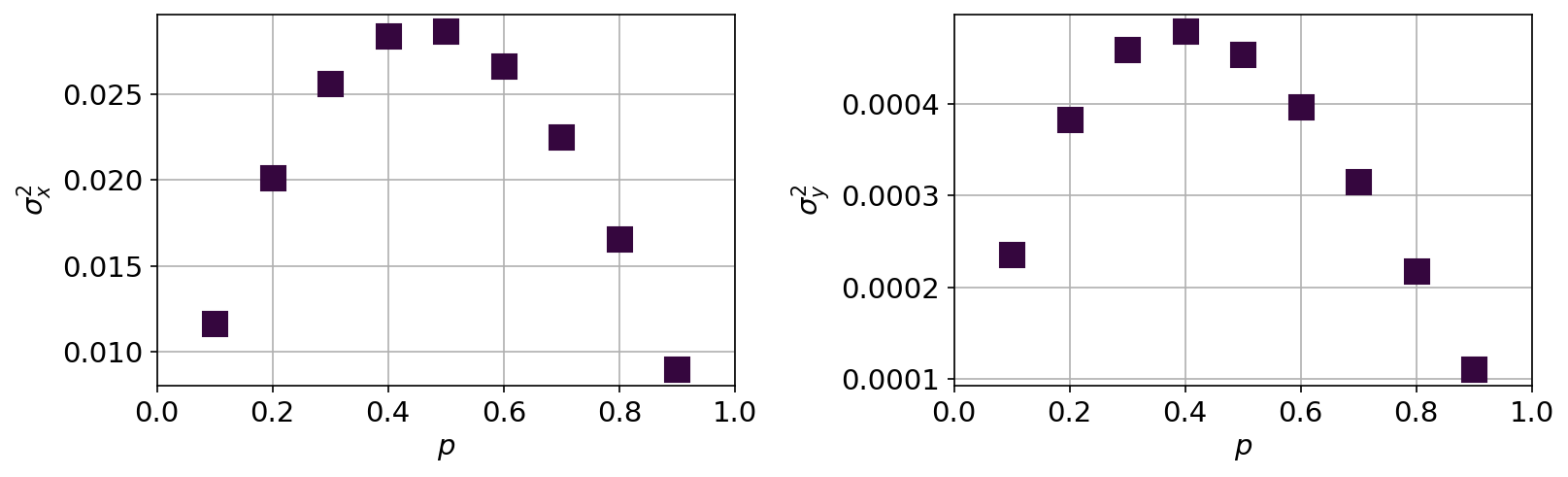}
    \caption{Disordered Hartree statistical averages (top row), qV trendline slopes (middle row), and variances (bottom row) for the binary alloy with interspecies site-energy splitting $\Delta = 2$, and whose filling increases with positive impurity concentration $p$ from $\langle n \rangle = 0.7$ to $\langle n \rangle = 0.75$ across the sample of data points shown.}
    \label{sfig:scanConc_nonIso}
\end{figure*}

\section{MULTI-BAND DISORDERED HARTREE THEORY} \label{ssec:multibandTheory}
In our perturbative disordered Hartree theory, we treat a single-band problem by taking our reference system to be a uniform, nearest-neighbor tight-binding model on a Bravais lattice with just a single orbital per atom per unit cell. The associated Hamiltonian may be written in the site ($i,j$) basis as
\begin{equation}
    \hat{H}_0 = - \sum_{\langle i,j \rangle} t \, \hat{c}_{i}^{\dag} \hat{c}_{j},
    \label{seq:H0_singleband}
\end{equation}
which sums only over the nearest-neighboring $i$-$j$ site-pairs, as denoted by angled brackets. This can be diagonalized in $\vb{k}$-space, e.g. upon the conventional momentization of its creation and annihilation operators

\begin{align}
\left.
\begin{aligned}
\hat{c}^{\dag}_i
&= \frac{1}{\sqrt{N}}
\sum_{\vb{k}} \hat{c}^{\dag}_{\vb{k}}\,
e^{-i\vb{k}\bigcdot\vb{r}_{i}}
\\
\hat{c}_i
&= \frac{1}{\sqrt{N}}
\sum_{\vb{k}} \hat{c}_{\vb{k}}\,
e^{i\vb{k}\bigcdot\vb{r}_{i}}
\end{aligned}
\right\}
\longrightarrow \notag
\\
\hat{H}_0
&= \sum_{\vb{k}}
\xi_{\vb{k}}\hat{c}^{\dag}_{\vb{k}}\hat{c}_{\vb{k}}
\qquad ; \notag
\\
\xi_{\vb{k}}
&= -t\sum_{\bm{\Delta}}
e^{-i\vb{k}\bigcdot\bm{\Delta}}
= -2t\sum_{\vb{a}}
\cos[\vb{k}\bigcdot\vb{a}].
\label{seq:FT_ccdag_singleband}
\end{align}

$\bm{\Delta}$ here represents nearest-neighbor displacement vectors, and by recognizing that these span all primitive translations together with their reflections ($\Delta \in \pm \vb{a}  $ with $ \vb{a} \in \{ \vb{a}_1, \vb{a}_2, ... , \vb{a}_d\}$, see \Cref{sfig:Bravais->Sublattice}'s left panel), we arrive at the standard cosinusoidal sum. From here, our theory proceeds as presented in \cref{sec:theory}. However, one may also consider what changes occur once multiple bands are involved, such as in the case where there are multiple sites per unit cell and/or multiple orbitals per site.

\begin{figure}[hbt!]
    \centering
    \includegraphics[trim={0 4.1cm 0 4.5cm}, clip, width=0.8\linewidth]{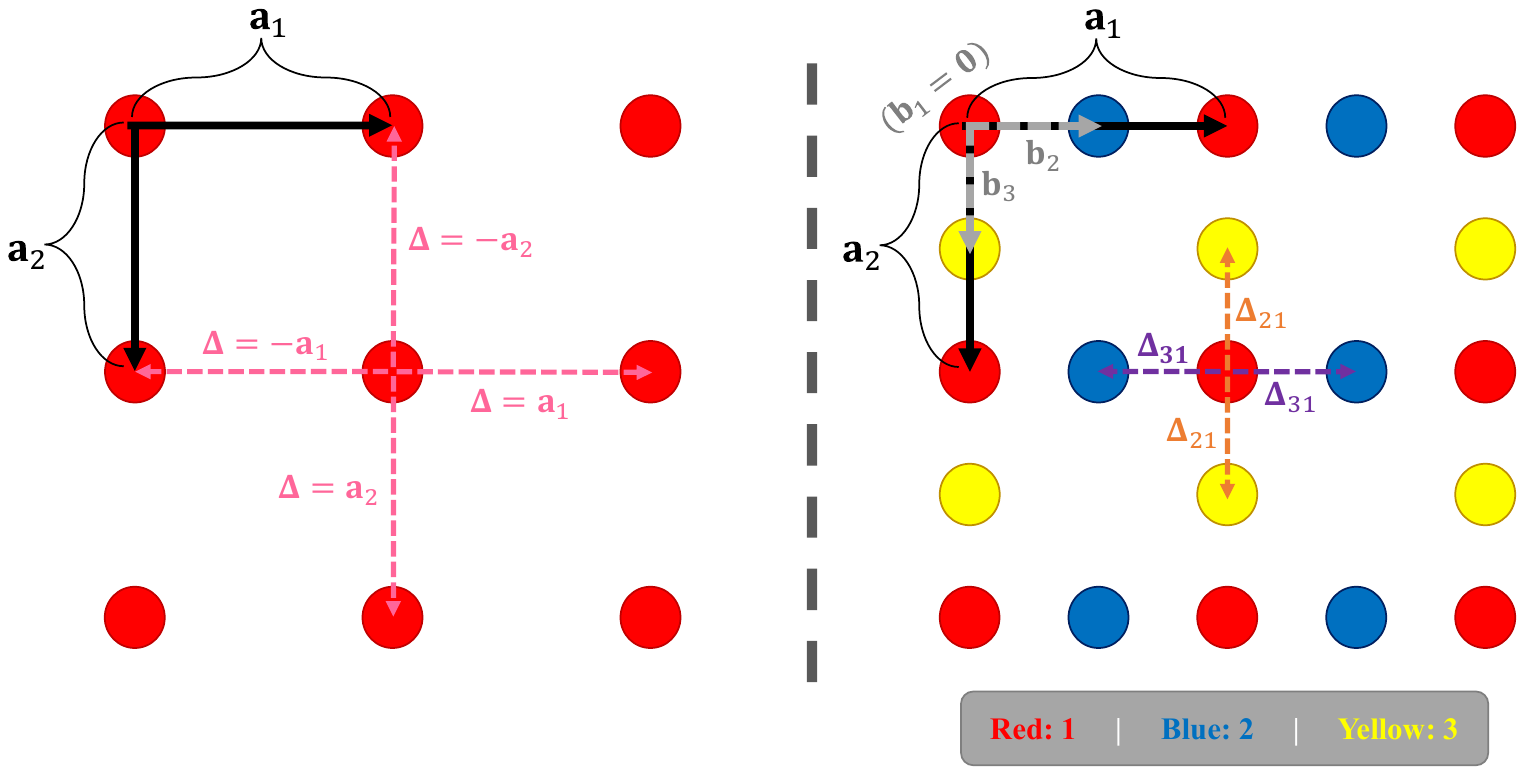}
    \vspace{-0.1cm}
    \caption{(Left panel) Square lattice with only one site per unit cell. (Right panel) Square lattice with a basis consisting of 3 different atoms, labeled \{1,2,3\} corresponding to a \{red, blue, yellow\} color-code. A choice of primitive translation vectors ($\vb{a}_1,\vb{a}_2$) are depicted in both cases. For the prior, the nearest-neighbor displacement vectors are shown to span all primitive vectors and their reflections ($\bm{\Delta} \in \pm \vb{a}$ with $\vb{a} \in \{\vb{a}_1,\vb{a}_2\}$). For the latter, the basis vectors ($\vb{b}_1,\vb{b}_2,\vb{b}_3$) and nearest-neighbor displacements $\bm{\Delta}_{\alpha\beta}$ for $\alpha \in\{2,3\}$ and $\beta = 1$ are also shown.}
    \label{sfig:Bravais->Sublattice}
\end{figure}

Examining for now just the prior case, we consider a nearest-neighbor tight-binding model for a crystal with some sublattice structure. Here, we may divide the lattice into identical unit cells ($i,j$ labels), each with multiple atomic species ($\alpha,\beta$ labels).
\begin{equation}
    \hat{H}_0 = - \sum_{\langle i \alpha,j \beta \rangle} t_{\alpha \beta} \, \hat{c}_{i\alpha}^{\dag} \hat{c}_{j\beta}.
    \label{seq:H0_multiband}
\end{equation}
The nearest-neighbor hopping amplitude $t_{\alpha \beta}$ is presumed to depend only on the identities of the adjacent atoms themselves, and we may apply a momentum transform, similarly as in the single-band problem. For an atom labeled $\alpha$ associated with basis vector $\vb{b}_{\alpha}$ that is in unit cell $i$ located at $\vb{nL}_i$, we write its position as $\vb{r}_{i\alpha} = \vb{nL}_{i} + \vb{b}_{\alpha}$ and have

\begin{align}
\left.
\begin{aligned}
\hat{c}^{\dag}_{i\alpha}
&= \frac{1}{\sqrt{N}}
\sum_{\vb{k}}
\hat{c}^{\dag}_{\vb{k}\alpha}\,
e^{-i\vb{k}\bigcdot\vb{r}_{i\alpha}}
\\
\hat{c}_{i\alpha}
&= \frac{1}{\sqrt{N}}
\sum_{\vb{k}}
\hat{c}_{\vb{k}\alpha}\,
e^{i\vb{k}\bigcdot\vb{r}_{i\alpha}}
\end{aligned}
\right\}
\longrightarrow \notag
\\
\hat{H}_0
&= \sum_{\vb{k}}
\sum_{\alpha,\beta}
\xi_{\vb{k}\alpha\beta}\,
\hat{c}^{\dag}_{\vb{k}\alpha}
\hat{c}_{\vb{k}\beta}
\qquad ; \notag
\\
\xi_{\vb{k}\alpha\beta}
&= -\sum_{\bm{\Delta}_{\alpha\beta}}
t_{\alpha\beta}\,
e^{-i\vb{k}\bigcdot\bm{\Delta}_{\alpha\beta}}.
\label{seq:FT_ccdag_multiband}
\end{align}

where $\bm{\Delta}_{\alpha \beta}$ spans nearest-neighbor displacements between atoms of type $\alpha$ and atoms of type $\beta$ (see \Cref{sfig:Bravais->Sublattice}, right panel). Note that the above recovers the single-band description once we reduce the atomic basis to one and erase all  $\alpha$ and $\beta$ labels.

Now this was just one particular example, but we can of course generalize to more complicated orbital arrangements -- e.g. by lumping their associated degrees of freedom into our species ($\alpha,\beta$) labels. We may also consider more expansive hopping hierarchies (next, next-next, etc. nearest-neighbors). In any case, such problems may always be expressed in this general form

\begin{equation}
\begin{aligned}
\hat{H}_0
&= \sum_{\vb{k}} {\hat{H}_0}_{\vb{k}}
\qquad ; \qquad
{\hat{H}_0}_{\vb{k}}
= \sum_{\alpha,\beta}
h_{\alpha\beta}(\vb{k})\,
\hat{c}^{\dag}_{\vb{k}\alpha}
\hat{c}_{\vb{k}\beta}
\\[4pt]
&=
\underbrace{
\left(
\begin{array}{ccc}
\hat{c}^{\dag}_{\vb{k}1} &
\hat{c}^{\dag}_{\vb{k}2} &
\cdots
\end{array}
\right)
}_{\hat{\Psi}^{\dag}_{\vb{k}}}
\\[-2pt]
&\quad\times
\underbrace{
\left(
\begin{array}{ccc}
h_{11}(\vb{k}) & h_{12}(\vb{k}) & \cdots \\
h_{21}(\vb{k}) & h_{22}(\vb{k}) & \cdots \\
\vdots & \vdots & \ddots
\end{array}
\right)
}_{{\hat{h}_0}_{\vb{k}}}
\underbrace{
\left(
\begin{array}{c}
\hat{c}_{\vb{k}1} \\
\hat{c}_{\vb{k}2} \\
\vdots
\end{array}
\right)
}_{\hat{\Psi}_{\vb{k}}}
\,\,\,\, .
\end{aligned}
\end{equation}

such that ${(\hat{H}_0)}_{ \vb{k} \alpha \beta} = {( {\hat{h}_0}_{\vb{k}} )}_{\alpha \beta} = h_{\alpha \beta}(\vb{k})$ are generic functions of momentum whose $\alpha,\beta$ labels denote the interspecies coupling. Thus, while in the single-band problem,  ${\hat{h}_{0}}_{\vb{k}} \rightarrow \xi_{\vb{k}}$ was taken simply to be a scalar, in the more general problem, there remains some matrix structure, e.g. due to the multi-atom basis and sublattice geometry in the previously considered example (\ref{seq:H0_multiband}). This requires we apply an additional transformation from the species basis ($\alpha,\beta$) to the band basis ($n$) in order to to diagonalize ${\hat{H}_0}_{\vb{k}}$ at each $\vb{k}$. We can express this as

\begin{equation}
    {\hat{H}_{0}}_{\vb{k}} = \hat{\Psi}^{\dag}_{\vb{k}} \, {\hat{h}_{0}}_{\vb{k}} \, \hat{\Psi}_{\vb{k}}
    =
    \overbrace{
        \hat{\Psi}^{\dag}_{\vb{k}} \hat{S}_{\vb{k}}
    }^{
        \hat{\Phi}^{\dag}_{\vb{k}}
    }
    \,
    \underbrace{
        \hat{S}^{-1}_{\vb{k}}  {\hat{h}_{0}}_{\vb{k}}  \,  \hat{S}_{\vb{k}}
    }_{
        {\hat{{\mathcal{H}}}_{0}}_{\vb{k}}
    }
    \overbrace{
        \hat{S}^{-1}_{\vb{k}}    \hat{\Psi}_{\vb{k}}
    }^{
        \hat{\Phi}_{\vb{k}}
    }
     = \sum_{n} \xi_{ n \vb{k} }
    \, \hat{c}^{\dag}_{n \vb{k} } \hat{c}_{n \vb{k}}
\end{equation}

by applying a similarity transformation $\hat{S}_{\vb{k}}$ that is typically constructed through appropriate organization of ${\hat{h}_0}_{\vb{k}}$'s eigenvectors -- i.e. if we consider $m$ degrees of freedom (atoms, orbitals, etc) per unit cell, such that ${\hat{h}_0}_{\vb{k}}$ is an $m \times m$ matrix, then we may express its $n^{\textrm{th}}$ eigenvector (appropriately orthonormalized and in column-format) as $\vb{v}_{n\vb{k}}$ such that the following will produce the diagonal form above.

\begin{equation}
\begin{aligned}
\left.
\begin{aligned}
\hat{S}_{\vb{k}}
&=
\left(
\begin{array}{ccc}
\vb{v}_{1\vb{k}} & \cdots & \vb{v}_{m\vb{k}}
\end{array}
\right)
\\
\hat{S}^{-1}_{\vb{k}}
=\hat{S}^{\dag}_{\vb{k}}
&=
\left(
\begin{array}{c}
\vb{v}^{\dag}_{1\vb{k}} \\
\vdots \\
\vb{v}^{\dag}_{m\vb{k}}
\end{array}
\right)
\end{aligned}
\right\}
&\quad \longrightarrow
\\[4pt]
\hat{\mathcal{H}}_{0\vb{k}}
&=
\hat{S}^{-1}_{\vb{k}}\,
\hat{h}_{0\vb{k}}\,
\hat{S}_{\vb{k}}
\\
&=
\left(
\begin{array}{ccccc}
\xi_{1\vb{k}} & 0 & 0 & \cdots & 0 \\
0 & \xi_{2\vb{k}} & 0 & \cdots & 0 \\
0 & 0 & \xi_{3\vb{k}} & \cdots & 0 \\
\vdots & \vdots & \vdots & \ddots & \vdots \\
0 & 0 & 0 & \cdots & \xi_{m\vb{k}}
\end{array}
\right)
\qquad ; 
\\[4pt]
\hat{\Phi}^{\dag}_{\vb{k}}
&=
\hat{\Psi}^{\dag}_{\vb{k}}\hat{S}_{\vb{k}}
=
\left(
\begin{array}{ccc}
c^{\dag}_{1\vb{k}} & \cdots & c^{\dag}_{m\vb{k}}
\end{array}
\right)
\\
\hat{\Phi}_{\vb{k}}
&=
\hat{S}^{-1}_{\vb{k}}\hat{\Psi}_{\vb{k}}
=
\left(
\begin{array}{c}
c_{1\vb{k}} \\
\vdots \\
c_{m\vb{k}}
\end{array}
\right).
\end{aligned}
\label{seq:H0_multiband_diag}
\end{equation}

Thus, $(\hat{H}_0)_{n\vb{k}} = ({\hat{\mathcal{H}}_0}_{\vb{k}})_{nn} = (\hat{S}_{\vb{k}}^{-1} {\hat{h}_0}_{\vb{k}} \hat{S}_{\vb{k}})_{nn} = \xi_{n\vb{k}}$  provides the dispersion relation for the Bloch band labeled by index $n$.

Now the (bare) Green's function is maximally diagonal in the band-momentum basis ($n\vb{k}$), since we may similarly construct for it an auxiliary object

\begin{equation}
\begin{aligned}
{\hat{\mathcal{G}}_0}_{\vb{k}}
&=
(\omega^+ - {\hat{\mathcal{H}}_0}_{\vb{k}})^{-1}
\\[3pt]
&=
\left(
\begin{array}{ccccc}
(\omega^+ - \xi_{1\vb{k}})^{-1}
& 0 & 0 & \cdots & 0
\\
0
& (\omega^+ - \xi_{2\vb{k}})^{-1}
& 0 & \cdots & 0
\\
0
& 0
& (\omega^+ - \xi_{3\vb{k}})^{-1}
& \cdots & 0
\\
\vdots & \vdots & \vdots & \ddots & \vdots
\\
0 & 0 & 0 & \cdots
& (\omega^+ - \xi_{m\vb{k}})^{-1}
\end{array}
\right).
\end{aligned}
\end{equation}

and show that a similarity transformation, this time in the reverse direction, may be applied at every $\vb{k}$-point to produce the Green's function elements in the momentum-species ($\vb{k}\alpha\beta$) basis. \vspace{-0.3cm}

\begin{align}
    \hat{g}_{\vb{k}}
    =
    \hat{S}_{\vb{k}} {\hat{\mathcal{G}}_0}_{\vb{k}} \hat{S}_{\vb{k}}^{-1}
    =
    \hat{S}_{\vb{k}} (\omega^+ - {\hat{\mathcal{H}}_0}_{\vb{k}} )^{-1} \hat{S}_{\vb{k}}^{-1}
    =
    \big( \omega^+ -
        \overbrace{ \hat{S}_{\vb{k}} {\hat{\mathcal{H}}_0}_{\vb{k}} \hat{S}_{\vb{k}}^{-1}}^{{\hat{h}_{0}}_{\vb{k}}}
    \big)^{-1}
    \\[0.2cm]
    \therefore
    (\hat{g}_{\vb{k}})_{\alpha\beta}
     =
    (\hat{S}_{\vb{k}} {\hat{\mathcal{G}}_0}_{\vb{k}} \hat{S}_{\vb{k}}^{-1})_{\alpha \beta}
    =
    (\omega^+ - {\hat{h}_{0}}_{\vb{k}} )^{-1}_{\alpha \beta}
    =    (
    \underbrace{ \omega^+ - {\hat{H}_{0}} }_{\hat{G}_0}
    )^{-1}_{\vb{k} \alpha \beta}
    =
    {G_0}_{\vb{k} \alpha \beta }.
\end{align}
Fourier transforming finally from momentum- to position-space, we obtain the real-space Green's function's matrix elements

\begin{equation}
        {G_0}_{i \alpha j \beta} = \frac{1}{N} \sum_{\vb{k}}     {G_0}_{ \vb{k} \alpha \beta}  \, e^{i \vb{k} \bigcdot (\vb{r}_{i\alpha} - \vb{r}_{j \beta}) },
\end{equation}
which come now carrying pairs of both of unit cell ($i,j$) and species ($\alpha,\beta$) labels. These shall form part of the building blocks we use to work through the multi-band problem, for which we now reformulate our perturbative disordered Hartree theory.

\clearpage
This multi-band theory may be developed almost identically to that of the single-band problem, as we now proceed to show%
\footnote{In fact, simply erasing all greek species labels ($\alpha, \beta, \gamma$), band indices ($n$), and associated objects/operations (e.g. species and band sums) from the following derivation will provide essentially the exact blueprint we follow in developing the single-band theory that is presented in the main body of this manuscript.}.%
We first define the full Hamiltonian for the chemically disordered alloy as
\begin{equation}
    \hat{H} = \hat{H}_0 + \hat{V}
    \qquad ; \qquad
    \left\{
    \begin{aligned}
        \\[-0.35cm]
        \hat{H}_0
        & = \sum_{n \vb{k} } \xi_{n \vb{k}} \,  \hat{c}^{\dagger}_{n \vb{k}}\hat{c}_{n \vb{k}}
        \\
        \hat{V}
        & = \sum_{i\alpha} \widetilde{\epsilon}_{i\alpha} \, \hat{c}^{\dagger}_{i\alpha} \hat{c}_{i\alpha}
    \end{aligned}
    \right.
    \label{seq:H_SM}
\end{equation}
where we take for our bare reference system a multi-band model $\hat{H}_0$, this perhaps corresponding to the tight-binding treatment (\ref{seq:H0_multiband})\,-\,(\ref{seq:H0_multiband_diag}) of a crystal with a multi-atom basis. We then apply local perturbations $\hat{V}$ which combine the effects of both random impurities ($\epsilon_{i\alpha}$) and the local Madelung fields ($\phi_{i\alpha}$).

\begin{gather}
    \widetilde{\epsilon}_{i\alpha} = \epsilon_{i\alpha} +\phi_{i\alpha}
    \label{seq:epsTilde}
    \\
    \phi_{i\alpha} = \sum_{j\beta \neq i\alpha } V_{i\alpha j\beta}^C (n_{j\beta} - \langle n_{\beta} \rangle )
    \qquad \qquad ; \qquad \qquad
    V_{i\alpha j\beta}^C = \frac{1}{|\vb{r}_{i\alpha} - \vb{r}_{j\beta}|}.
    \label{seq:MadPot}
\end{gather}

The Green's functions for both the bare and perturbed systems may then be constructed, similarly as in \cref{subsec:PTgen} of the main text, and provide for us the local charge distribution associated with either problem

\vspace{-0.4cm}
\begin{subequations}
\begin{minipage}[t]{0.45\textwidth}
    \begin{equation}
        \hat{G}_{0} = (\omega^+ - \hat{H}_0)^{-1}
        \hspace{2.25cm}
    \end{equation}
\end{minipage}
\hfill
\begin{minipage}[t]{0.45\textwidth}
    \begin{align}
        \hat{G} = (\omega^+ - \hat{H}_0)^{-1}
        \hspace{1.cm}
    \end{align}
\end{minipage}
\end{subequations}

\vspace{-0.2cm}
\begin{subequations}
\begin{minipage}[t]{0.45\textwidth}
    \begin{equation}
        {n_0}_{i\alpha} =  -\frac{1}{\pi} \int_{-\infty}^{\mu} \textrm{Im} [  {G_0}_{i\alpha i\alpha} ] \, \dd \omega = \langle n_{\alpha} \rangle
        \label{seq:n0i}
    \end{equation}
\end{minipage}
\hfill
\begin{minipage}[t]{0.45\textwidth}
    \begin{equation}
        {n}_{i\alpha} =  -\frac{1}{\pi} \int_{-\infty}^{\mu} \textrm{Im} [  {G}_{i\alpha i\alpha} ] \, \dd \omega
        \label{seq:n}
    \end{equation}
\end{minipage}
\end{subequations}  \\

\noindent These are related through Dyson's equation
\begin{equation}
    \hat{G} =  \hat{G}_0 + \hat{G}_0 \hat{V }  \hat{G}_0 + \hat{G}_0 \hat{V}  \hat{G}_0 \hat{V}  \hat{G}_0 + \cdots
\end{equation}
which we project to real-space by assuming a matrix form, attaching appropriate unit cell ($i,j,k$) and species ($\alpha,\beta,\gamma$) indices, and using the local property of $\hat{V}$ to filter out any extraneous labels (i.e. $ (\hat{V})_{i\alpha j\beta} = \widetilde{\epsilon}_{i\alpha} \delta_{i\alpha = j \beta} $).

\begin{equation}
\begin{aligned}
G_{i\alpha i\alpha}
={}& {G_0}_{i\alpha i\alpha}
\\
&+ \sum_{j\beta}
{G_0}_{i\alpha j\beta}
\widetilde{\epsilon}_{j\beta}
{G_0}_{j\beta i\alpha}
\\
&+ \sum_{j\beta,k\gamma}
{G_0}_{i\alpha j\beta}
\widetilde{\epsilon}_{j\beta}
{G_0}_{j\beta k\gamma}
\widetilde{\epsilon}_{k\gamma}
{G_0}_{k\gamma i\alpha}
+\cdots
\end{aligned}
\end{equation}
Treating explicitly only the terms up to linear order in $\widetilde{\epsilon}$s, we first separate out the local ($i\alpha=j\beta$) component in the first sum above, then perform the frequency integral on the imaginary part to obtain the following result.

\begin{align}
G_{i\alpha i\alpha}
={}& {G_0}_{i\alpha i\alpha}
+ {G_0}_{i\alpha i\alpha}
\widetilde{\epsilon}_{i\alpha}
{G_0}_{i\alpha i\alpha}
\nonumber\\
&+ \sum_{j\beta\neq i\alpha}
{G_0}_{i\alpha j\beta}
\widetilde{\epsilon}_{j\beta}
{G_0}_{j\beta i\alpha}
+ \mathcal{O}[\widetilde{\epsilon}^2]
\nonumber\\[-0.15cm]
&\hspace{1.5cm}
-\frac{1}{\pi}
\int_{-\infty}^{\mu}
\textrm{Im}
\left[\;\Bigg\downarrow\;\right]
\,\dd\omega
\nonumber\\[-0.05cm]
n_{i\alpha}
={}&
\langle n_{\alpha}\rangle
\nonumber\\
&+
\underbrace{
\left(
-\frac{1}{\pi}
\int_{-\infty}^{\mu}
\textrm{Im}
[{G_0}^{2}_{i\alpha i\alpha}]
\,\dd\omega
\right)
}_{A_{\alpha}}
\widetilde{\epsilon}_{i\alpha}
\nonumber\\
&+
\sum_{j\beta\neq i\alpha}
\underbrace{
\left(
-\frac{1}{\pi}
\int_{-\infty}^{\mu}
\textrm{Im}
[{G_0}^{2}_{i\alpha j\beta}]
\,\dd\omega
\right)
}_{B_{i\alpha j\beta}}
\widetilde{\epsilon}_{j\beta}
\nonumber\\
&+
\mathcal{O}[\widetilde{\epsilon}^2].
\label{seq:Dyson2DH}
\end{align}

Here, we use equations (\ref{seq:n0i}) and (\ref{seq:n}) and also
exploit $\hat{H}_0$'s time reversal symmetry (implying ${G_0}_{i\alpha j\beta} = {G_0}_{j\beta i\alpha}$) to reveal the perturbative expansion coefficients $A_{\alpha}$ and $B_{i\alpha j\beta}$ above. Additionally, we note both $\langle n_{\alpha} \rangle$ and $A_{\alpha}$'s lack of a unit cell index, this reflecting the pure crystal translational invariance of these local quantities in the bare/unperturbed problem. Now this last line recovers what is analogous to formulas (\ref{eq:DH_eqn1})\,-\,(\ref{eq:coefficientBijquantum}) provided in the main text, and may be reorganized to follow (\ref{eq:DH_eqn1})'s format

\begin{equation}
    \delta n_{i\alpha} =
    n_{i\alpha} - \langle n_{\alpha} \rangle
    = A_{\alpha}  \widetilde{\epsilon}_{i\alpha}
    + \sum_{j\beta \neq i\alpha} B_{i\alpha j\beta}
    + \mathcal{O}[ \widetilde{\epsilon}^2 ].
\end{equation}
Truncating off now those expansion terms beyond linear order in $\widetilde{\epsilon}$s, what remains together with (\ref{seq:epsTilde}) and (\ref{seq:MadPot}) -- the latter of which we notice contains $\delta n_{j\beta} = n_{j\beta} - \langle n_{\beta} \rangle$ as an input function -- establishes a self-consistent set of equations we must work through in order to solve the current multi-band problem.

Fourier transforming the three prior-addressed equations, we now exchange any unit cell indices ($i,j$) for a common momentum ($\vb{k}$) index, applying the convolution theorem where appropriate. Note that, as this is done, any species indices ($\alpha,\beta$) remain nonparticipant spectators, and will thus endure the momentization procedure.
\begin{equation}
\begin{aligned}
\left.
\begin{aligned}
{\delta n}_{i\alpha}
&= A_{\alpha}\widetilde{\epsilon}_{i\alpha}
+ \sum_{j\beta\neq i\alpha}
B_{i\alpha j\beta}\widetilde{\epsilon}_{j\beta}
\\[0.1cm]
\phi_{i\alpha}
&= \sum_{j\beta\neq i\alpha}
V^C_{i\alpha j\beta}\delta n_{j\beta}
\\[0.1cm]
\widetilde{\epsilon}_{i\alpha}
&= \epsilon_{i\alpha}+\phi_{i\alpha}
\end{aligned}
\right\}
&\quad\longrightarrow
\\[0.2cm]
\left\{
\begin{aligned}
\delta n_{\vb{k}\alpha}
&= A_{\alpha}\widetilde{\epsilon}_{\vb{k}\alpha}
+ \sum_{\beta}
B_{\vb{k}\alpha\beta}
\widetilde{\epsilon}_{\vb{k}\beta}
\\[0.1cm]
\phi_{\vb{k}\alpha}
&= \sum_{\beta}
V^C_{\vb{k}\alpha\beta}
\delta n_{\vb{k}\beta}
\\[0.1cm]
\widetilde{\epsilon}_{\vb{k}\alpha}
&= \epsilon_{\vb{k}\alpha}
+\phi_{\vb{k}\alpha}
\end{aligned}
\right.
&
\end{aligned}
\label{seq:FT_sysEqns}
\end{equation}

Comparing this result with the corresponding equation (\ref{eq:FT_sysEqns}) for the single-band problem in \cref{subsec:linSC} of the main text, we see that our retention of the species index means that the current issue is no longer one that is simply algebraic in $\vb{k}$-space. Rather, the equations displayed on the right-hand side above bear a mixture of more matrix- and vector-valued forms.

Now to consolidate the various species-indexed quantities into more compact data structures, we establish below the following vector objects which depend only on momentum $\vb{k}$. Again, presuming that there are $m$ degrees of freedom per unit cell, we define
\begin{equation}
\begin{aligned}
\bm{\delta n}_{\vb{k}}
&=
\left(
\begin{array}{c}
\delta n_{\vb{k}1} \\
\vdots \\
\delta n_{\vb{k}m}
\end{array}
\right)
\qquad ; \qquad
\bm{\phi}_{\vb{k}}
=
\left(
\begin{array}{c}
\phi_{\vb{k}1} \\
\vdots \\
\phi_{\vb{k}m}
\end{array}
\right)
\\[0.3cm]
\bm{\epsilon}_{\vb{k}}
&=
\left(
\begin{array}{c}
\epsilon_{\vb{k}1} \\
\vdots \\
\epsilon_{\vb{k}m}
\end{array}
\right)
\qquad ; \qquad
\bm{\widetilde{\epsilon}}_{\vb{k}}
=
\left(
\begin{array}{c}
\widetilde{\epsilon}_{\vb{k}1} \\
\vdots \\
\widetilde{\epsilon}_{\vb{k}m}
\end{array}
\right).
\end{aligned}
\end{equation}

Similarly, the local and nonlocal part of the Lindhard function are promoted to matrices (the prior being strictly diagonal), as is the Fourierized Coulomb interaction -- given, perhaps, by a multi-band generalization of the Ewald sums discussed in the earlier \cref{ssubsec:Ewald}. Their elements, thus, are all doubly indexed according to species.

\begin{equation}
    \hat{A} =
    \left(\begin{array}{ccccc}
         A_1 & 0 & \cdots & 0
         \\
         0 & A_2 & \cdots & 0
         \\
         \vdots & \vdots &\ddots & \vdots
         \\
         0 & 0 & \cdots & A_m
    \end{array} \right)
    \qquad ; \qquad
    \hat{B}_{\vb{k}} =
    \left(\begin{array}{ccccc}
         {B_{\vb{k}}}_{11} & {B_{\vb{k}}}_{12}  & \cdots & {B_{\vb{k}}}_{1m}
         \\
         {B_{\vb{k}}}_{21} &  {B_{\vb{k}}}_{22}  & \cdots & {B_{\vb{k}}}_{2m}
         \\
         \vdots & \vdots &\ddots & \vdots
         \\
         {B_{\vb{k}}}_{m1} & {B_{\vb{k}}}_{m2}  &  \cdots & {B_{\vb{k}}}_{mm}
    \end{array} \right)
    \qquad ; \qquad
    \hat{V}^C_{\vb{k}} =
    \left(\begin{array}{ccccc}
         {V_{\vb{k}}}_{11}^C & {V_{\vb{k}}}_{12}^C& \cdots & {V_{\vb{k}}}_{1m}^C
         \\[0.1cm]
         {V_{\vb{k}}}_{21}^C & {V_{\vb{k}}}_{22}^C & \cdots & {V_{\vb{k}}}_{2m}^C
         \\
         \vdots & \vdots &\ddots & \vdots
         \\
         {V_{\vb{k}}}_{m1}^C & {V_{\vb{k}}}_{m2}^C & \cdots & {V_{\vb{k}}}_{mm}^C
    \end{array} \right).
\end{equation}

While $\hat{A}$ and $\hat{B}_{\vb{k}}$'s matrix elements are obtained through appropriate momentization of their real-space counterparts, identified in (\ref{seq:Dyson2DH}), those of the lattice interactions may be obtained directly by Ewald sums, as described in the previous \cref{ssubsec:Ewald}.

With these, the system of equations derived on (\ref{seq:FT_sysEqns})'s right-hand side may be reexpressed in the following way (compare again with (\ref{eq:FT_sysEqns}))

\begin{align}
    \left\{
    \begin{aligned}
    \bm{\delta n}_{\vb{k}} & =
    \hat{A} \bm{\widetilde{\epsilon}}_{\vb{k}} + \hat{B}_{\vb{k}} \bm{\widetilde{\epsilon}}_{\vb{k}}
    \\
    \bm{\phi}_{\vb{k}} & = \hat{V}_{\vb{k}}^C \bm{\delta n}_{\vb{k}}
    \\
    \bm{\widetilde{\epsilon}}_{\vb{k}} & = \bm{\epsilon}_{\vb{k}} + \bm{\phi}_{\vb{k}}
    \end{aligned}
    \right. ,
\end{align}
their $\vb{k}$-space solution -- analogously to their single-band counterparts obtained through standard (linear) algebra.

\begin{align}
    \bm{\delta n}_{\vb{k}}
     & = \hat{M}_{\vb{k}} \bm{\epsilon}_{\vb{k}}
    & \quad   ;  \quad &&
    \hat{M}_{\vb{k}}
    & = \left( \hat{ \mathds{1} } - (\hat{A}+\hat{B}_{\vb{k}}) \hat{V}_{\vb{k}}^C \right)^{\! -1}
    (\hat{A}+\hat{B}_{\vb{k}})
    \label{seq:δnk}
    \\
    \bm{\phi}_{\vb{k}}
    & = \hat{ \widetilde{M} }_{\vb{k}} \bm{\epsilon}_{\vb{k}}
    & \quad  ;  \quad &&
    \hat{ \widetilde{M} }_{\vb{k}}
    & = \hat{V}^C_{\vb{k}} \hat{M}_{\vb{k}}
    =  \left( { (\hat{V}^C_{\vb{k}}) }^{-1}- (\hat{A}+\hat{B}_{\vb{k}}) \right)^{\! \! -1}
    (\hat{A} + \hat{B}_{\vb{k}} ).
    \label{seq:φk}
\end{align}
Here, $\hat{ \mathds{1} }$ is an $m \times m$ identity matrix, and we see that the linear response functions $\hat{M}$ and $\hat{\widetilde{M} }$ now possess some matrix structure

\begin{equation}
\begin{aligned}
\hat{M}_{\vb{k}}
&=
\left(
\begin{array}{cccc}
{M_{\vb{k}}}_{11} & {M_{\vb{k}}}_{12}
& \cdots & {M_{\vb{k}}}_{1m}
\\
{M_{\vb{k}}}_{21} & {M_{\vb{k}}}_{22}
& \cdots & {M_{\vb{k}}}_{2m}
\\
\vdots & \vdots & \ddots & \vdots
\\
{M_{\vb{k}}}_{m1} & {M_{\vb{k}}}_{m2}
& \cdots & {M_{\vb{k}}}_{mm}
\end{array}
\right)
\qquad ;
\\[0.3cm]
\hat{\widetilde{M}}_{\vb{k}}
&=
\left(
\begin{array}{cccc}
\widetilde{M}_{\vb{k}11} &
\widetilde{M}_{\vb{k}12} &
\cdots &
\widetilde{M}_{\vb{k}1m}
\\
\widetilde{M}_{\vb{k}21} &
\widetilde{M}_{\vb{k}22} &
\cdots &
\widetilde{M}_{\vb{k}2m}
\\
\vdots & \vdots & \ddots & \vdots
\\
\widetilde{M}_{\vb{k}m1} &
\widetilde{M}_{\vb{k}m2} &
\cdots &
\widetilde{M}_{\vb{k}mm}
\end{array}
\right).
\end{aligned}
\end{equation}

Thus, (\ref{seq:δnk}) and (\ref{seq:φk}) may be rewritten in a component-wise fashion, then reverse Fourier transformed back over to real-space

\begin{alignat}{3}
  \delta n_{\vb{k}\alpha} & = \sum_{\beta} M_{\vb{k}\alpha \beta} \epsilon_{\vb{k}\beta}
	  &
    \qquad \qquad
    \longrightarrow
    \qquad \qquad
    &&
    \delta n_{i\alpha} & = \sum_{j\beta} M_{i\alpha j\beta} \epsilon_{j\beta}
	   \\
    \phi_{\vb{k}\alpha} & = \sum_{\beta} \widetilde{M}_{\vb{k}\alpha \beta} \epsilon_{\vb{k}\beta}
    &
    \qquad \qquad
    \longrightarrow
    \qquad \qquad
    &&
    \phi_{i\alpha} & = \sum_{j\beta} \widetilde{M}_{i\alpha j\beta} \epsilon_{j\beta}
    \label{tag}
\end{alignat}

providing a direct multi-band analogy to equations (\ref{eq:δni}) and (\ref{eq:φi}) of the main text -- the only formal contrast lying in the presence of some additional species ($\alpha,\beta$) indices.

\end{appendices}


\bibliographystyle{apsrev}
\bibliography{references}

\end{document}